\documentclass[10pt]{article}
\pdfoutput=1
\usepackage{amsmath}             
\usepackage{amsthm}
\usepackage{amssymb}
\usepackage{amsfonts}
\usepackage{graphicx}
\usepackage{geometry}

\renewcommand{\vec}[1]{ {\bf{#1}}}

\def\be{\begin{equation}}
\def\ee{\end{equation}}

\def\XXint#1#2#3{{\setbox0=\hbox{$#1{#2#3}{\int}$}
     \vcenter{\hbox{$#2#3$}}\kern-.5\wd0}}

\def\dist{\ell}
\def\distC{{d}}

\begin{document}
\title{Results for Capacitances and Forces in cylindrical systems}

\author{%%%% Author details
Giampiero Paffuti\\
{\small 
Dipartimento di Fisica - Universit\`a di Pisa and INFN sezione di Pisa,}\\
 {\small Largo Pontecorvo 3, I-56127 Pisa, Italy
}}
\date{}
%%%%%%%%% Insert author address here
%\address{$^1$Dipartimento di Fisica - Universit\`a di Pisa and INFN sezione di Pisa,  Largo Pontecorvo 3, I-56127 Pisa, Italy
%}

\maketitle

\begin{abstract}
In this paper we report on some new results concerning the behavior of forces between two equal circular electrodes with finite thickness. We show that for close electrodes different scenarios can result, depending on the thickness and on the
ratio of charges on the conductors. Attractive or repulsive forces can appear depending on the parameters and on the separation of the electrodes.
We give a unified description of cylindrical systems using an high precision
method based on a Galerkin expansion and we check our results with a quite sophisticated boundary element method (BEM).
We perform a preliminary study of a single cylinder (both full and hollow) to check the method and to improve
several existing results and compute some relevant parameters as the quadrupole moment and the polarizability.
\end{abstract}

\section{Introduction}
In this work we present some results on the capacitance matrix and the forces in a system of two parallel circular
conductors, taking into account their thickness. 
For close electrodes the magnitude and the sign of the force depends on the geometry and of course on the charges $Q_1, Q_2$ 
of the two conductors.
For planar discs and squares it has been shown\cite{paf,paf2,mac2,JAP}
 that at close distance the force is repulsive and logarithmically divergent, except the particular case $Q_2=-Q_1$.
 This is at variance with the case of two spheres, where the force is attractive\cite{lek1}.
 In \cite{paf2,mac2,JAP} it is discussed in an elementary way how this different behavior depends on the difference between the quadrupole-like charge distribution present for discs and the polarizability effect in the case of spheres, and absent for flat discs.

The case of planar discs has a mathematical interest but for physical applications it is important to investigate the forces
in the realistic case of non-zero thickness.
In this paper we found that the force presents a great variety of behaviors, depending on the distance, the thickness and the ratio of the two charges, $Q_1/Q_2$. One or more equilibrium points can appear and some of them can be stable.
A repulsive force is obtained at short distance in a region of the parameters., also in the case of unlike charges.

The problem of forces between two close conductors has an obvious importance in the physics of nanoelectromechanical systems
(NEMS). The measure of subtle effects as Casimir forces, requires a control on various effects, the main being
the elastic stresses and the electrostatic forces, see for example \cite{mems1,mems2}.
A precise control on the electrostatic force is helpful in these cases.

To have accurate results we have constructed a numerical/analytical 
framework working for different  problems with cylindrical symmetry.

In our computation we used two different procedures: a Galerkin projection method and a boundary elements method (BEM),
as a check for the first. The Galerkin method is an extension of similar approaches existing in literature.
We computed almost all the necessary matrix elements analytically, this improves accuracy and allows the study of
forces at very short distances.

The version of BEM we used is relatively sophisticated and we give some details on this procedure in the text.

A proper understanding of the forces between two electrodes requires the knowledge of the intrinsic parameters
for the single conductors, i.e. their capacity, quadrupole moment and polarizability.
We perform the necessary computations for a single cylinder, both for the solid case  and for the hollow cylinder.
This problem has an old tradition, starting with Maxwell\cite{maxwCyl}, and we are able to give an high precision computation
of the relevant parameters, improving some asymptotic estimates known in the literature.
The parameters of the isolated electrodes determine the long distance behavior of capacity and forces at large 
separations\cite{mac},
we verify this feature in our calculations.

The paper is organized as follows. In section \ref{sezphysdisc} we 
define the geometrical and physical quantities for the general case of a two conductor system. 

In section \ref{seznummethods} we discuss in some detail the version of the Galerkin method
and BEM used in this work. 

Section \ref{sezsingolocond} contains the numerical and partly analytical results for 
systems 
with a 
single conductor and for a particular
configuration of two flat discs. 

In section \ref{seznumris} we give our results on the capacitance matrix for the case
of two electrodes. We perform the computations for 14 thickness and for several distances between electrodes.
We collect the main numerical results in tables \ref{tabtot1}-\ref{tabtot5} at the end of the paper.

In section \ref{sezforze} we analyze the problem of forces between two cylindrical conductors.
A phase diagram showing the  separation between attractive and repulsive regime at very short distances is 
constructed.

Section \ref{sezlongdist} is devoted to the study of the capacitance matrix and forces for large separations of the electrodes.

In section \ref{sezseries} we present an application of our methods to an classic problem: the effective capacity of 
to capacitors in series. It is amusing to find that the usual textbook answer needs a correction which can be computed
easily in our framework.

In section \ref{sezmatrici} we collect 
the matrices used in the computation. Whenever possible we give an analytical closed form for the matrix elements, in particular all the matrices used in the case of a single conductor have been computed analytically.

Section \ref{sezconclusioni} gives a brief summary of our results, some possible extensions are proposed. 

\section{Physical parameters and geometry of the system\label{sezphysdisc}}
In a system of conductors the charges $Q_i$ and the potentials $V_i$ are related by the symmetric capacitance matrix $C_{ij}$
\be Q_i = \sum_j C_{ij} V_j\,;\qquad V_i = \sum_j M_{ij} Q_j \,.\label{sez1.1}\ee
$M_{ij}$, the potential matrix, being  the inverse of the matrix $C_{ij}$. In terms of these matrices the energy takes the form
\be W = \frac12\sum_{ij} C_{ij}V_i V_j = \frac12 \sum_{ij} M_{ij} Q_i Q_j\,. \label{sez1.2}\ee
In the following we specialize the above relations to the case of two equal, parallel, thick discs of radius $a$ and thickness $b = 2 h$. 
The symmetry of the problem implies $C_{11} = C_{22}$ and an analogous relation holds for $M_{ij}$.
The separation between nearest surfaces (bases of the cylinders) will be denoted by $\ell$.
In the following, when possible, we will use adimensional variables, taking $a$ as the fundamental unit of length:
$\kappa = \ell/a$, $\tau = b/a$.
We will need to compare different configurations for the system, when necessary we will write the functional dependence on $\tau, \kappa$, in the form $C_{ij}[\tau, \kappa]$. When $\tau =0$ the conductors degenerate in flat discs.

In \cite{paf,paf2,mac2,JAP} it has been pointed out that the capacitance coefficients can be organized in a hierarchy according to 
their behavior for $\kappa\to 0$. In particular the sum $C_{11}+ C_{12}$ stays finite in the limit $\kappa\to 0$, then it is useful to consider as independent quantities
\be C = \dfrac{C_{11} - C_{12}}{2}\,;\qquad C_{g_1} = C_{11}+ C_{12}\,. \label{sez1.3}\ee
$C$ is the usual relative capacity, i.e. the quotient between the charge of one conductor and the potential difference when
the two conductors are oppositely charged.
The total sum of the matrix elements $C_{ij}$ tends, in the limit $\kappa\to 0$, to the capacity of the conductor obtained by the ``fusion'' of the two elements: in the present case a cylinder of thickness $2\tau$. We will use the notation $C_1$ for this capacity, then the above statement reads
\be C_g[\tau,\kappa] = C_{11} + C_{12} + C_{21} + C_{22} \equiv 2 C_{g_1}[\tau,\kappa]
\mathop\rightarrow_{\kappa\to 0} C_1[2\tau] \,.\label{sez1.limCg}\ee

In terms of the above quantities the force between the two electrodes can be written as
\be F = - \frac{\partial}{\partial\dist} W =
\frac14 \dfrac{(Q_1+Q_2)^2}{C_{g_1}^2}\frac{\partial}{\partial \dist}C_{g_1}
+ \frac18 \dfrac{(Q_1-Q_2)^2}{C^2}\frac{\partial}{\partial \dist}C\,.
\label{sez1.forza}
\ee
The generalization for two different electrodes can be found in \cite{mac2}.
In the works mentioned above it is shown that the second term in \eqref{sez1.forza} produces a constant attractive force at short distances, for planar contacts. The first term is in general repulsive, and in the particular case of two equal discs it is logarithmically divergent\cite{paf2,mac2} for $\kappa\to 0$:
\be F_{discs}(Q_1,Q_2,\kappa) = - \frac{(Q_1-Q_2)^2}{2 a^2}  - \frac{(Q_1+Q_2)^2}{8 a^2}(1 + \log\frac\kappa\pi) \,;\qquad
\kappa\ll 1
\label{forzaQris}\ee
 In the following we will show how this behavior is smoothened by the thickness.

The results for planar discs ($\tau=0$) follow from the analytical known behavior for $\kappa\to 0$:
\be C[0,\kappa]\to  
a\left\{ \frac{1}{4\kappa} + \frac{1}{4\pi}\left[ \log\left(16\pi \frac 1 \kappa\right) - 1\right]\right\}
+ a\left\{\frac1{16\pi^2} \kappa\left[ \left(\log\frac{\kappa}{16\pi}\right)^2-2\right] \right\}\,,
\label{corrShaw}
\ee
and
\be C_{g_1}[0,\kappa] \to a\left[ \frac1\pi + \frac{\kappa}{2\pi^2}\left(1 - \log(\frac{\kappa}{\pi})\right)\right]\,.\label{valorecgasint}\ee
For the discussion of these results and the comparison with numerical computations we refer to the above references, a summary is given in \cite{paf2}, where references on the various terms of this expansion are also given.

For the case $\tau\neq 0$,
to our best knowledge,
the only known result is the original computation of Kirchhoff\cite{Kirchh,nist},
confirmed in \cite{Shaw}:
\be
C_K[\tau,\kappa]\to a\left\{
\frac1{4\kappa} -\frac1{4\pi}\left[1+\log\frac{\kappa}{16\pi} - \left(1 + \frac{\tau}{\kappa}\right)\log\left(1+\frac{\tau}{\kappa}\right)
+\frac{\tau}{\kappa}\log\frac{\tau}{\kappa}
\right]\right\} \equiv a\,f_K(\kappa,\tau)
\label{sez1.Kirch1}\ee
while for $C_{g_1}$ we have, from \eqref{sez1.limCg} 
\be C_{g_1}[\tau,\kappa]\mathop\rightarrow_{\kappa\to 0} \frac12 C_{1}[2\tau]\,. \label{sez1.cyl}\ee
Formula \eqref{sez1.Kirch1} exhibits clearly the problem introduced by the thickness. In the region
$ b\ll \dist \ll a$, i.e. $\tau \ll \kappa \ll 1$, the system is to all effects equivalent to a capacitor composed by two planar discs,
for $\kappa\ll \tau \ll 1$ the approach for $\kappa\to 0$ changes. In effect from \eqref{sez1.Kirch1} for $\tau\to 0$ 
\be C_K[\tau,\kappa] \to C[0,\kappa] \sim a\left\{ \frac1{4\kappa} - \frac1{4\pi}\left(1 + \log\frac{\kappa}{16\pi}\right)
\right\}\label{sez1.limite1}\ee
in agreement with \eqref{corrShaw}. In the limit $\kappa\to 0$ at fixed $\tau$ we have instead
\be C_K[\tau,\kappa] \to a\left\{ \frac1{4\kappa} -\frac1{2\pi}\log\kappa + \frac1{4\pi}\log\frac{\tau}{16\pi}\right\}
\label{sez1.limite2}\ee
It is worth to note that the logarithmic correction to the geometric capacitance in \eqref{sez1.limite2} has a 
doubled coefficient
with respect the analogous term in \eqref{corrShaw}.
The two limits do not commute, as far as the short distance behavior is concerned, a circumstance noticed in \cite{paf,paf2} for the analogous case of discs of different radii.
 The numerical computations below will confirm some of these expectations,
but we also find a need for an additional term in \eqref{sez1.Kirch1}, vanishing for $\tau\to0$.

Let us note that the sum of the coefficients of the logarithmic terms in \eqref{sez1.limite2} reproduces
the coefficient of $\log\kappa$ in \eqref{sez1.limite1}. This is expected as the two asymptotic regions
$\kappa\ll \tau\ll 1$ and $\tau\ll \kappa\ll 1$ must have a match and the second region on physical grounds must be described by
\eqref{sez1.limite1}. This kind of matching mechanism will appears also in the analysis of $C_{g_1}$.

In the case of a system of two identical conductors with a plane of symmetry the study of the coefficients $C, C_{g_1}$ brings also 
some simplifications from the mathematical point of view. From \eqref{sez1.1} it follows that for the determination of the two coefficients it suffices to consider the case in which the conductors are at the same potential or at the opposite potential, having in the two cases
\be Q_1 = (C_{11} + C_{12}) V = C_{g_1} V,;\qquad Q_1 = (C_{11}- C_{12}) V = 2 C V \label{defq1}\ee
The problem is reduced to the computation of $Q_1$ in these two configurations.

For a system with a symmetry plane the charge distribution at equilibrium is symmetric or antisymmetric with respect to this plane in the two cases, i.e. if on conductor 1 there is a charge density $\sigma(\vec x)$, on the reflected point $\vec x_R$ on conductor 2 there is a charge density $\sigma(\vec x_R) = \pm \sigma(\vec x)$.
The equilibrium charge in the two cases is then compited by solving two similar integral equations
\be \int_{\vec y\in S_1} \left[\dfrac{1}{|\vec x-\vec y|} \pm \dfrac{1}{|\vec x-\vec y_R|}\right]\sigma(\vec y) = V \label{defq2}\ee
The variable $\vec y$ runs on the conductor 1, together with the reflected coordinate $\vec y_R$ on the second conductor. Once computed $\sigma$ on the conductor 1 its integration gives $Q_1$ and thus the two coefficients $C_{g_1}$ and $C$ 
using \eqref{defq1}.

Let us consider the case under study, two equal discs, and choose for the upper cylinder the coordinate axis $z$ centered in the body
and positive upwards, with range $-h \leq z \leq + h$.
In our configuration the density depends only on $z$ and on the distance $r$ from the symmetry axis.
If we label the points of the 
second discs with $z$ pointing downward, in the same interval, and with the origin in medium plane of the second disc we have, by symmetry, for the two charge densities:
\[ \sigma^{(1)}(r,z) = \sigma(r, z)\,;\quad \sigma^{(2)}(r,z) = \pm \sigma(r,z) \]
i.e. the same function describes both densities. The explicit implementation of this procedure for the discs is given in section
\eqref{seznummethods}.

A similar and even simpler approach can be used for a single conductor, a single cylinder in our case. This allows the determination of the superficial charge density and the parameters $C_1$, the capacitance of the isolated conductor, and its 
longitudinal 
quadrupole moment for unit charge:
\be d_{zz} = \frac1{Q} \int_{\text{surf}} \sigma\, (2 z^2 - (x^2 + y^2)) \label{defquad}\ee
  The presence of a uniform electric field ${\cal E}$ directed along the $z$ axis do not destroy the symmetry of the problem: the equilibrium charge density induced on the cylinder will be antisymmetric with respect the median plane of the body and such to cancel the external potential. For a cylinder this means that 
the charge density $\sigma$ must determine a constant potential $\pm {\cal E} h$ on the basis and ${\cal E} z$ on the lateral surface.
Once found the charge density we can immediately found the longitudinal polarization by computing the induced dipole moment
\be \alpha_{zz}\, {\cal E} = \int_{\text{surf}} z \;\sigma \label{defalpha} \ee
In the following we put for brevity $\alpha = \alpha_{zz}, D = d_{zz}$. Apart for their intrinsic importance these parameters
are relevant for the large distance behavior of the capacitance coefficients\cite{mac}, then their precise values can help
to test the consistency of the procedure.

For a single conductor we will use often the symbol $L = 2 h \equiv a x $ to denote the length of the cylinder, while for two conductors the reader has to be careful to distinguish between the distance between nearest surfaces, denoted by $\dist = \kappa a$ as stated above, and the distance between the centers of the cylinders, denoted by $d$ in the formulas below
and in section 9.
Clearly
$ d = \dist + b \equiv \dist + 2 h \,$.
The explicit formulas for the matrix elements will be given in terms of $d$ and $h$.

\section{Numerical methods\label{seznummethods}}
In this work we present an approximate solution of equation \eqref{defq2} for the capacitor composed by two
thick discs with two different methods: the Galerkin method and a particular form of the Boundary Element Method (BEM).

\subsection{Galerkin method}
Consider the electrostatic problem \eqref{defq2} and let us expand the unknown charge density in a complete basis of functions:
\[ \sigma(x) = \sum_{n=1}^\infty c_n f_n(x) \]
Inserting in \eqref{defq2} and performing the integral we transform the integral equation in an infinite set of equations
\be \sum_{n=1}^\infty  A_n(x) c_n = V\,;\qquad A_n(x) = \int_{y} K(x,y) f_n(y) \label{gal1.1}\ee
$K$ stays for the integral kernel and $x$ is a generic point on the surface of the conductors.
Projecting again the system \eqref{gal1.1} on the basis functions we convert the initial problem in an infinite sets of linear equations
\be \sum_n A_{m n} c_n = V_m\,;\qquad A_{mn} = \int_x f_m(x) A_n(x)\,\quad V_m = \int_x f_m(x) V \label{gal1.2}\ee
Except for very particular cases the system \eqref{gal1.2} does not admit an analytical known solution. The Galerkin approximation consists in a truncation of the system, i.e. we consider only functions with $n\leq N$, in this way the problem reduces, in principle, to the solution of a system of  $N$ linear equations.

In our case we have to determine three different functions: the surface densities on the two basis, $\sigma_B^{(1)}(r), \sigma_B^{(2)}(r)$ which depend only on $r$, and the density on the lateral surface, $\sigma_L(z)$, which depends only on $z$.
For the problem at hand we choose a slight generalization of the basis suggested in \cite{FPV,Verolino}
\be \sigma_L = a V \sum a_n g_n(z)\,;\qquad\sigma_B = a V \sum_n b_n f_n(r) \,;  \label{svilupposigma}\ee
where
\begin{align}
& f_n(r) = \frac1{a^2}\Bigl(1- \frac{r^2}{a^2}\Bigr)^{p-1}  \dfrac{\Gamma(n)}{2^{p-1} \Gamma(n-1+p)} P_{n-1}^{(0,p-1)}
\Bigl(1- 2 \frac{r^2}{a^2}\Bigr)\\
& g_n(z) = \frac{1}{2\pi a h} \Bigl( 1 - \frac{z^2}{h^2}\Bigr)^{s-1/2}
\dfrac{ 2^s \Gamma(s) \Gamma(n)}{ \Gamma(n-1+2 s)} C_{n-1}^s\Bigl(\frac{z}{h}\Bigr)
\label{gegenb}
\end{align}
$P_n^{(\alpha,\beta)}$ and $C_n^s$ are respectively Jacobi and Gegenbauer polynomials.
The coefficients for the two basis will be denoted by $b_n^{(1)}, b_n^{(2)}$ while $a_n$ are the coefficients for the lateral surface.
The parameters $s, p$ are in principle arbitrary, but clearly it is better to choose them in such a way the functions embody the expected singular behavior at the edges. As the surfaces intersect at ninety degrees the expected singularity is 
of the form $\xi^{-1/3}$, where $\xi$ is the distance form the edges, see \cite{LL} (problem 3 of \textsection~3) or 
\cite{mex}. This is realized by the choice $s=1/6, p=2/3$, the choice for different systems will be specified below.
We will use this freedom in the choice of the parameters $s, p$ studying the asymptotic behaviors of the capacities: in certain schemes the matrices are dominated by the diagonal elements and 
this allows 
a simpler evaluation of the asymptotic form of the results.

The axial symmetry of the problem allows the use of a simplified form of the general Coulombic kernel in \eqref{defq2}:
\be G(r,r_0,x) = \int_0^{2\pi} \dfrac{d\Phi}{\sqrt{r^2 + r_0^2 - 2 r r_0 \cos\Phi + x^2}} \label{green1}\ee
and in terms of this kernel it is immediate to see that the condition of constant potential on the three surfaces can be written
in the form
\begin{subequations}\label{equazionigenerali}
\begin{align}
\text{Lateral surface:}
\quad& \int_{-h}^h \!\!d\mu_{z_0} \,\sigma_L(z_0)\left[ G(a,a,z-z_0) \pm G(a,a,\distC + z + z_0)\right]\nonumber\\
& +   \int_0^a \!\!d\mu_{r_0}\, \sigma^{(1)}_B(r_0) \left[G(a,r_0,z+h) \pm G(a,r_0,z+\distC-h)\right]\nonumber\\
& +   \int_0^a \!\!d\mu_{r_0}\, \sigma^{(2)}_B(r_0) \left[G(a,r_0,h-z) \pm G(a,r_0,z+\distC+h)\right] = V \label{equazioneA1}
\end{align}
\begin{align}
\text{Base 1:}\quad&  \int_{-h}^h \!\!d\mu_{z_0} \, \sigma_L(z_0) \left[G(r,a,z_0 + h) \pm G(r,a,\distC- h + z_0)\right] \nonumber\\
&+ \int_0^a \!\!d\mu_{r_0}\, \sigma^{(1)}(r_0)
\left[ G(r,r_0,0) \pm G(r,r_0,\distC -2 h)\right] \nonumber\\ 
&+ \int_0^a \!\!d\mu_{r_0}\, \sigma^{(2)}(r_0)
\left[ G(r,r_0,2h) \pm G(r,r_0, \distC )\right] = V \label{equazioneA2}
\end{align}
\begin{align}
\text{Base 2:}\quad&  \int_{-h}^h \!\!d\mu_{z_0} \, \sigma_L(z_0) \left[G(r,a,h-z_0) \pm G(r,a, \distC +h + z_0)\right] \nonumber\\
& \int_0^a \!\!d\mu_{r_0}\, \sigma^{(1)}(r_0)
\left[G(r,r_0,2h) \pm G(r,r_0,\distC)\right]\nonumber\\
&+ \int_0^a \!\!d\mu_{r_0}\, \sigma^{(2)}(r_0)
\left[ G(r,r_0,0) \pm G(r,r_0, \distC + 2h )\right] = V\label{equazioneA3}
\end{align}
\end{subequations}
In these equations and in the following we will use the symbols $d\mu_r = r dr$ and $d\mu_z = a dz$ for the measures in cylindrical coordinates.

The direct projection of these equations on the basis is not convenient due to the bad behavior of the integrand, it is better
to transform the kernel \eqref{green1} with a Fourier transform or with an Hankel transform:
\begin{align}
&G(r,r_0,x) = 
4 \int_{0}^\infty\!\! d\omega\, \cos(\omega x) I_0(\omega r) K_0(\omega r_0) =
2 \int_{-\infty}^\infty\!\! d\omega\, e^{-i \omega x} I_0(\omega r) K_0(|\omega| r_0)\,;\;\; r_0 \geq r
\label{rapgreen1}\\
& G(r,r_0,x) = 2\pi \int_0^\infty\!\! d\omega\,  J_0(\omega r) J_0(\omega r_0) e^{-\omega |x|}
\label{rapgreen2}
\end{align}
The first step in the procedure is then to substitute the expansion \eqref{svilupposigma} in \eqref{equazionigenerali} and to transform
the kernel with \eqref{rapgreen1} where appears $\sigma_L$ and with \eqref{rapgreen2} where appears $\sigma_B$. The integrals appearing in \eqref{equazionigenerali} can be performed using the known properties of Gegenbauer and Jacobi polynomials

\be 
\int_{-h}^h d\mu_z g_n(z) e^{i \omega z} =  i^{n-1}\dfrac{J_{n-1+s}(\omega h)}{(\omega h)^s}
\,;\qquad
\int_0^a d\mu_r f_n(r) J_0(\omega r) =  \dfrac{ J_{2n-2+p}(\omega a)}{(\omega a)^p}\,.
\label{relazionifg}\ee
These relations hold also with an analytic continuation in $\omega$ then can be used also for real exponential functions.
At this stage the remaining dependence of equation \eqref{equazioneA1} is on $z$ while the other two depend on $r$.
Projecting the first on $g_m(z)$ and the remaining two on $f_m(r)$ with the measures $d\mu_z$ and $d\mu_r$ we have finally the 
sought linear system
\begin{subequations}\label{sistfin0}
\begin{align}
& \sum_n A^{(0)}_{m n} a_n + \sum_nA^{(1)}_{m n} b^{(1)}_n + \sum_nA^{(2)}_{m n} b^{(2)}_n = \frac{1 }{2^s\Gamma(1+s)} \delta_{m1}
\equiv t_s \delta_{m1}
\\
&\sum_nB^{(0)}_{m n} a_n + \sum_nB^{(1)}_{m n} b^{(1)}_n + \sum_nB^{(2)}_{m n} b^{(2)}_n = \frac{1}{2^p\Gamma(p+1)}  \delta_{m1}
\equiv t_p \delta_{m1}\\
&\sum_nC^{(0)}_{m n} a_n + \sum_nC^{(1)}_{m n} b^{(1)}_n + \sum_nC^{(2)}_{m n} b^{(2)}_n = \frac{1}{2^p\Gamma(p+1)}  \delta_{m1}
\equiv t_p \delta_{m1}
\end{align}
\end{subequations}
Here and in the following we will use often for short the notation
$ t_x =1/{2^x \Gamma(1+x)}$.
Each matrix in \eqref{sistfin0} is in correspondence with the analogous term in \eqref{equazionigenerali}. The explicit calculation is tedious but straightforward, we present the final results in section \ref{sezmatrici}. 
We note that in the case of a single conductor this procedure has been developed in \cite{FPV} and \cite{Verolino}. On the numerical side the generalization given in this paper concerns
only the case of two conductors, while for a single conductor the only improvement consists in the analytical computation of the matrix elements.

Once obtained the solution the charge on the conductor 1 is 
\be Q_1 = 2\pi \left(\int_{-h}^h d\mu_z \sigma_L(z) + \int_0^a d\mu_r (\sigma_B^{(1)} + \sigma_B^{(2)})\right)
\ee
Using
\be \int_{-h}^h g_n(z) d\mu_z = \frac{1}{2^s\Gamma(1+s)}\delta_{n1}
\equiv t_s \delta_{n1}
\,;\quad \int_0^a f_n(r) d\mu_r = \frac{1}{2^p \Gamma(p+1)} \delta_{n,1}\,\equiv t_p\delta_{n1}; 
 \label{integcost}\ee
one easily finds
\be \frac1a\frac{Q_1}{V} = 2\pi \left( t_s a_1 + t_p (b^{(1)}_1 + b^{(2)}_1)\right)
\ee
As explained in the previous section $Q_1/V$ gives $C_{g_1}$ for equations with both conductors at the same potential and $2 C$ for 
conductors at opposite potential. 

Almost all matrices in \eqref{sistfin0} are given in an explicit analytic form, reported in section \ref{sezmatrici}.
We checked our computations also using a quite accurate version of Boundary Element Method, presented in the 
subsection \ref{subsBEM}.

In all our computations we take the same number $N$ of polynomials, both for the  coordinate $z$ than for the variable $r$.
In usual applications a relatively small $N$ (less than 10) gives accurate results, but as we will be interested in exploring a 
quite extreme regime, at very small distances between the electrodes, we used $N$ up to 50 at small values of $\kappa$.
In the last case the computation can be quite time-consuming and as an help we used a variant of the interpolation method
suggested in \cite{Norgren}. At ``large'' values of $\kappa$, surely for $\kappa \gtrsim 0.1$, the results acquire quickly stability with growing $N$ then the error between the computation at small $N$ and the stable value can be estimated. We noticed that the errors
all lie on a universal curve depending on $N\sqrt{k}$. In the original method, used also in \cite{paf2} the dependence was on $N k$.
This observation allows, adapting \cite{Norgren}, the following extrapolation procedure. 
\begin{itemize}
\item[a)] We order the values of $\kappa$ to be computed in decreasing order: $\kappa$: $\kappa_1, \kappa_2 \ldots$.
The computation is performed with a maximum number $N_{max}$ of polynomials in each variable.
Let us call $S_i(N)$ the numerical result obtained  for the capacity for the $i$-th term in the above sequence of $\kappa$'s using 
$N < N_{max}$ polynomials.
\item[b)] For each $i$ the best numerical result is $S_i(N_{max})$. 
For small $\kappa$ the extrapolated value $C_i$ for the capacitance is given by
\be C_i = S_i( N_{max}) + \left[C_{i-1} - S_{i-1}\left(\sqrt{\frac{\kappa_{i}}{\kappa_{i-1}}} N_{max}\right)\right] \label{proceduraN}\ee
The necessary values $S_i$ can be obtained, if needed, by linear extrapolation.
\end{itemize}
This method is particularly easy to use in the Galerkin method: once the matrix with $N= N_{max}$ has been computed, the matrices
for lower $N$ are simply extracted by dropping the appropriate lines and columns, and as the computation of the matrix is the 
time-consuming part of the calculus the advantages are clear. 
In our case the extrapolation procedure only affects the values of $C$ at very small $\kappa$, as will be shown in the next section.
In the case of the solution of Love's equation in \cite{paf2}, an extrapolation similar to \eqref{proceduraN} was mandatory as  $N$, in that case, was
the number of collocation points, and a limit $N\lesssim 50000$ was imposed by memory allocation limits.
In Galerkin method the matrices are always relatively small then the memory allocation is not a problem: the extrapolation procedure then is not mandatory but it is useful to save computer-time. In this paper the extrapolated data are mainly used to give an estimation
of the error in the computation.

Finally we note that for a part of the matrices $A^{(0)}, A^{(2)}$ in \eqref{sistfin0} we have not been able to find a closed analytical form, then the relative integrals
have been computed numerically. The integrand can be highly oscillating and after several experiments we found that the Levin algorithm~\cite{levin}, as implemented in @Mathematica\cite{math}, works quite well.

\subsection{Single conductor}
The applications of the full system \eqref{sistfin0} will be given in  
section \ref{seznumris}, while in section \ref{sezsingolocond} we present the results 
for the capacity (and  polarizability)
for a single conductor.
In this case the problem is evidently even (odd) in $z$ then it is more efficient to use a basis with Gegenbauer polynomials of definite parity.
For the computation of $C_1$ and $D$ one can use the basis
\be
 g_n(z) = \frac{1}{2\pi a h} \Bigl( 1 - \frac{z^2}{h^2}\Bigr)^{s-1/2}
\dfrac{ 2^s \Gamma(s) \Gamma(2n-1)}{ \Gamma(2n-2+2 s)} C_{2n-2}^s\Bigl(\frac{z}{h}\Bigr)\label{gegenbPari}
\ee
The system of equations takes the form
\be
\begin{split}
\sum_n A^{(0)}_{mn} a_n + \sum_n A^{(1)}_{mn} b_n = t_s \delta_{m1}\\
\sum_n B^{(0)}_{mn} a_n + \sum_n B^{(1)}_{mn} b_n = t_p  \delta_{m1}
\end{split}
\label{equazionepari}
\ee
The matrices are given in section \ref{sezmatrici}.
The capacity and the quadrupole moment are given by
\begin{subequations}\label{defc1d}
\begin{align}
\frac{C_1}{a} =& 2\pi \left( t_s a_1 +  2 t_p b_1\right)\\
\frac{C_1}{a}\frac{D}{a^2} =& 
\frac{2\pi}{2^s \Gamma(1+s)}\left[- a_1 + \left(\frac{h}{a}\right)^2 \left( \frac{a_1}{1+s} + 
\frac{a_2}{(1+s)(2+s)}\right)\right]+\\
& \frac{4\pi}{2^p\Gamma(1+p)}\left[b_1\left(2 \left(\frac{h}{a}\right)^2 - \frac{1}{1+p}\right) + \frac{b_2}{(1+p)(2+p)}
\right]\nonumber
\end{align}
\end{subequations}
For the computation of the polarizability one proceed as follows. An external electric field ${\cal E}$ directed along $z$ produces on the surface of the conductor a potential $- {\cal E} z$, the point $z=0$ being at the center of the conductor. The surface density must produce the opposite potential, i.e. the right hand side of \eqref{defq2} must be $+{\cal E} z$. For a unitary electric field the induced dipole
gives directly the polarization, see \eqref{defalpha}. The charge density is obviously odd in $z$, then only the odd Gegenbauer polynomials are needed. 
The basis of odd functions is enumerated by
\be
 g_n(z) = \frac{1}{2\pi a h} \Bigl( 1 - \frac{z^2}{h^2}\Bigr)^{s-1/2}
\dfrac{ 2^s \Gamma(s) \Gamma(2n)}{ \Gamma(2n-1+2 s)} C_{2n-1}^s\Bigl(\frac{z}{h}\Bigr)\label{gegenbdisPari}
\ee
while the linear system, projecting the modified equation \eqref{defq2}, has the form
\be
\begin{split}
&\sum_n A^{(0)}_{mn} a_n + \sum_n A^{(1)}_{mn} b_n = \frac{1 }{2^{1+s}\Gamma(1+s)} \frac{h}{a}\delta_{m1}\\
&\sum_n B^{(0)}_{mn} a_n + \sum_n B^{(1)}_{mn} b_n = \frac{1}{2^p\Gamma(p+1)} \frac{h}{a} \delta_{m1}
\end{split}
\label{equazionedispari}
\ee
The matrices are given in section \ref{sezmatrici}
and the polarizability is
\be \frac{\alpha}{a^3} = 2\pi \frac{h}{a} \frac{1 }{2^{1+s}\Gamma(2+s)}  a_1 + 4\pi \frac{h}{a} \frac{1}{2^p\Gamma(p+1)} b_1 
\label{espressionealpha}
\ee

The previous equations \eqref{sistfin0},
\eqref{equazionepari} and \eqref{equazionedispari} can be specialized to describe particular sub-systems, in particular they can describe two hollow cylinders or one hollow cyclinder, described by the matrix $A^{(0)}$, or the system of two flat discs, where only matrices of the $B$-type occurs. 
For hollow cylinders
 only the Gegenbauer polynomials are needed while for discs only Jacobi polynomials are used. The edge of these systems require a charge density with a behavior $\xi^{-1/2}$, where $\xi$ is the distance from the edge, see references \cite{LL, mex} cited above, then for the two systems the natural choice for the parameters of the polynomials is $s=0, p=1/2$ respectively.
 
To assist the reader we list in section \ref{sezmatrici} the matrices adapted to these particular cases.

\subsection{Boundary element method\label{subsBEM}}
In the boundary element method (BEM) the surfaces of the conductors are divided  into elementary domains (plaquettes) of area $A_i$ and to each plaquette is assigned a charge $q_i$. The distribution of charge must reproduce the assigned potential on the body, i.e. a constant potential for a single conductor. The problem then reduces to a system of linear equations
\be V_i = \sum_{j} K_{ij} q_j \label{bem1.1}\ee
the sum runs on all plaquettes.
Various implementations of the method differ in the choice of the kernel $K_{ij}$. We choose to consider the charges uniformly distributed on the plaquettes and we perform a mean on the $i$-th plaquette where the potential is computed, i.e. the system
\eqref{bem1.1} takes the form
\be V_i = \sum_j \frac1{A_i A_j} \int_{\vec x \in S_i} \int_{\vec y \in S_j} \dfrac{1}{|\vec x - \vec y|}\, q_j 
\label{bem1.2}\ee
This choice of the kernel is equivalent to a variational calculation with  piecewise constant functions, as it is easily seen by considering the energy of the system, see \cite{JAP}. This implies that the computed capacities are a lower bound of the true result
and that the numerical estimates must be growing for finer grids of plaquettes.

For axial systems like cylinders we have two kind of surfaces: discs and cylindrical lateral surfaces. The discs are divided in
annuli from a starting radius $r_i$ to $r_i + dr_i$, the lateral surface is divided into rings from $z_i$ to $z_i+dz_i$.
Here and in the following the axis $z$ is directed along the symmetry axis of the system.

In actual calculations we choose all rings equal, i.e. $dz_i = dz\;\forall i$, and all annuli with the same area.
We have then three different basic forms for the kernel: interaction between two rings of the lateral surface, $K^{(LL)}$,
interaction between two annuli of the bases, $K^{(BB)}$, and finally the mixed term, $K^{(BL)}$. In the class of problems under study all single conductors have the same radius $a$, we can always assume $a=1$ for dimensional reasons, all formulas below are
given in these units.

A straightforward calculation gives
\be
K^{(LL)}_{ij} = \frac1{2\pi\,dz^2} \int_0^{dz} d\xi \int_0^{dz} d\eta \,\dfrac{4}{\sqrt{(z_i- z_j + \xi - \eta)^2}}
\textbf{K}\left[- \frac{4}{(z_i + \xi - z_j - \eta)^2}\right]
\label{KLL}
\ee
$\textbf{K}$ is the elliptic integral. The two rings start at coordinates $z_i, z_j$. For $K^{(BB)}$:
\be
K^{(BB)}_{ij} = \frac2\pi \frac1{a_i}\frac1{a_j}
\int_{0}^{dr_j}d\xi\,\left[\Phi(r_j+\xi,r_i+dr_i,z) - \Phi(r_j+\xi,r_i,z)\right]
\label{KBB}
\ee
where
\[ a_i = {2 r_i dr_i + dr_i^2}\,.\]
$\Phi(x,R,z)$ is proportional to the electrostatic potential of a disc of radius $R$ at a point at distance $x$ from the axis and coordinate $z$ along the symmetry axis:
\begin{align} \Phi(x,R,z)
= 
&2 \pi  x \left(-\frac{\left| z\right|  \left(\sqrt{x^2+z^2}+x\right)}{\sqrt{2 x
   \left(\sqrt{x^2+z^2}+x\right)+z^2}}  \right.\label{potphi}\\
   &\left.
   +\frac{\left(R^2-x^2-z^2\right) \pmb{K}\left(\frac{4 R
   x}{(R+x)^2+z^2}\right)+\left((R+x)^2+z^2\right) \pmb{E}\left(\frac{4 R
   x}{(R+x)^2+z^2}\right)}{\pi  \sqrt{(R+x)^2+z^2}}
   \right.\nonumber\\ &\left. 
   +\frac{\left(z^2-(R+x)
   \left(\sqrt{x^2+z^2}-x\right)\right) \pmb{\Pi} \left(\frac{2 x
   \left(\sqrt{x^2+z^2}-x\right)}{z^2}|\frac{4 R x}{(R+x)^2+z^2}\right)}{\pi 
   \sqrt{(R+x)^2+z^2}}+ \right.\nonumber
   \end{align}
   \begin{align}
  & \left.
   \frac{\left((R+x) \left(\sqrt{x^2+z^2}+x\right)
   +z^2\right) \pmb{\Pi}
   \left(-\frac{2 x \left(x+\sqrt{x^2+z^2}\right)}{z^2}|\frac{4 R
   x}{(R+x)^2+z^2}\right)}{\pi  \sqrt{(R+x)^2+z^2}}\right)\nonumber
\end{align}
$\pmb{\Pi, E, K}$ are elliptic integrals, defined by
\[ \pmb{K}(z) = \int_0^{\pi/2}\dfrac{dt}{\sqrt{1-z\sin^2t}}\,;\quad
\pmb{E}(z) = \int_0^{\pi/2}dt\,{\sqrt{1-z\sin^2t}}\]
\[ \pmb{\Pi}(n|m) = \int_0^{\pi/2}\dfrac{dt}{(1-n\sin^2t)\sqrt{1-m\sin^2t}} \]

The case of planar annuli is given by $z=0$.
The prefactor $2\pi x$ in \eqref{potphi} comes from the integration measure in \eqref{KBB}.

Finally the factors $K^{(BL)}$ are given by
\be
K^{(BL)}_{ij} = \frac1{\pi dz} \frac1{2 r_i dr_i + dr_i^2}
\int_0^{dz}d\eta 
\left[\Phi(1,r_i+dr_i,z_j+\eta) - \Phi(1,r_i,z_j+\eta)\right]
\label{KBL}
\ee
Here $z_j$ is the distance between the base and the lateral ring, extending from $z_j$ to $z_j+dz$. The first argument $1$ in $\Phi$ is due to our choice of units ($a=1$).

The integrals in \eqref{KLL}, \eqref{KBB} and \eqref{KBL} have to be done numerically: in our computations we used a gaussian integration routine. Actually the elements $K^{(BB)}_{ij}$ for $z=0$ can be computed analytically but we omit here the long resulting expression as a numerical computation of the integrals gives satisfactory results.

Starting from the elements given above it is easy to assemble the whole matrix $K_{ij}$. 

For a single conductor one has to solve the system \eqref{bem1.2} with $V_i=1,\;\forall i$, in this case the sum of charges $q_i$ gives directly the capacity.

For two conductors it is simpler to break the matrix elements $K_{ij}$ in two parts, $A_{ij}$ for the self interaction of a conductor,
$B_{ij}$ for the mutual interaction, i.e. $B$ is the part of the matrix $K$ which depends on the distance between the conductors.
In this way we realize the decomposition \eqref{defq2} and solving the two systems
\be (A_{ij} \pm B_{ij}) q_j = 1 \label{sistemaBemAB}\ee
we can compute $C$ and $C_g$ for the system, as explained in section~\ref{sezphysdisc} and in parallel with the strategy used with the Galerkin apprach.

In each computation we choose $N_r$ annuli and $N_z$ lateral rings (when both are present). To avoid the introduction of
new parameters all computations for full cylinders have been done with a fixed ratio $N_r/N_z = 10$.
As always for BEM the final results have to be extrapolated by intermediate calculations with growing $N$, we used a quadratic 
fit in $1/N$, here $N$ means $N_r$ for cylinders and discs and  $N_z$ for hollow cylinders.

The BEM is reasonably accurate and will be a constant check of our calculations with Galerkin method. It has to be noted that for particular systems, like an hollow cylinder, the solution of the problem requires only a number of operations proportional to $N_z$,  in this case the method is very fast.

\section{Results for a single conductor\label{sezsingolocond}}
In this section we present the results obtained for three different systems: a cylinder of radius $a$ and length $L=2 h$, an hollow cylinder with the same geometry and a system of two parallel flat discs of radius $a$ at distance $L=2 h$, held at the same potential. The purpose is twofold: to check the accuracy of the procedure and to provide some slightly improved 
results  for the asymptotic behaviors these widely studied systems.

The capacities for the systems will be denoted respectively by $C, C^{(H)}, C^{(D)}$. Once divided by $a$ all quantities depends only on the ratio $x = 2h/a$. As the surface of the hollow cylinder and of the two discs are subsets of the surface of the full cylinder, then the Dirichlet principle implies that, at fixed $a$:
\be  C(x) \geq  C^{(H)}(x)\,;\qquad  C(x) \geq  C^{(D)}(x)\label{numris1.1}\ee
$C^{(D)}$ is the total capacitance of two flat discs, i.e. adapting the notation used in \cite{paf,paf2}
\be C^{(D)}(x) = C_g(x) = 2 C_{g_1}(x) \label{numris1.2}\ee
$C_{g_1}$ is the charge on a single disc in the  chosen configuration, and it is the result of the computation performed by solving the linear system with the matrix \eqref{matricedischi}.

Before starting let us remind some elementary facts which can be useful as a guide in the following results, at least for the 
non-expert readers. The system composed by two equal discs at the same potential clearly collapse into a single disc as their distance goes to zero, while when the discs are far apart their energy is identical to two point-like charges at distance $d$. This allows
an immediate determination of the capacity of the system in the two limit cases, using the known capacity of a single disc of radius $a$: $C_1 = 2a/\pi$, see equations \eqref{risnum1.10b} and \eqref{risnum1.10a} below.
The second kind of conductors considered in the sequel have the geometry of a cylinder. Since the pioneering work of Maxwell~\cite{maxwCyl}
it is known that the equilibrium charge density per unit length distribution of such a systems for great lengths $L=2h$ (or equivalently small radius) is approximatively constant $\lambda\simeq Q/L$ along the axis of the body, that we call $z$ axis. The potential can be estimated be computing it at the center and consequently computing the leading order of the capacity:
\be V = \int_{-h}^h \frac{Q}{L}\dfrac{dz}{\sqrt{a^2 + z^2}} 
\frac{Q}{L} 2\log\left[\dfrac{h+\sqrt{a^2+h^2}}{a}\right] 
\simeq \frac{Q}{L}2 \log\frac{L}{a}\,;\quad\Rightarrow\quad C\simeq \dfrac{L}{2\log\frac{L}{a}}
\label{stima1}\ee
The same density gives for the quadrupole moment for unit charge
\be D = \frac1Q\int_{-h}^h \frac{Q}{L} (2z^2 - (x^2+y^2)) dz
\sim\,\frac{L^2}{6}\label{stima2}\ee
Likewise if the system is subjected to an electric field ${\cal E}$ along $z$ the necessary linear charge distribution suited to 
cancel the longitudinal component of the field along the conductor can be assumed a linear function of $z$, $\lambda= k z$. The resulting field, at $z\sim 0$ is
\[ \int_{-h}^h dz \,{k \, z} \dfrac{z}{(a^2+ z^2)^{3/2}}\simeq 2 k \log(L/a) \]
This field must cancel the external field ${\cal E}$, then we obtain $k = {\cal E}/(2 \log(L/a)$. Computing the resulting induced dipole 
we find the polarizablity
\be d = \int_{-h}^h (k z) z \, dz = \frac1{24} \dfrac{L^3}{\log(L/a)}\,{\cal E}
\,;\quad\Rightarrow\quad \alpha \simeq \frac1{24} \dfrac{L^3}{\log(L/a)}\label{stima3}\ee
These three  estimates, (\ref{stima1},\,\ref{stima2},\,\ref{stima3}) must be obviously reproduced by our computations, both numerical and analytical. We remembered here these elementary facts to show that the physics beyond the formulas is quite clear.

We performed a computation using Galerkin method on a quite large range of distances $x = 5\times 10^{-6}$ to $x=500$, 
spanned with 155 values for $x$.
We have checked the results at selected points using the BEM method. Some of these results are collected in table~\ref{tabella1}.
A particular attention has been devoted to the case of an hollow cylinder, where we performed the computations up to the 
rather unrealistic value of $x = 10^5$ to check the asymptotic form of the capacity.

\begin{table}[!ht]
\begin{center}
\begin{tabular}{lrrrr}
$x=L/a$ & $C/a$\;\;\; & $C^{(H)}/a$ & $C^{(D)}/a$ & Love's Eq.\\
 500. & 42.76810 & 42.75412 &   &   \\
 400. & 35.59436 & 35.57994 &   &   \\
 300. & 28.16081 & 28.14578 &   &   \\
 200. & 20.34812 & 20.33215 &   &   \\
 160. & 17.06563 & 17.04910 &   &   \\
 100. & 11.87275 & 11.85490 &   &   \\
 75. & 9.563945 & 9.545201 &   &   \\
 50. & 7.112831 & 7.092673 &   &   \\
 30. & 4.980934 & 4.958717 &   &   \\
 25. & 4.408806 & 4.385767 &   &   \\
 20. & 3.812776 & 3.788663 &   &   \\
 17.5 & 3.503351 & 3.478557 &   &   \\
 15. & 3.184355 & 3.158738 &   &   \\
 12.5 & 2.853574 & 2.826927 &   &   \\
 10. & 2.507702 & 2.479711 &   &   \\
 7.5 & 2.141370 & 2.111499 &   &   \\
 5. & 1.744592 & 1.711773 &   &   \\
 2.5 & 1.293504 & 1.254713 &   &   \\
 1. & 0.9639434 & 0.9121775 & 0.8800721688 & 0.8800721688 \\
 0.5 & 0.8281367 & 0.7569050 & 0.7895926357 & 0.7895926356 \\
 0.1 & 0.6894760 & 0.5446842 & 0.6823068816 & 0.6823068816 \\
 0.01 & 0.6441727 & 0.3892495 & 0.6434688952 & 0.6434688952 \\
 0.001 & 0.6376071 & 0.3028482 & 0.6375371187 & 0.6375371188 \\
 0.0001 & 0.6367396 & 0.2478364 & 0.6367348250 & 0.6367348250 
\end{tabular}
\end{center}
\caption{ Capacity for the cylinder (C), the hollow cylinder (H) of length $L$ and radius $a$ and for a two discs system (D) of radius $a$ at distance $L$. The last column gives the values of $C^{D}$ computed by solving a kind Love's equation, see\cite{paf2}.
\label{tabella1}}
\end{table}
The results in table \ref{tabella1} satisfy the bound  \eqref{numris1.2}.

As a side-product of the computation we give the values of the longitudinal component of the quadrupole moment for unit charge and of the 
longitudinal polarization for the cylinder and the hollow cylinder in table~\ref{rabelldalpha}.

\begin{table}[!ht]
\[
\begin{array}{lllll}
x = L/a & D/a^2 & D^{H}/a^2 & \alpha/a^3 & \alpha^{H}/a^3\\
 500. & 45687.70 & 45651.89 & 1.167744\times 10^6 & 1.166463\times 10^6 \\
 400. & 29410.08 & 29381.29 & 630204.2 & 629342.9 \\
 300. & 16682.71 & 16660.95 & 285764.2 & 285246.4 \\
 200. & 7518.342 & 7503.643 & 94618.77 & 94364.02 \\
 160. & 4854.461 & 4842.602 & 51774.29 & 51601.06 \\
 100. & 1938.144 & 1930.570 & 14752.92 & 14675.05 \\
 75. & 1107.552 & 1101.776 & 6918.874 & 6870.668 \\
 50. & 505.0383 & 501.0748 & 2421.176 & 2396.278 \\
 30. & 188.8376 & 186.3425 & 667.2881 & 656.1230 \\
 25. & 133.0866 & 130.9631 & 425.7766 & 417.3096 \\
 20. & 86.76685 & 85.01794 & 247.7982 & 241.7154 \\
 17.5 & 67.17089 & 65.61072 & 180.1042 & 175.0915 \\
 15. & 49.97016 & 48.60001 & 125.2074 & 121.1802 \\
 12.5 & 35.18392 & 34.00543 & 82.01799 & 78.88933 \\
 10. & 22.83773 & 21.85311 & 49.40666 & 47.08602 \\
 7.5 & 12.96784 & 12.18025 & 26.19374 & 24.58639 \\
 5. & 5.631251 & 5.045814 & 11.13360 & 10.13920 \\
 2.5 & 0.9372851 & 0.5636154 & 2.888484 & 2.400048 \\
 1. & -0.4950776 & -0.7483552 & 0.6081524 & 0.3859648 \\
 0.5 & -0.6971090 & -0.9373046 & 0.2188476 & 0.09746500 \\
 0.1 & -0.7123570 & -0.9974994 & 0.02956084 & 0.003924835 \\
 0.01 & -0.6761311 & -0.9999750 & 0.002561868 & 3.926955\times10^{-5} \\
 0.001 & -0.6680985 & -0.9999997 & 0.0002507951 &
   3.926990\times10^{-7} \\
 0.0001 & -0.6668541 & -1.000000 & 0.00002500811 &
   3.926991\times10^{-9}\\\end{array}
\]
\caption{Quadrupole moment and polarization for the cylinder and for the hollow cylinder (denoted by the suffix $(H)$).
\label{rabelldalpha}}
\end{table}

We give now a few details on the parameters used in the computation. For the hollow cylinder and for the discs we have only a ``one block'' matrix and the only parameter is the dimension, $N_1\times N_1$, fixed by the number of polynomials used. For the whole cylinder we have a freedom in varying the dimension of the block relative to the radial variable, $N_R$, and the dimension relative to the lateral surface, $N_L$, for a total dimension of the matrix
$(N_L+N_R)\times (N_L+N_R)$. In all the computation we have chosen $N_L = N_R$. 
For all practical purposes a low value of $N_1, N_R$, say in the range 5-10, is sufficient, but as we want to explore some extreme regimes and compare the results with available theoretical expectations we push these parameters to higher values. The data shown in tables have been obtained in general with $N_R = 20$ (and at least 25 for small thickness) for the cylinder, and
$N_1$ between 10 and 40
 for the hollow cylinder 
with $x\leq 1$, and $N_1 = 30$ for $x>1$. For two discs we used
$N_1 = 500$. We have verified that with these choices all 
the digits given in the tables appear to stabilize with growing $N$.

As an additional check we have compared the results with the simpler BEM method, based on an entirely different algorithm and within the errors of this last approach the two methods give identical results.

All the matrices involved in this computation can be computed both numerically and analytically, we have checked again the agreement and we chose to work with the analytical version in order to avoid any possible numerical error induced by the computation of integrals. When necessary all special functions appearing in the matrix elements have been computed with high precision arithmetic using the software @Mathematica\cite{math}.

\begin{figure}[ht]
\begin{center}
\includegraphics[width=0.8\textwidth]{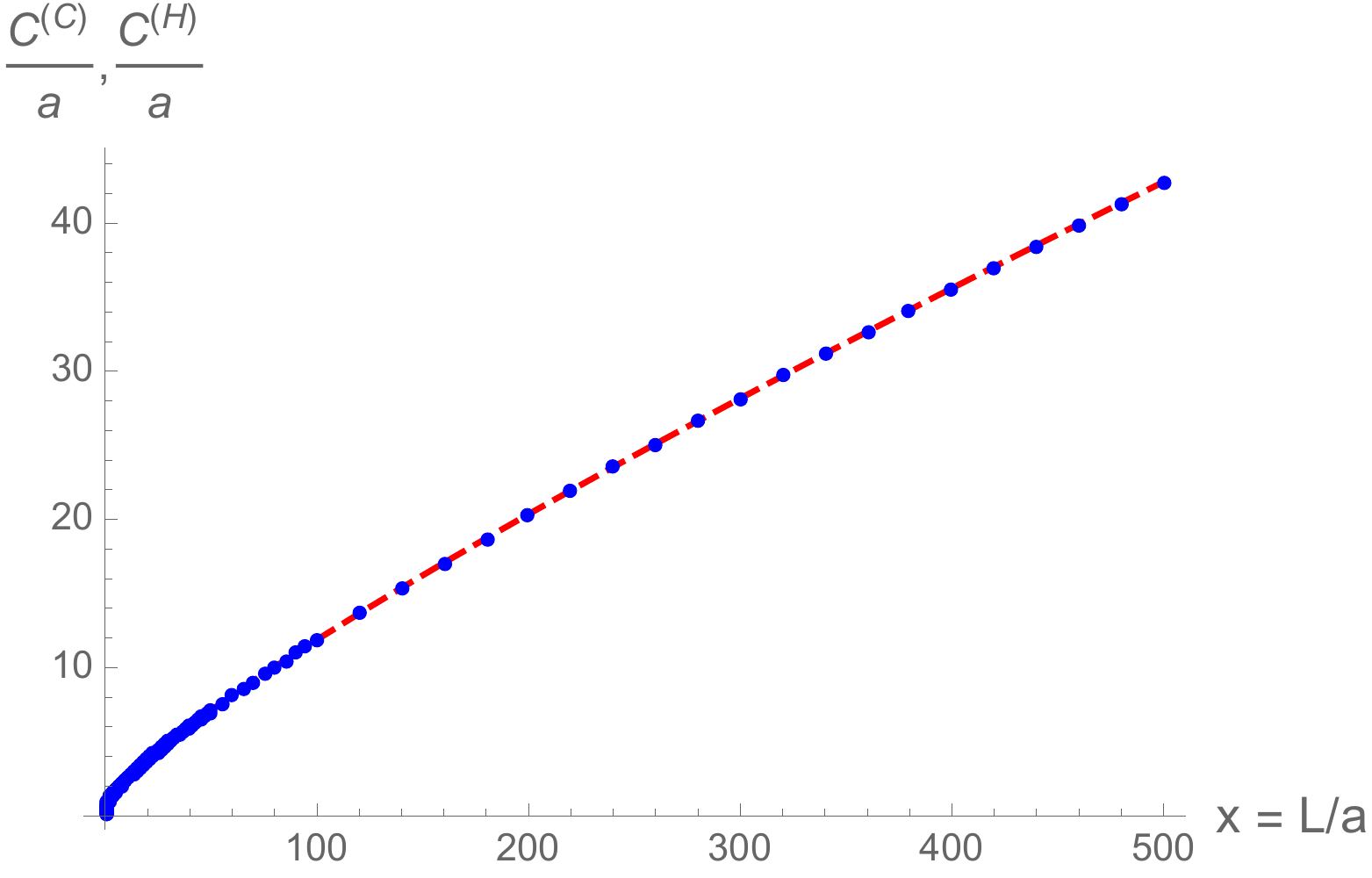}
\caption{
An overall view of the capacities for the cylinder (points) and the hollow cylinder (dashed line). On this scale the two results
are practically identical.
\label{figTot}}
\end{center}
\end{figure}

Finally we have to choose the parameters $s, p$ appearing in the expansions in Gegenbauer polynomials and Jacobi polynomials.
In principle the choice is arbitrary but clearly the best choice is the one which reproduces the correct edge behavior for the charge densities. For flat surfaces, i.e. hollow cylinder and discs, this means $s=0, p=1/2$ respectively, while for the whole cylinder the boundaries meet at an angle of $\pi/2$ and the expected singular is reproduced with the couple $(s=1/6, p=2/3)$.
We have in any case tested on selected points that the results are the same with different values of these parameters, but as expected the rate of convergence of the results with growing $N$ worsens.
An overall view of the results for capacities is given in figure~\ref{figTot}.

Let us now briefly discuss the relevant results for the separate systems.

\subsection{Two flat discs}
The results are in complete agreement with the results obtained by the solution of Love's equations, see\cite{paf2}, and at very small distances the results are even slightly more accurate, using $N_1 = 500$. We do not need, for this reason, to repeat here the comparison with theoretical expectations at 
small distances:
\be
\frac1aC_{g1} = \frac1{2a} C^{(D)} \sim 
\frac1\pi +\frac{x}{2\pi^2}\left(\log\frac\pi{x} + 1\right)\,.
\label{risnum1.10b}\ee
and
at
large distances:
\be
\frac1a C_{g1} = \frac1{2a} C^{(D)} \sim 
\frac{2}{\pi } -\frac{4}{\pi^2 x} +\frac{8}{\pi ^3 x^2} +\frac{8 \left(\pi ^2-6\right)}{3 \pi ^4 x^3}\,.
\label{risnum1.10a}\ee
It is instead instructive to investigate whether these asymptotic behaviors can be obtained directly from the matrix equations.
For the two discs system the problem is reduced to a matrix ${\cal M}_{ij}$ with a single block, given in \eqref{matricedischi}.
In both limits the matrix is dominated by the diagonal part and a simple iterative solution of the system
\[ \sum_j {\cal M}_{ij} b_j  = t_p \delta_{i1} \]
can be given. At the leading order $b_1= t_p/{\cal M}_{11} $ and $b_\alpha = 0\,\forall \alpha >1$. Substituting the value
of $b_1$ in the equation for $b_\alpha$ one  easily obtains the correction to the leading result and finally
\be C_{g_1} = 2\pi b_1 = 2\pi t_p^2 \dfrac{1}{{\cal M}_{11} - \sum_{\alpha>1} {\cal M}_{1\alpha}^2/{\cal M}_{\alpha\alpha}}
\label{cg1viaM}\ee
At large distances we used $p=1/2$ and the relevant matrix elements are
\begin{align*}
&{\cal M}_{mm} \simeq 
\left\{\frac{60 x^4-40 x^2+64}{15 x^5}+2 \pi ,\frac{32}{75 x^5}+\frac{2 \pi
   }{5},\frac{2 \pi }{9},\frac{2 \pi }{13},\frac{2 \pi }{17},\frac{2 \pi
   }{21} \,, \ldots\right\} \to \dfrac{2\pi}{4m - 3}\\
&{\cal M}_{1 i} = \left\{\frac{60 x^4-40 x^2+64}{15 x^5}+2 \pi ,\frac{8 \left(7 x^2-20\right)}{105
   x^5},\frac{32}{315 x^5}\,\;\ldots\right\}
\end{align*}
Substituting in \eqref{cg1viaM} and expanding in powers of $1/x$ we obtain immediately \eqref{risnum1.10a}.
For the opposite limit the situation is more difficult.
By expanding in $x$ the matrix we have, for $x\to 0$:
\[ {\cal M}_{11} \simeq 4\pi -2 x \left(1 + \log2 + \log{x}\right) \]
and the leading order of \eqref{cg1viaM}
\be \frac1a C_{g_1} \simeq \frac1\pi + x\left(\log\frac1x + \log2 + 1\right) \ee
which has the correct leading behavior but a \textit{wrong} subleading constant term. The na\"ive expectation that the correction term
\eqref{cg1viaM} gives the correct answer fails as the sum of leading asymptotic terms diverges. We have found no easy way to
compute the known $\log\pi$ constant in \eqref{risnum1.10b}.
\subsection{Hollow cylinder}
This is one of the most studied systems  in electrostatics. An important work on this topic goes back to Kapitza et al.\cite{Kap} where the basis
for an analytical work is first proposed. Approximate formulas for the capacity has been developed in \cite{vain1,vain2} and subsequently rediscovered and reworked in \cite{jack1}. 
In our approach we have to solve the by now familiar kind of system of linear equations
\[ \sum_j {\cal M}_{ij} a_j = t_s \delta_{i1} \]
where the matrix ${\cal M}$ is given in \eqref{uncilindrovuoto}, with $s=0$, to satisfy the correct edge behavior. A practically identical system has been considered in \cite{Verolino}, where the matrix elements were computed by numerical integration.
The asymptotic analysis runs parallel to the derivation of \eqref{cg1viaM}, i.e.
\be C^{(H)} = 2\pi a_1 = 2\pi t_s^2 \dfrac{1}{{\cal M}_{11} - \sum_{\alpha>1} {\cal M}_{1\alpha}^2/{\cal M}_{\alpha\alpha}}
\label{ChshortdaM}\ee
\subsubsection*{Short cylinder}
For small values of $x$ a straightforward expansion of the matrix gives
\begin{subequations}\label{mHsmallx}
\begin{align}
 {\cal M}_{11} &= 2 \log(\frac{32}{x}) + \frac{x^2}{32}(2- \log(\frac{32}{x}) \,;\\
 {\cal M}_{n n} & = \left\{\frac{1}{384}
   \left(x^2+192\right),\frac{x^2+960}{3840},\frac{x^2}{13440}+\frac{1}{6},\frac{
   x^2}{32256}+\frac{1}{8},\frac{x^2+6336}{63360}, \ldots\right\}\,;\quad n \geq 2\\
  {\cal M}_{1n} &= \left\{x^2 \left(\frac{1}{128} \log
   \left(\frac{x}{32}\right)+\frac{5}{256}\right),{\cal O}(x^4),{\cal O}(x^4),\ldots\right\}\,;\quad n\geq 2
\end{align}
\end{subequations}
and substitution in \eqref{ChshortdaM} gives
\be \frac1a C^{(H)} \mathop{\sim}_{x\to 0}\;\; \dfrac{\pi}{\log(32/x) + \frac{x^2}{64}(2-\log\frac{32}{x})}\,.\label{risnum1.hollowth1impr}\ee
This result has been first obtained by Lebedev and Skal'skaya \cite{leb} using the technique of dual integral equations.

It is tempting to consider the limit $x\to 0$ as a regularization for a circular conducting wire of radius $R = a$, 
\eqref{risnum1.hollowth1impr} suggests that the capacitance for this system is
\be C_{wire} \simeq \pi  R /\log\frac R\mu \label{capacitafilo}\ee
where $\mu$ is a cutoff depending on the section of the wire. Equation \eqref{capacitafilo} is supported by two different models
\begin{itemize}
\item[a)] We can regularize the wire by introducing a cutof $\mu$ in the Coulomb law, in this approach the potential for a
ring of radius $R$ is
\[ V = \frac{Q}{2\pi R} \int_0^{2\pi} \dfrac{R d\varphi}{\sqrt{2 R^2(1-\cos\varphi) + \mu^2}}\]
and the leading order as $\mu\to 0$ gives \eqref{capacitafilo} for the capacity $C = Q/V$.
\item[b)] On physical ground one expects that the leading behavior for the capacity of the wire is identical
to the one for a infinitely long cylinder of length $L = 2\pi R$ and radius $\mu$, this amounts to neglect the small 
effect of boundary charges. 
The value \eqref{capacitafilo} is obtained by substituting $\mu = a$
 in the asymptotic formula \eqref{chasint} given below and considering a cylinder of length $L = 2\pi R$.
\end{itemize}
For the particular case of a toroidal conductor, with central radius $R$ and a circular section of radius $\rho$, the 
capacity is given by\cite{hicks,dyson,loh}
\be C = 4\pi\sqrt{R^2-\mu^2}\left( \frac12 \dfrac{Q_{-1/2}(R/\rho)}{P_{-1/2}(R/\rho)}
+
\sum_{n=1}^\infty \dfrac{Q_{n-1/2}(R/\rho)}{P_{n-1/2}(R/\rho)}
\right)
\label{ctoro}\ee
$Q_s, P_s$ are Legendre functions.
The asymptotic expansion of \eqref{ctoro}, dominated by the first term, gives,
in the limit $R\gg \rho$
\be C\simeq \dfrac{\pi R}{\log(8 R/\rho)} \label{ctoro2}\ee
confirming again equation \eqref{capacitafilo}. The accuracy of the limit \eqref{ctoro2} is confirmed by the numerical
evaluation of \eqref{ctoro}.

\subsubsection*{Long cylinder}
Let us consider now the limit of a very long cylinder, i.e. large $x = L/a$.
For large $x$ the matrix is tractable only in the scheme $s=1/2$. In this case one find
\begin{subequations}\label{valoreMperH}
\begin{align}
{\cal M}_{11} \mathop{\sim}_{x\to\infty} &\frac4x\,\Omega\,;\qquad {\cal M}_{1n} \mathop{\sim}_{x\to\infty} \frac4x \dfrac{1}{2n^2-3n+1}\,,\quad n\geq 2\,; \\
 {\cal M}_{nn} \mathop{\sim}_{x\to\infty} 
&-\frac{4 \left(\frac{2}{4 n-3}+2 \psi\left(2 n-\frac{3}{2}\right)+2 \log
   \left(\frac{1}{x}\right)+2 \gamma_E +\log (4)\right)}{(4 n-3) x} \nonumber\\
   \mathop{\sim}_{x\to\infty}&
\dfrac{4\Omega}{x}\frac1{4n-3} (1 + {\cal O}(1/\Omega))\,,\quad n \geq 2\,.
\end{align}
\end{subequations}
$\psi$ is the logarithmic derivative of the $\Gamma$-function and $\gamma_E$ is the Euler's constant.
We used the notation introduced in \cite{vain1}
\be \Omega = 2 (\log(2x-1)\,. \label{notazionevain}\ee
Inserting in \eqref{ChshortdaM} we find
\be \frac1a{C^{(H)}}\mathop{\sim}_{x\to \infty}\;\; \dfrac{x}{\Omega +\frac1\Omega(\frac{\pi^2}{3}-4)}\bigl(1 + {\cal O}(1/\Omega)
\bigr)
\label{chasint}\ee
Expanding in $x$ one reproduces the results of 
Vainshtein\cite{vain1} and Jackson\cite{jack1},
we keep the unexpanded form \eqref{chasint} as this is the natural form for the approximation in our scheme.

\begin{figure}[!h]
\begin{center}
\includegraphics[width=0.8\textwidth]{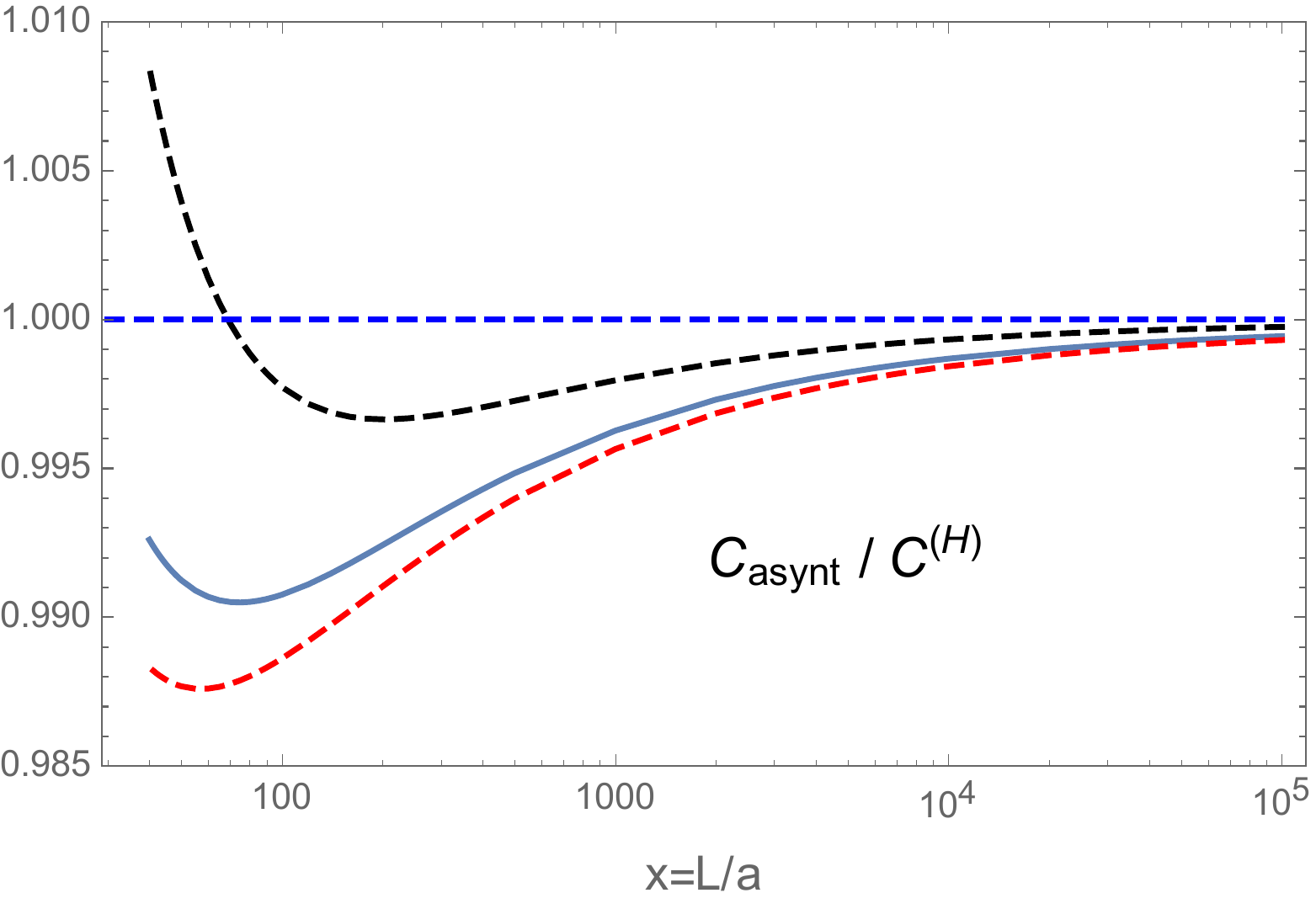}
\caption{ The ratio between \eqref{chasint} and the numerical values of $C^{(H)}$ up to distances $L/a=10^5$. The dashed line is the
approximation exposed in \cite{jack1}, i.e. the expansion of \eqref{chasint} in powers of  $1/\log(x)$. The dashed black line (upper curve) is the approximation \eqref{chasintbis}.
\label{rapportoCCtot}}
\end{center}
\end{figure}

In figure~\ref{rapportoCCtot} we show the ratio between the asymptotic approximation \eqref{chasint} and the numerical values of $C^{(H)}$ up to distances $L/a = 10^5$, for comparison we show also the Jackson form of the approximation. The approximation can be improved by expanding to second order in $1/\Omega$ the full matrix element ${\cal M}_{nn}$ in \eqref{valoreMperH}
and summing numerically the resulting series. We give here the result for reference, but this form has to be used only for large $x$, say $x>50$.
\be \frac1a{C^{(H)}}\mathop{\sim}_{x\to \infty}\;\; x\Bigl/
\left\{\Omega +\frac1\Omega\left[(\frac{\pi^2}{3}-4) - \frac{2.929591}{\Omega} - \frac{12.61970}{\Omega^2}
\right]
\right\}
\label{chasintbis}\ee

Let us now  consider the quadrupole moment. From \eqref{defc1d} it follows, in this case, where only the lateral surface matters:
\be \frac{D}{a^2} = -1 + \frac{x^2}{4}\left(\frac1{1+s} + \dfrac{a_2/a_1}{(1+s)(2+s)}\right)
\label{quadH1}\ee
We used the definition of $C_1$, see \eqref{defc1d} and \eqref{ChshortdaM}, and $h/a = x/2$. The obvious result $D\to -a^2$ for $x\to 0$ is automatic. For large $x$ we have seen that if we do the computations with a matrix with a diagonal dominance $a_2\ll a_1$ we can predict the order of magnitude of $D$. In the present case we have diagonal dominance for $s=1/2$ and  the order of magnitude
predicted by \eqref{quadH1} is
\[ \frac{D}{a^2} \sim \frac16 x^2 \]
which is a rough approximation of the values given in table \ref{rabelldalpha}. The result reproduces the elementary estimate
\eqref{stima2} 
 and of course immediately
explains the change of sign of $D$ from short to large dimensions of the cylinder.
 For a matrix with diagonal dominance we have, as
discussed above
\be a_1 {\cal M}_{12} + {\cal M}_{22} a_2 = 0\quad  \Rightarrow \quad \frac{a_2}{a_1} = - \dfrac{{\cal M}_{12}}{{\cal M}_{22}} 
\label{relazionem12m11}\ee
Taking the matrix elements form \eqref{valoreMperH} we have (with $s = 1/2$)
\be \frac{D}{a^2} \mathop{\sim}_{x\to\infty}\;\; -1 + \frac{x^2}{4}\left(\frac23 - \dfrac{20}{168-45 \Omega}\right)
\simeq \frac{x^2}{4}\left(\frac23 - \dfrac{20}{168-45 \Omega}\right)
\label{quadrupoloH}
\ee
This expression reproduce the numerical values with a relative precision better than 2\%  in the range $100<x<10^5$.

For small $x$, using \eqref{mHsmallx} and again \eqref{quadH1} and \eqref{relazionem12m11} (with $s=0$) we have
\be \frac{D^{(H)}}{a^2} \mathop{\sim}_{x\to0}\;\; - 1 +\frac{x^2}{4}\left( 1 -\frac{x^2}{128}\bigl(\frac52 + \log\frac{x}{32}\bigr)
\right)\label{quadshort}\ee
which describes with good accuracy the numerical data even at $x\sim 3$.

For the polarization one can repeat step by step the above procedures. In the limit of diagonal dominance of the matrix
$\alpha^{(H)}$ can be approssimated by (we remember that $x = 2 h/a$):
\be \frac1{a^3}\alpha^{(H)} = 2\pi \left[ \frac1{2^{1+s}\Gamma(2+s)}\right]^2
\frac{x^2}{4}\,\dfrac{1}{{\cal M}_{11} - \sum_{\alpha>1} {\cal M}_{1\alpha}^2/{\cal M}_{\alpha\alpha}}
\ee
The matrix ${\cal M}$ is given in equation \eqref{equazionialpha1cylH} and using as usual $s=0$ for small $x$ and $s=1/2$ for large $x$ we easily obtain
\begin{figure}[!ht]
\begin{center}
\includegraphics[width=0.8\textwidth]{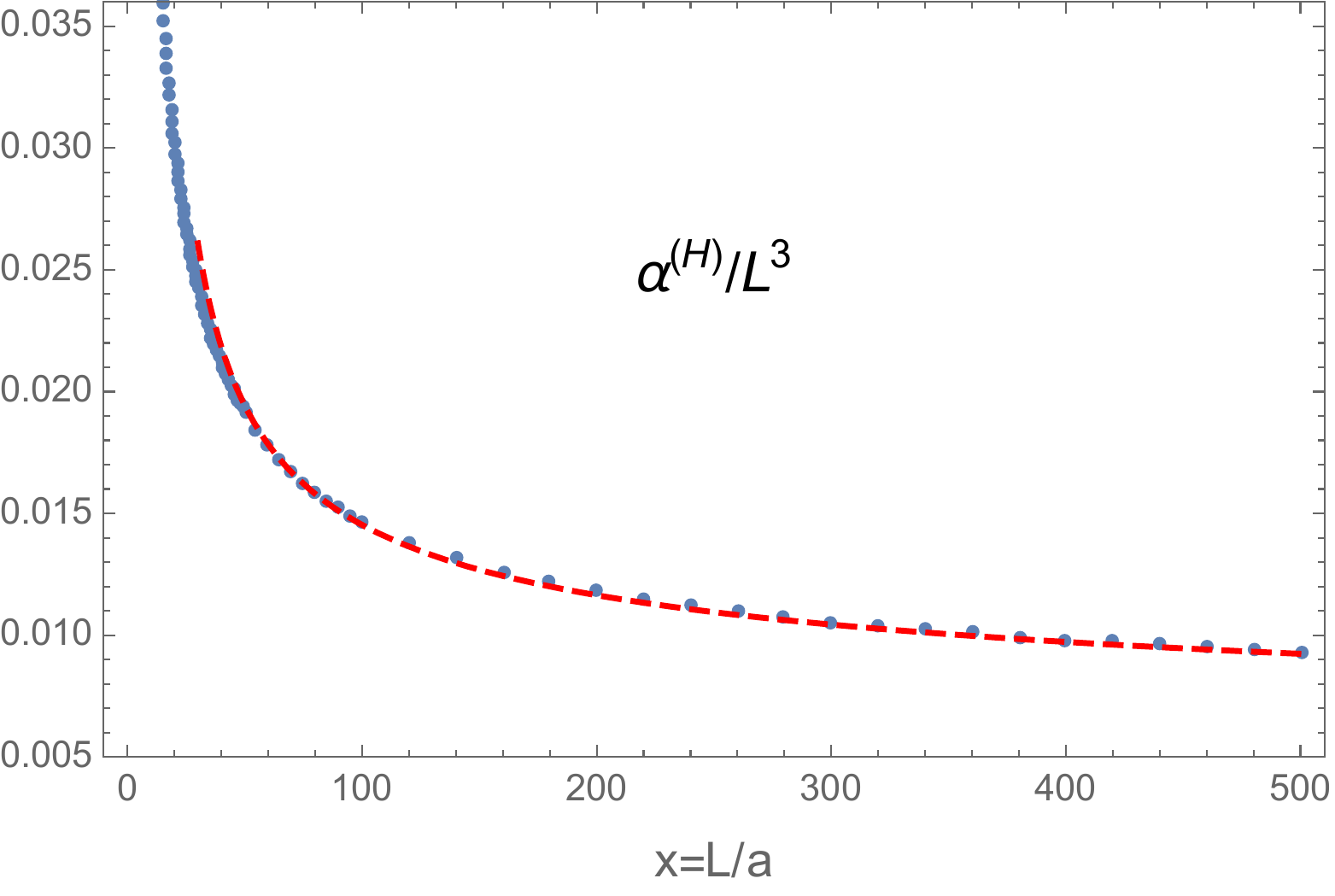}
\caption{ The ratio $\alpha^{(H)}/L^3$ for an hollow cylinder. The dashed line is the asymptotic prediction 
\eqref{alphaHlargeL}.
\label{figalphaH}}
\end{center}
\end{figure}
\begin{subequations}
\begin{align}
&\frac{\alpha^{(H)}}{a^3} \mathop{\sim}_{x\to0}\;\; \frac{1}{8} \pi  x^2 \left(1+\frac{1}{256} x^2 \left[4 \log
   \bigl(\frac{x}{32}\bigr)+9\right]\right)\label{alphaHshortL}\\
&\frac{\alpha^{(H)}}{a^3} \mathop{\sim}_{x\to\infty}\;\;
\frac{x^3}{12} \frac{1}{\Omega_2} \dfrac1{1 + \dfrac{3\pi^2 - 40}{9}\dfrac1{ \Omega_2^2}}\,,\quad
\Rightarrow\quad
\dfrac{\alpha^{(H)}}{L^3} \sim \frac1{12} \frac1{\Omega_2}\dfrac1{1 + \dfrac{3\pi^2 - 40}{9}\dfrac1{ \Omega_2^2}}
\label{alphaHlargeL}
\end{align}
\end{subequations}
We remember that $L=a x$. To simplify the comparison with existing literature we introduced the same notations of \cite{vain1}
\be \Omega_2 = 2(\log(2x) - \frac7 3) \ee
Expanding the denominator in \eqref{alphaHlargeL} we obtain the result of \cite{vain1}, apart a numerical error in that work.
The polarizability controls also the Rayleigh scattering on the conductor. In \cite{wat2}
a result for $\alpha^{(H)}$ is obtained in this context, and agrees with the series expansion of the result \eqref{alphaHlargeL}.

The numerical results and the asymptotic behavior for $\alpha^{(H)}$ are shown in figure~\ref{figalphaH}.
We note that on the scale of the figure the corresponding polarizability for the cylinder is indistinguishable from $\alpha^{(H)}$.

\subsection{Cylinder}
The last system considered in this section is a single solid cylinder, of radius $a$ an length $L = 2 h$.
The first systematic approach to the numerical computation of $C$ dates back to Smythe\cite{smytheC}, to our best knowledge. A recent
investigation, in principle identical to our apart the analyticity of the matrix elements and the range of lengths considered, is
given in \cite{FPV}.
The numerical data for small $L$ displayed in tables \ref{tabella1} and \ref{rabelldalpha} are simply explained.
The capacity tends toward the capacity of a single disc of radius $a$, $C^{(D)} = 2 a/\pi \simeq 0.63662 a$, more precisely the
numerical data follow the two-flat discs law \eqref{risnum1.10b} apart a small ${\cal O}(x)$ correction
\be \frac{C}{a} \mathop{\sim}_{x\to 0}\; \frac2{\pi} + \frac1{\pi^2}\,x\left( \log\frac1x + 2.852\right) + {\cal O}( x^2\log^2x)
\label{Csmalldist}
\ee
In the range $0.001 < x < 16$ our results reproduce, with a slight improvement in precision, the results obtained in \cite{smytheC} 
and we verified that in this region the interpolation formula proposed in that work reproduce the data within $0.2\%$.

The quadrupole  goes to  the analogous value for a disc $D_{zz} = -2/3 a^2$. 

The polarizability tends to zero and its behavior is easily estimated.
In an external potential $- {\cal E} z$ two opposite charges are separated on the two basis. The potential difference is $2 h {\cal E} = {\cal E} L$ and, as the capacity at small distance for two discs at distance $L$ is  $C = a^2/4L$ the corresponding charge on each disc is $Q = {\cal E} a^2/4$. The corresponding induced dipole is $d = Q L = {\cal E} a^2/4 L$ and the polarizabiity then is, as $L\to 0$ approximatively
\be \alpha \simeq \frac14 a^2 L = \frac14 a^3 x \ee
and this can be easily checked directly on table~\ref{rabelldalpha}. For large $L$, $\alpha$ is very similar to $\alpha^{(H)}$, as already noticed. The polarizablity of a cylinder for few selected values was computed in \cite{Taylor}, we checked that the results agree. 

\begin{figure}[!ht]
\begin{center}
\includegraphics[width=0.8\textwidth]{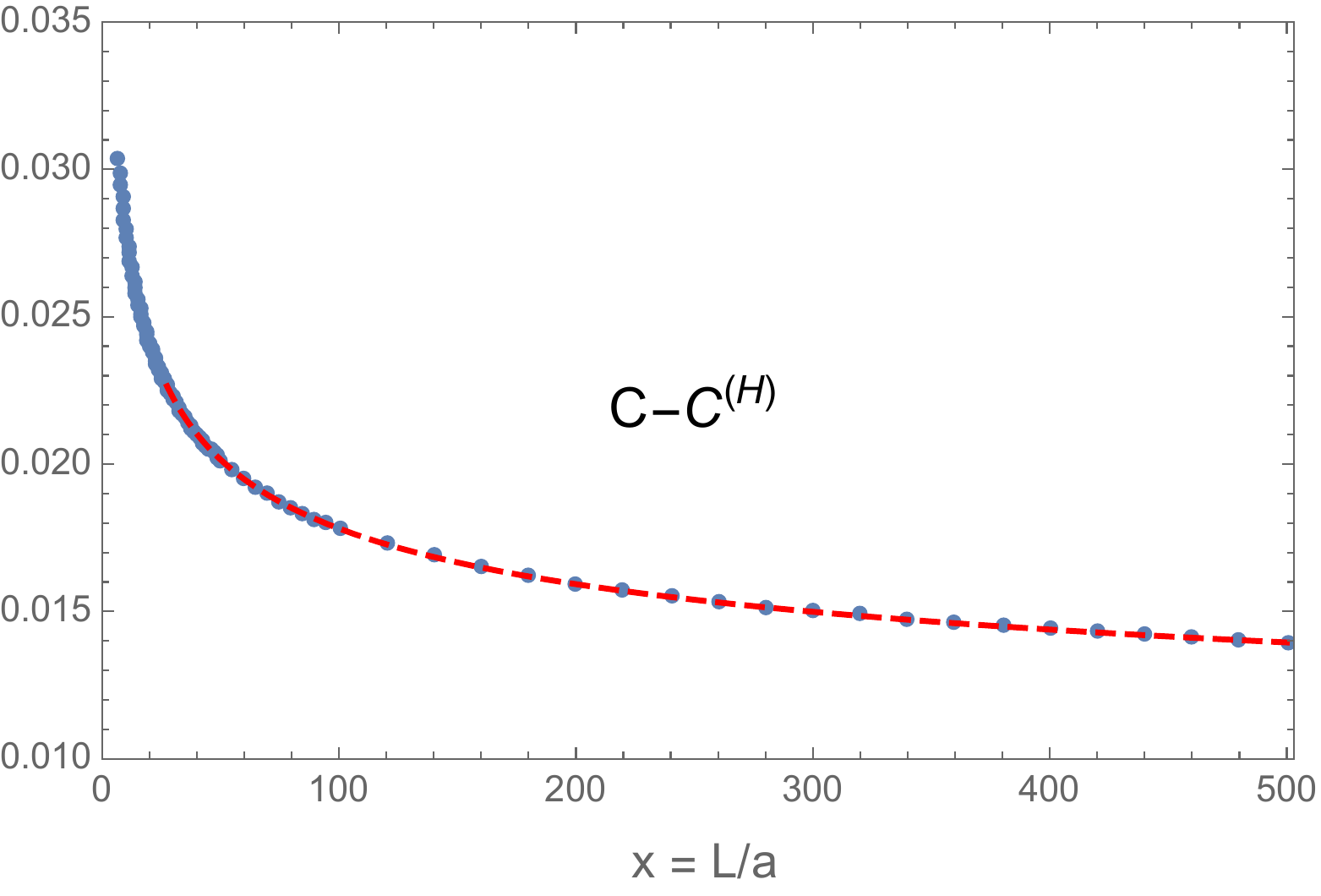}
\caption{ Difference between the capacity of a solid cylinder and of an hollow cylinder ($C^{(H)}$). The dashed line is the fit 
\eqref{fitdiffCCH}.
\label{figdiffCCH}}
\end{center}
\end{figure}

For $x=L/a\gg 1$ table \ref{tabella1} renders quite evident that the capacity for an hollow and a full cylinder are very similar 
already from $x\sim 5$. In figure~\ref{figdiffCCH} the difference $\delta C = C- C^{(H)}$ is shown in the range $x\geq 7$. In this range
all data are fitted  by
\be \delta C = \frac1a(C- C^{(H)}) \simeq\frac{0.0999991}{\log (x)}-\frac{0.0827803}{\log ^2(x)}
\label{fitdiffCCH}\ee
with a maximum absolute error of $0.0008$.
It is possible to give an argument from diagonal matrix dominance that $C- C^{(H)}\propto 1/\log x$ but we have not been able to compute the coefficients in \eqref{fitdiffCCH}. The numerical result \eqref{fitdiffCCH} is quite impressive from the point of view of the asymptotic expansions: it implies that all terms proportional to $x/\log^n(x)$ in \eqref{chasint} are identical for $C$ and $C^{(H)}$. This result has a simple physical interpretation. The capacities are the charges on the conductors for $V=1$, \eqref{fitdiffCCH} implies that the ``bulk'' charge of the two cylinders are the same and the differences come only from ``edge charges'', i.e. the charges on the bases of the cylinder and the different accumulation of lateral surface charges near the edges,
due to the different edge-singularities in the two cases. The fraction $\delta C/C$ is just the fractional charge at the edges, and as $C\to x/2\log(x)$ the result \eqref{fitdiffCCH} amount to say that the fraction of charge in the difference is of the order
\[ \frac{\delta Q}{Q} \sim \frac{0.2}{x} = 0.2 \frac{a}{L} \]
As expected the edge charges vanishes as $L\to\infty$ and \eqref{fitdiffCCH} shows that they vanish proportional to the inverse of the length of the cylinder. Below we give some other evidence for this effect.

We can not refrain from asking how our results behave compared to the classical problem of the charge distribution on a long cylinder, i.e. a wire. The problem has a long and interesting history, admirably synthesized in \cite{jack2}.

\begin{figure}[!ht]
\begin{center}
\includegraphics[width=0.49\textwidth]{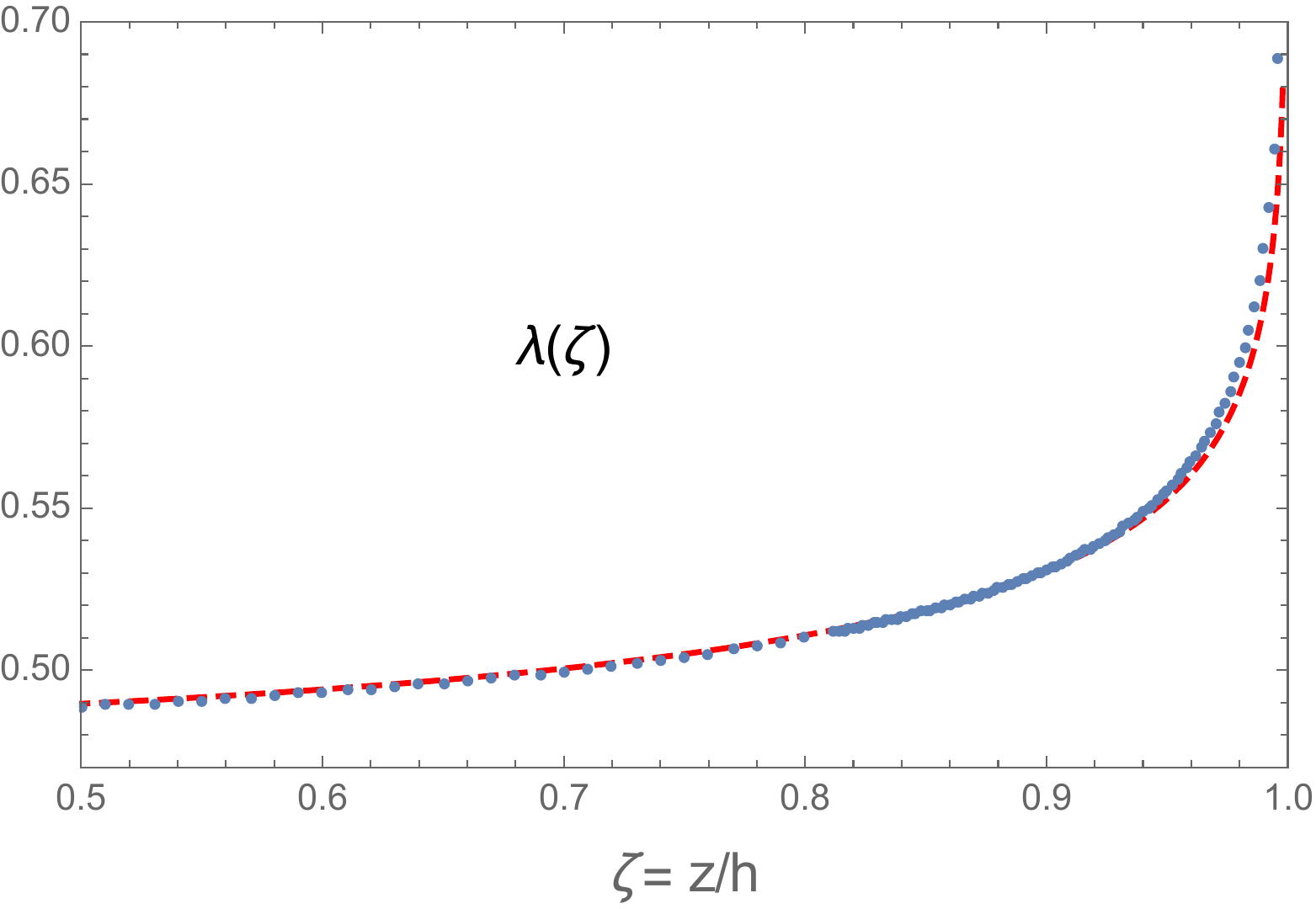}
\includegraphics[width=0.50\textwidth]{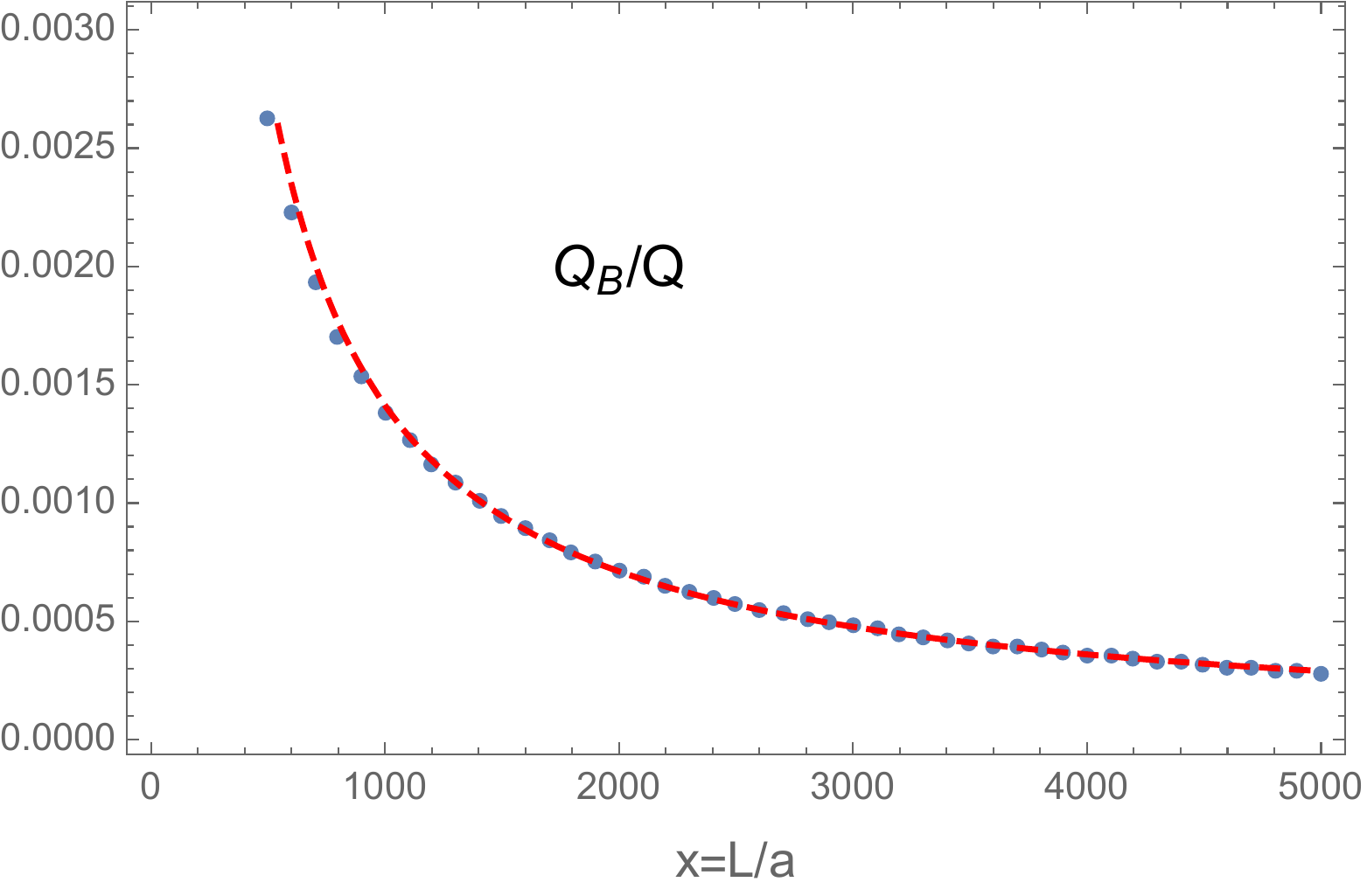}
\caption{Left panel: Linear density along the cylinder for two lengths,  $L/a=22000$ corresponding to
$\Lambda\simeq20$ in \eqref{lambdaJack}. The curves are normalized to unit charge in half the cylinder.
The dashed line is the prediction
\eqref{lambdaJack}. Right panel: fraction of charge on a basis for a full cylinder, as a function of its length, the dashed curve
is $1.4/x$,
\label{figsigmaH}}
\end{center}
\end{figure}

In \cite{jack1} the following form of the distribution in the region of 
large $x= L/a$, is derived:
\be \lambda(\zeta) = \lambda_0\left\{
1 - \frac1{\Lambda}\log(1-\zeta^2) + \frac1{\Lambda^2}
\left[ \left(\log(1-\zeta^2)\right)^2 + \frac12 \left[\log\Bigl(\frac{1+\zeta}{1-\zeta}\Bigr)\right]^2
-\frac{\pi^2}{6}\right]
\right\}\label{lambdaJack}
\ee
where, using the notations of \cite{jack1},  $\Lambda = 2 \log(x)$ and $\zeta = z/h$.
The distribution \eqref{lambdaJack} is in excellent agreement with the data except near the edge of the cylinder. In
figure \ref{figsigmaH} we plot the numerical data and the prediction \eqref{lambdaJack} for $x=22000$.
For this computation we used a basis of 60 Gegenbauer polynomials, the intermediate steps for the linear
density have to be done in high precision arithmetic. We note also, for the interested reader, that 
to compute $\lambda(z)$
the limit $s\to 0$ of the functions
\eqref{gegenb} must be performed.

The computation in \eqref{lambdaJack} refers to an hollow cylinder and give rise to an asymptotic formula for the capacity
consistent with \cite{vain1} and \eqref{chasint}. It is widely believed that the addition of caps to the cylinder do not change
qualitatively the picture and in particular that the fractional charge deposited on the bases goes to zero as $L\to \infty$.
With our data we can explicitly verify this statement. In the second figure in \ref{figsigmaH} we give the fractional charge on a basis, $Q_B/Q$, as a function of $x=L/a$. The points lie on a curve of the form
\[ \frac{Q_B}{Q} \mathop{\sim}_{x\to\infty} \frac{1.4}{x} \]
\begin{figure}[!ht]
\begin{center}
\includegraphics[width=0.49\textwidth]{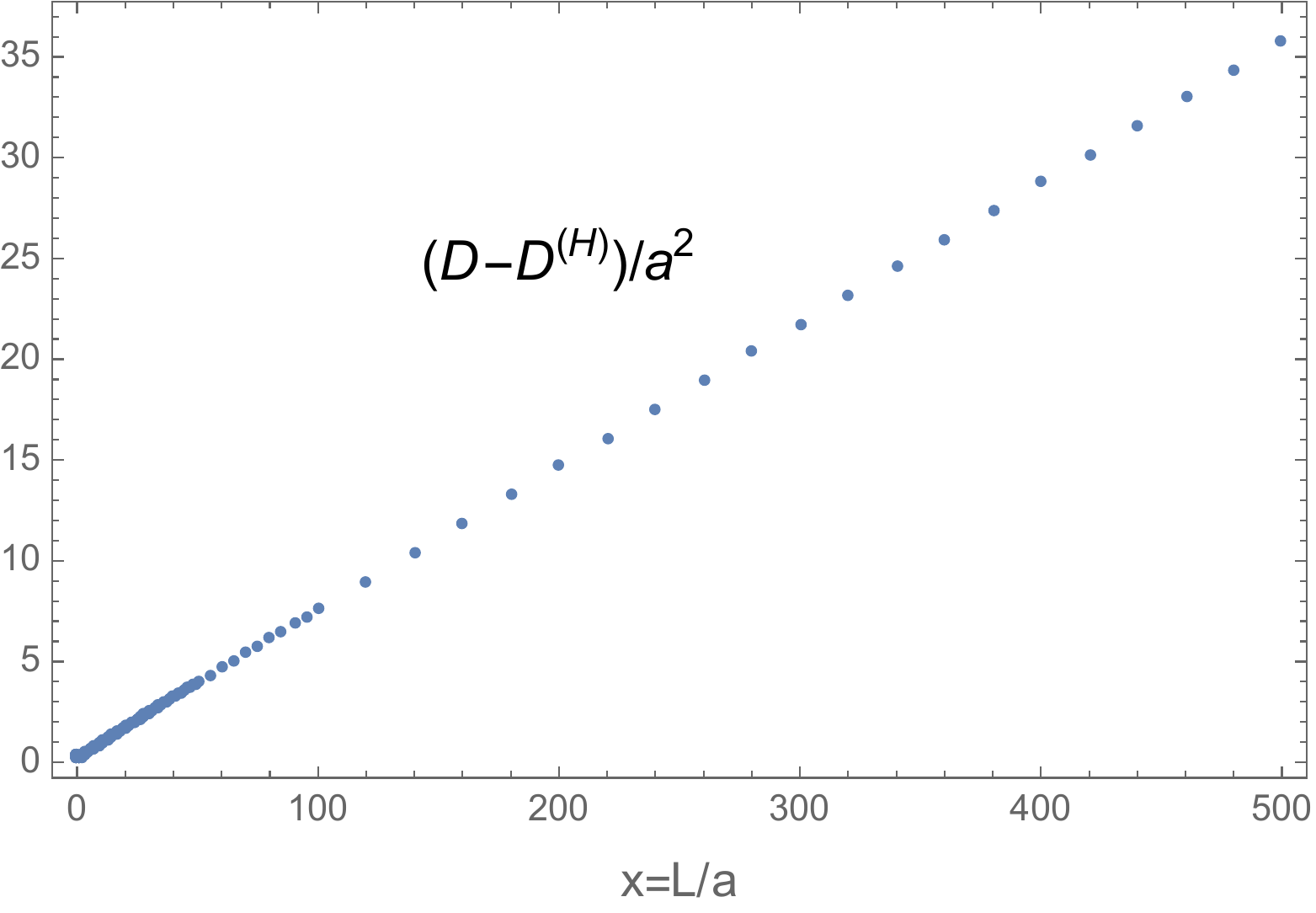}
\includegraphics[width=0.49\textwidth]{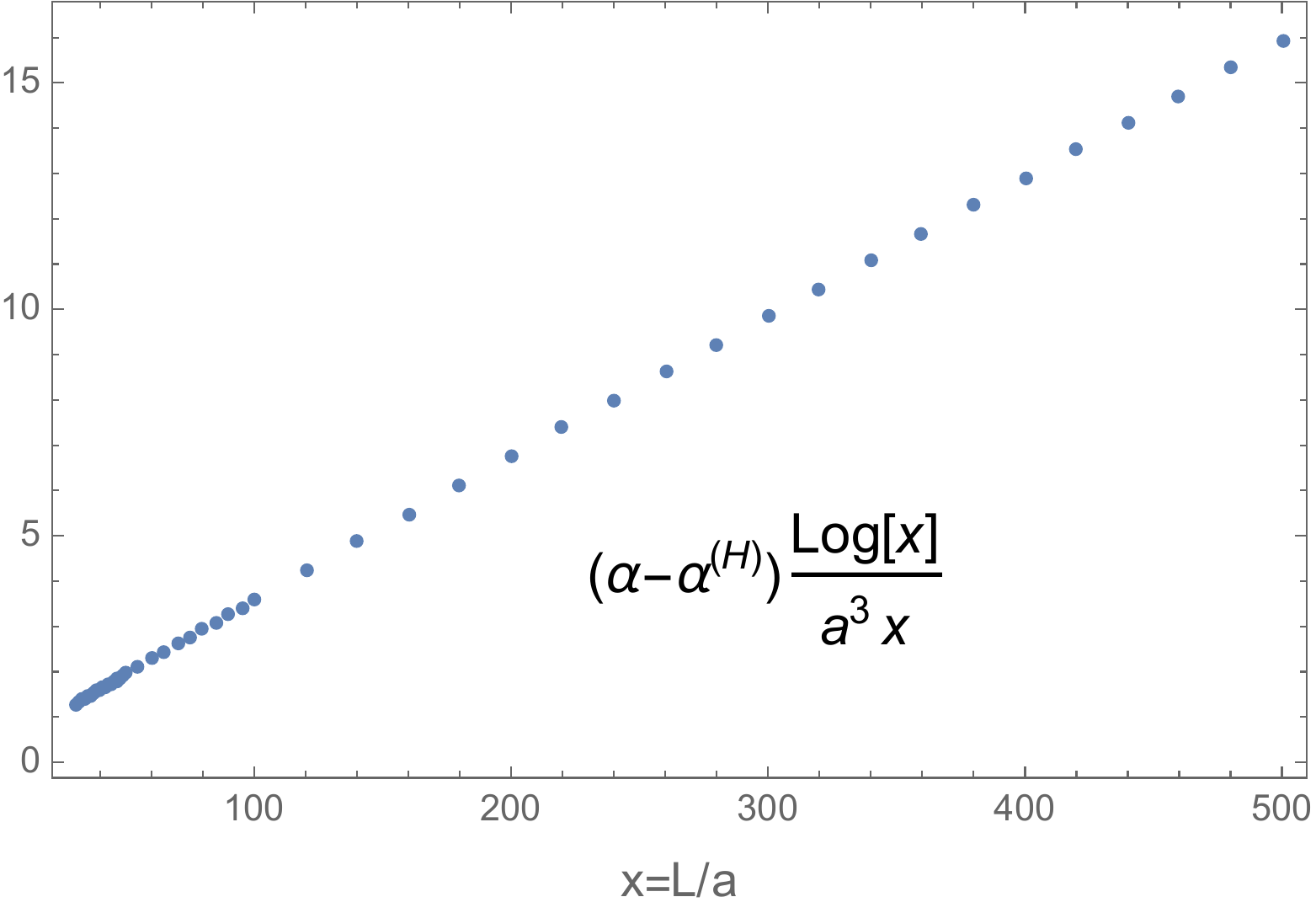}
\caption{Difference between quadrupole moments (left panel) and polarizabilities (right panel) between a solid cylinder and an hollow cylinder of the same length.
\label{qbsuq}}
\end{center}
\end{figure}
It is interesting to see how the ``edge charges'' mentioned above discussing the capacity,
affect the simple estimates given at the beginning of this section, \eqref{stima1}--\eqref{stima3}. 

A residual edge charge, proportional to $1/L$ affects on the quadrupole monopole with a term of order $L^2/L = L$. Analogously the usual
polarizability can be considered as composed by a couple of opposite charges of order $L^2/\log(L)$ separated by a distance of order $L$, a fractional excess of charge proportional to $1/L$ must give rise to a correction of order $L^2/\log(L)$ to $\alpha$. These effects can be be qualitatively exposed by comparing the quadrupole moments and polarizability for a solid cylinder and an hollow cylinder, the data are consistent with the fact that only edge effects distinguish the two systems. This is shown in figure \ref{qbsuq}, where the measured quantities
$D- D^{(H)}$ and $(\alpha - \alpha^{(H)})(\log(L)/L)$ are reported as a function of $L$. Both must linear functions for large  $L$'s, and indeed this is what happens.

\section{Two electrodes: numerical results for $C$ and $C_{g_1}$\label{seznumris}}
The main computation in this work concerns two equal circular electrodes with radius $a$ and thickness $b = \tau a$.
We performed calculations for 14 thickness in the range $0.001\leq\tau\leq0.3$ and for $0.0001\leq\kappa\leq1$, for the short distances regime. We present the data in tables~\ref{tabtot1}-\ref{tabtot5} at the end of the paper. In figures
below we used $\tau = 0.001, 0.01, 0.1$.
In the lower part of the tables
we give the extrapolated results obtained using \eqref{proceduraN}. We see that $C_{g_1}$ is practically unaffected 
while only the first values of $C$ have a relatively significant change. Let us examine the results for $C$ and $C_{g_1}$ separately.

\subsection{Mutual capacitance $C$}
\begin{figure}[!ht]
\begin{center}
\includegraphics[width=0.49\textwidth]{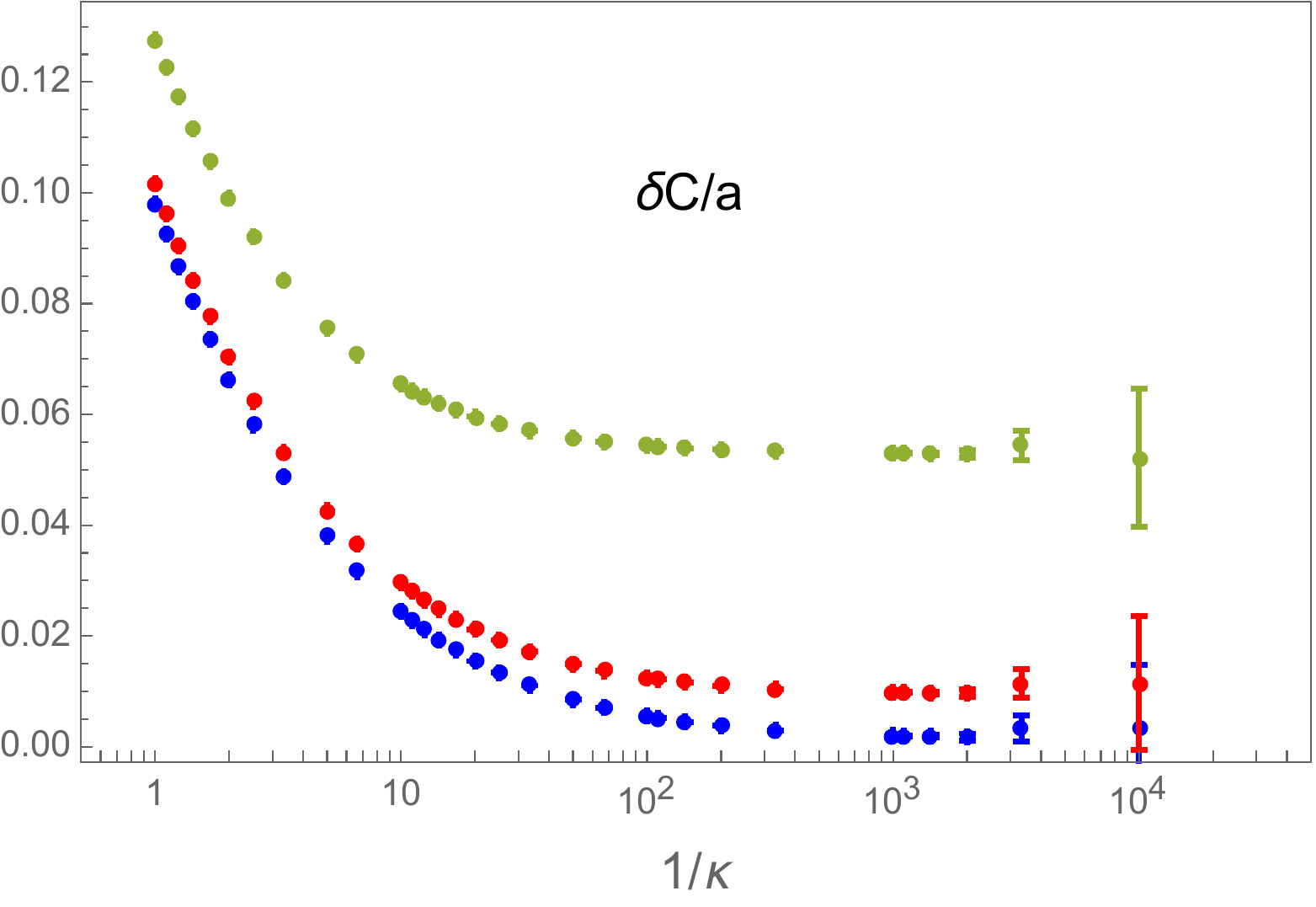}
\includegraphics[width=0.49\textwidth]{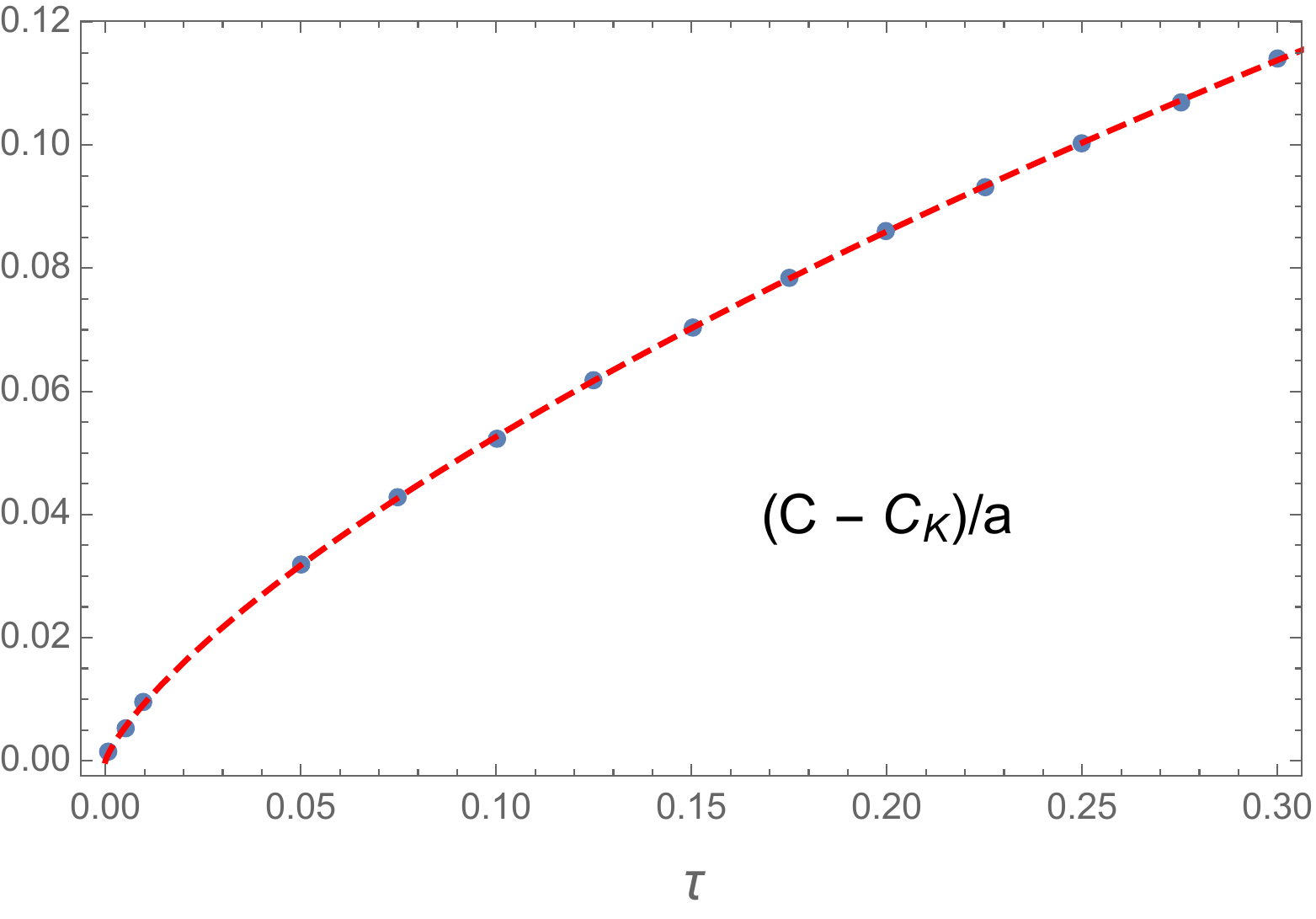}
\caption{Left panel:
difference between the computed mutual capacity $C$ and the values predicted by the Kirchhoff approximation.
The errors are estimated, as an order of magnitude, by the difference from computed and extrapolated values.
The three set of points refer to $\tau=0.001, 0.01, 0.1$, from lower to upper part of the figure. 
Right panel: the same difference computed at fixed $\kappa = 0.001$ as a function of $\tau$. The dashed line is
the fit \eqref{cckvstau}.
\label{p2figCmenok}}
\end{center}
\end{figure}

The only theoretical result, to  our best knowledge, is Kirchhoff calculation \eqref{sez1.Kirch1}. There is in general a good but non completely satisfactory agreement between \eqref{sez1.Kirch1} and our data. To clarify the point let us plot the differences
between the numerical values $C[\tau,\kappa]$ and the ones predicted in \eqref{sez1.Kirch1}:
\be \delta C = C[\tau,\kappa] - C_K[\tau,\kappa] \label{diffC1}\ee
This difference is expected to vanish for small $\kappa$. The results are shown in the left part of figure~\ref{p2figCmenok}.
It is apparent an offset depending on $\tau$, i.e. the data suggest that a correction, constant in $\kappa$, is required 
for the Kirchhoff approximation $ C[\tau,\kappa] \sim C_K[\tau,\kappa] + a\,\delta f_K[\tau]$ for $\kappa\to 0$.
We estimate this correction by performing the computation of $C$ for 14 different values of thickness in the region $0.001\leq\tau\leq 0.3$, excluding from the analysis the values $\kappa < 0.0005$ to avoid any effect due to the extrapolation procedure.
The results can be roughly described by the interpolation

\be \frac1a(C - C_K) \equiv \delta f_K(\tau)  \simeq 0.1052 \tau +0.0132 \tau \log ^2(0.0347 \tau) \label{cckvstau}\ee
The right panel of figure~\ref{p2figCmenok} is a summary of our calculations, the dashed line is
\eqref{cckvstau}.

\begin{figure}[!h]
\begin{center}
\includegraphics[width=0.9\textwidth]{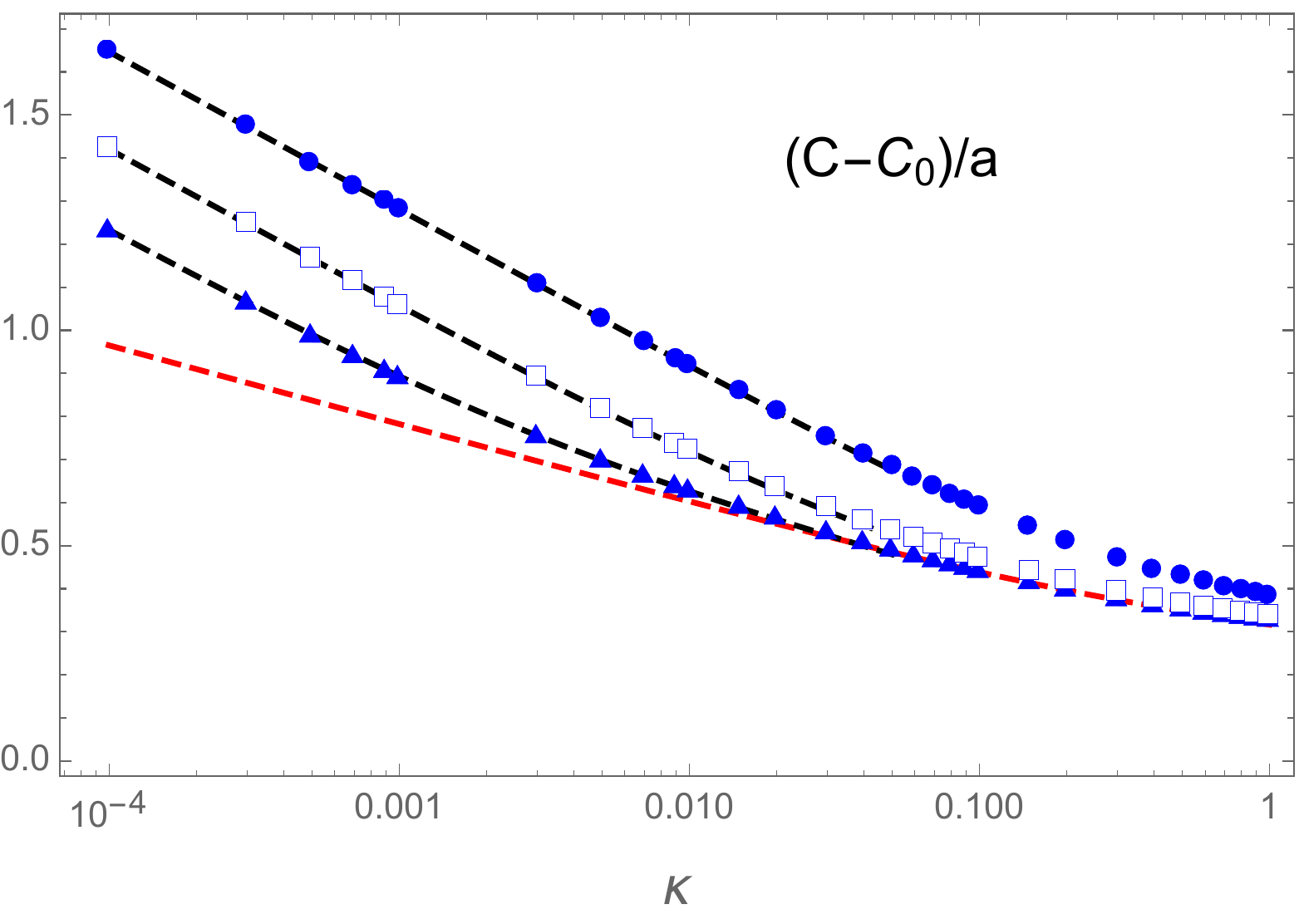}
\caption{
Difference between the computed mutual capacity $C$ and the geometrical capacitance $C_0 = a/(4\kappa$ for thickness
0.1 (circles), 0.01 (empty squares), and 0.001 (triangles). The dashed black curves have been computed with \eqref{Capprox},
the dashed lower curve is the limiting case of flat discs, \eqref{corrShaw}.
\label{p2multiC}}
\end{center}
\end{figure}

Of course \eqref{cckvstau} is just a compact way to express our results, we have no result on the analytical form of the correction
as a function of $\tau$.
It is interesting to have a look at the numerical data isolating the edge corrections, defined here as the difference between
the computed $C$ and the geometrical value $C_0 = a/(4 \kappa)$:
\[ E(\kappa,\tau) = \dfrac{C - C_0}{a} \]
 The data are shown in figure~\ref{p2multiC}. It is quite clear how for small $\tau$ the function $E$ tends to the limiting behavior given by \eqref{corrShaw} (the lower dashed red curve in the figure) when $\kappa\gtrsim\tau$, while for $\kappa\ll \tau$
a clear common logarithmic behavior appears, but with a slope different from the case of flat discs, and coincident with the
prediction of the Kirchhoff approximation, \eqref{sez1.Kirch1} and \eqref{sez1.limite2}. The vertical distance between the data, however, is not completely described by the term $\kappa$-independent in \eqref{sez1.limite2}. The dashed black lines in
figure \eqref{p2multiC} are given by computing $E$ with
\be \frac{C}{a} \simeq f_K(\kappa,\tau) +  \delta f_K(\tau) \label{Capprox}\ee
with $\delta f_K$ defined in \eqref{cckvstau} and describe quite accurately both the slope and the offset between the points.

As the numerical results reported in tables~\ref{tabtot1}-\ref{tabtot5} can be of some interest in the analysis of realistic systems we give here some comment on these numbers. The precision can  be estimated by observing the approach to the asymptotic value for large $N$, the maximum numbers of polynomials used. Our conservative estimate is a possible error of 1 part in $10^6$. This is of the same order
of the ideal accuracy for a modern analog-digital converter device with a 20-bit scale, then the results can be used both for checking the measurements and for defining the accuracy of such devices. In principle the precision of the numerical data can be improved raising the value of $N$. We used $N=50$ for the two smallest values of $\kappa$, $N=40$ for $0.0005\leq \kappa\leq 0.003$ and $N=30$ for higher values of $\kappa$.

Usually measurements are expressed in SI units, we remember that if lengths are measured in $mm$ our numbers, multiplied by the conversion factor $K_\epsilon \simeq 0.1112650056$, give the capacities expressed in pF.

In usual applications it can be useful to have a simple interpolating formula expressing the results. One possibility is to use
an interpolation of the form:
\be \frac{C}{a} = f_K(\kappa,\tau) + \delta f_K(\tau) + 
\frac{\kappa}{16\pi^2}\log\frac\kappa{16\pi}\log\Bigl(\frac{\kappa}{16\pi} + \frac{\tau}{\pi}\Bigr)
\label{empC}\ee
Equation \eqref{empC} reproduces our data for all considered thickness with a maximum relative error of $0.03\%$ in the range $0\leq \kappa\leq0.1$ and within $0.6\%$ for $0.1< \kappa \leq 1 $.
With respect to our numerical data \eqref{empC} appears more accurate than the results achievable with other methods of which we are aware, like  the method proposed in \cite{lynch} and further reworked in \cite{kam}.

\subsection{Coefficient $C_{g_1}$}
An accurate determination of $C_{g_1}$ is necessary to study the short distance behavior of the forces between electrodes.
For $\kappa\to 0$ this coefficient approach half the value of capacity $C_1$ of the cylinder obtained by the fusion of the two electrodes, i.e. $C_1(2\tau)$:
\be C_{g_1}(\tau,\kappa) \mathop{\rightarrow}_{\kappa\to 0}
\frac12 C_{1}(2\tau) \label{limiteCg1}\ee
For the three thickness presented here, $b = a (0.001, 0.01,0.1)$ we have, using the results of 
section \ref{sezsingolocond}
%\cite{pafcyl1}:
\[ \frac12 \frac{C_1}{a} = \left(
0.3192273, 0.3251698, 0.3646827
\right) \]
and the agreement with \eqref{limiteCg1} can be directly verified on the tables~\ref{tabtot1}-\ref{tabtot5}.

The forces depend on the derivative of $C_{g_1}$ with respect to $\kappa$, then on the corrections to \eqref{limiteCg1}.
The na\"ive expectation
\be C_{g_1}(\tau, \kappa) \mathop{\rightarrow}_{\kappa\to 0} \frac12 C_1(2\tau) + a B(\tau)\,\kappa \label{corrcg1.a1}\ee
breaks down for flat discs, a term proportional to $\kappa\log\kappa$ appears and this induce a logarithmically divergent repulsive force between the two discs.
\begin{figure}[!h]
\begin{center}
\includegraphics[width=0.9\textwidth]{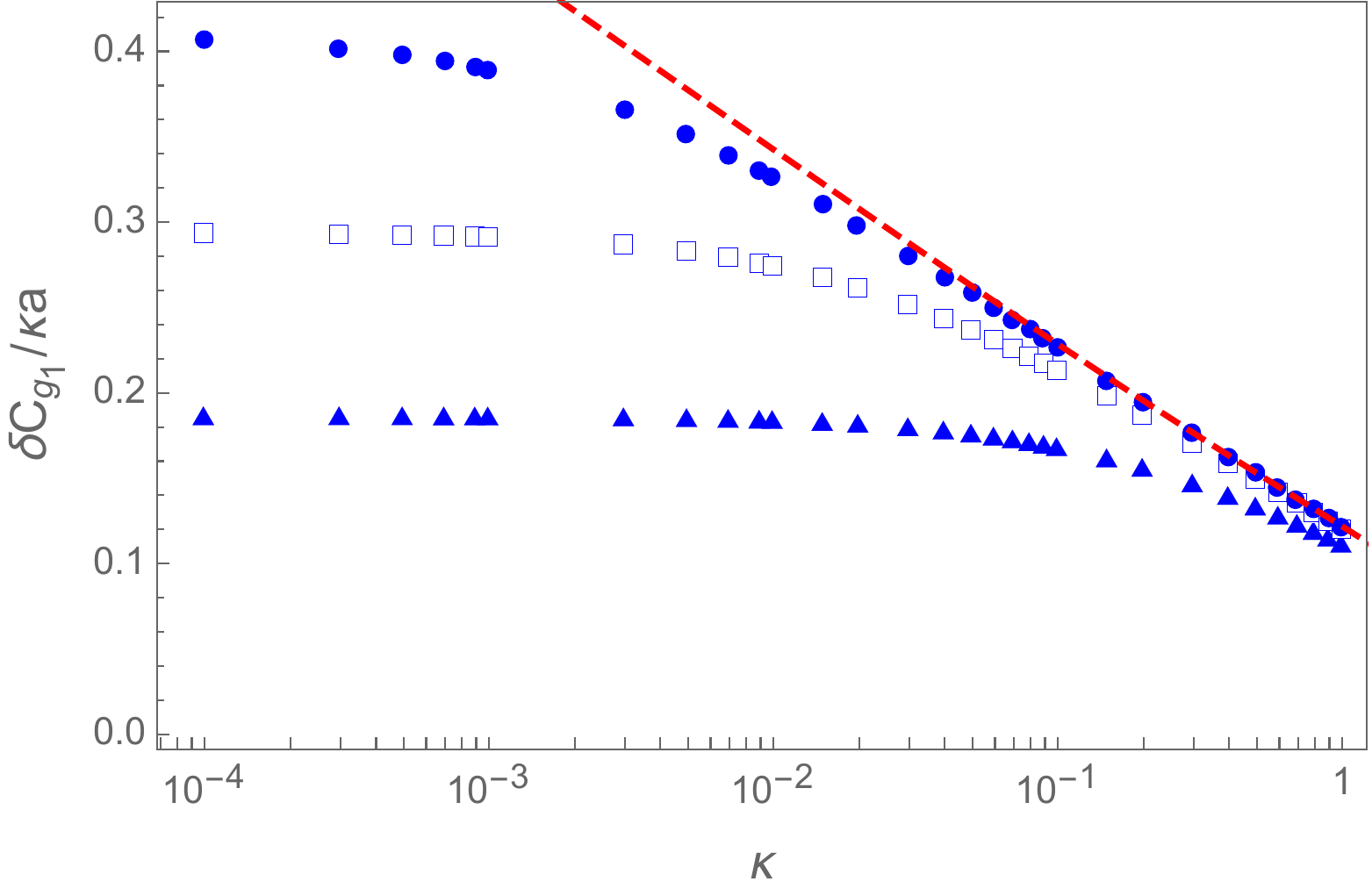}
\caption{
Numerical results for $\delta C_{g_1}/a\kappa$, equation \eqref{corrcg1.a1}. The three set of points refer to the three tichness
$\tau = (0.001, 0.1, 0.1)$ from top to bottom. The dashed line is the result for two flat discs.
\label{p2figCgsuk}}
\end{center}
\end{figure}
As explained in \cite{paf,paf2} we expect that any deviation from the ideal case of two flat discs causes the logarithmic divergence to be smoothed and the result \eqref{corrcg1.a1} recovered. This has been explicitely verified, analitically and numerically for the case of two discs of different radii, in \cite{paf2}.
Here the problem is similar. The thickness $b$ introduces another scale in the problem then the distance $\ell$ must be compared both to $a$ and to $b$. On physical grounds we expect that in the region
\be b \ll \ell \ll a \,;\qquad \text{i.e.}\qquad \tau \ll \kappa \ll 1\label{region1}\ee
the system ``does not know'' the scale $b$, then the behavior is the same as the behavior of two flat discs. 
Instead for
\be \ell \ll b \ll a \,;\qquad \text{i.e.}\qquad \kappa \ll \tau \ll 1\label{region2}\ee
the existence of $b$ affects the asymptotic behavior and the naive expectation \eqref{corrcg1.a1} is recovered.

To test this picture we consider the quantity
\be \frac{C_{g_1} - \frac12 C_1(2\tau)}{a \kappa} \equiv \frac1{a \kappa}\delta C_{g_1} \label{cg1suk.1}\ee
Every deviation from the asymptotic value is enhanced by a factor $1/\kappa$, than this test is a very sensible one. A behavior like two flat discs must produce a straight line in a logarithmic $\kappa$ scale, the behavior \eqref{corrcg1.a1} an horizontal line.
\begin{figure}[!h]
\begin{center}
\includegraphics[width=0.65\textwidth]{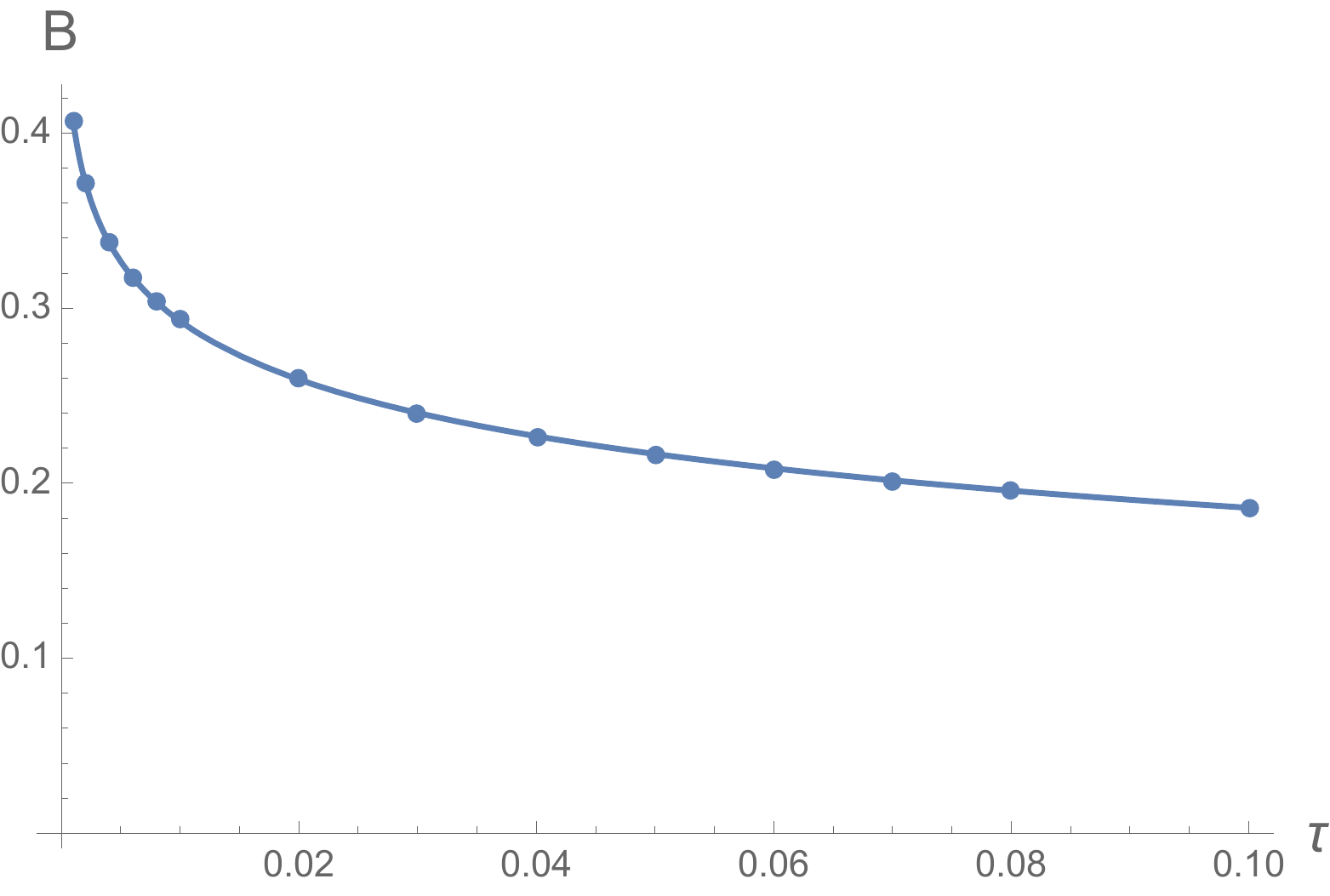}
\caption{
Numerical results the constant $B$ as a function of $\tau$.\label{p2figB}}
\end{center}
\end{figure}
The results are summarized in figure~\ref{p2figCgsuk}. It is apparent that the flat discs result acts as an envelope for the
results with finite thickness and that exists always a region, for sufficiently small $\tau$, where the variation of $C_{g_1}$, and consequently the behavior of the force, is well approximated by the two flat discs configuration, in perfect agreement with the
subdivision of the variation of $\kappa$ in two ranges, as indicated in \eqref{region1}-\eqref{region2}.
The consequences of these results for the forces will be investigated in the next section.

Some considerations can be done on the dependence of the constant $B$ on $\tau$. The logarithmic divergence divergence present for
$\tau=0$, see equation \eqref{valorecgasint}, must now be caused by an analogous divergence in $\tau$, this is the meaning of the envelope in figure~\ref{p2figCgsuk}.
This means that for $\tau\to 0$, $B(\tau)$ must be described by
\be B(\tau) \mathop{\sim}_{\tau\to 0} = \frac1{2\pi^2} \left(\log\frac1\tau + \beta\right) \label{andamentoB}\ee
We estimated approximatively $B$ by computing the ratio
\[ \dfrac{C_{g_1}(\kappa) - \frac12 C_{1}(2\tau)}{a\kappa} \]
for $\kappa = 0.0001$ for a series of $\tau$. The results are plotted in figure~\ref{p2figB}.
At small $\tau$ the points follow perfectly the expected trend \eqref{andamentoB}. The line in the figure is an interpolation of the form
\be B(\tau) = \frac1{2\pi^2} \left(\log\frac1\tau + 1.127\right) +\frac1{2\pi^2}\,\tau
\left(\log\frac1\tau + 0.0957\right)\,.\label{andamentoB1}\ee
\section{Forces between electrodes\label{sezforze}}
The general expression for the force between two equal circular, isolated, electrodes is given in \eqref{sez1.forza} and can be reorganized in the form
\be 
\begin{split}
F &= \frac1{a^2}\left(Q_1+Q_2\right)^2
\left( \frac{a}{4 C_{g_1}^2}\frac{\partial}{\partial\kappa}{C_{g_1}}+
R \frac{a}{8 C^2}\frac{\partial}{\partial\kappa}C
\right) \\
&\equiv \frac1{a^2}\left(Q_1+Q_2\right)^2 \left( f_1 + R f_2\right)
\equiv \frac1{a^2}\left(Q_1+Q_2\right)^2 f(R,\kappa)
\end{split}
\label{forzaexp2}
\ee
where $R$ is the dimensionless  ratio
\be R = \frac{(Q_1-Q_2)^2}{(Q_1+Q_2)^2}\,. \label{defR}\ee
The terms $f_1, f_2$ are simply the dimensionless terms in the parenthesis in \eqref{forzaexp2}, and can be computed putting directly $a=1$ in the relevant formulas.
We note that for like charges $R<1$ while for unlike charges $R>1$.

The results of the previous section allow us to answer two questions which can be of some experimental relevance
\begin{itemize}
\item[1)] In the limit $\ell\to 0$, i.e. for almost touching electrodes, the limit force is attractive or repulsive?

In this limit the sign of $F$ is determined by the $\kappa\to 0$ limit of the combination $f_1 + R f_2$, which in this limit
is a function only of $\tau$ and $R$. 
It is easy to show that $C_{g_1}$ is an increasing function of $\kappa$ for small $\kappa$, then $f_1>0$, i.e.  a repulsive force, 
while at short distances $C$ is dominated by the geometric contribution to the capacity, then $f_2<0$, i.e. an attractive force.
The balance between this two competing factors determines the nature of the force.
The line $f_1 + R f_2 =0$ must divide the ``phase space'' $\tau-R$ distinguishing the attractive
and the repulsive domains.
\item[2)] Which is the form of $F$ as a function of the distance and of $R$,
for a given thickness ? 

This question is particularly
interesting for unlike charges: at large distances they surely attract each other, then a repulsive force at short distances implies
the existence of a stationary point. Viceversa for like charges if the attractive part $f_2$ dominates.
\end{itemize}

\begin{figure}[!ht]
\begin{center}
\includegraphics[width=0.8\textwidth]{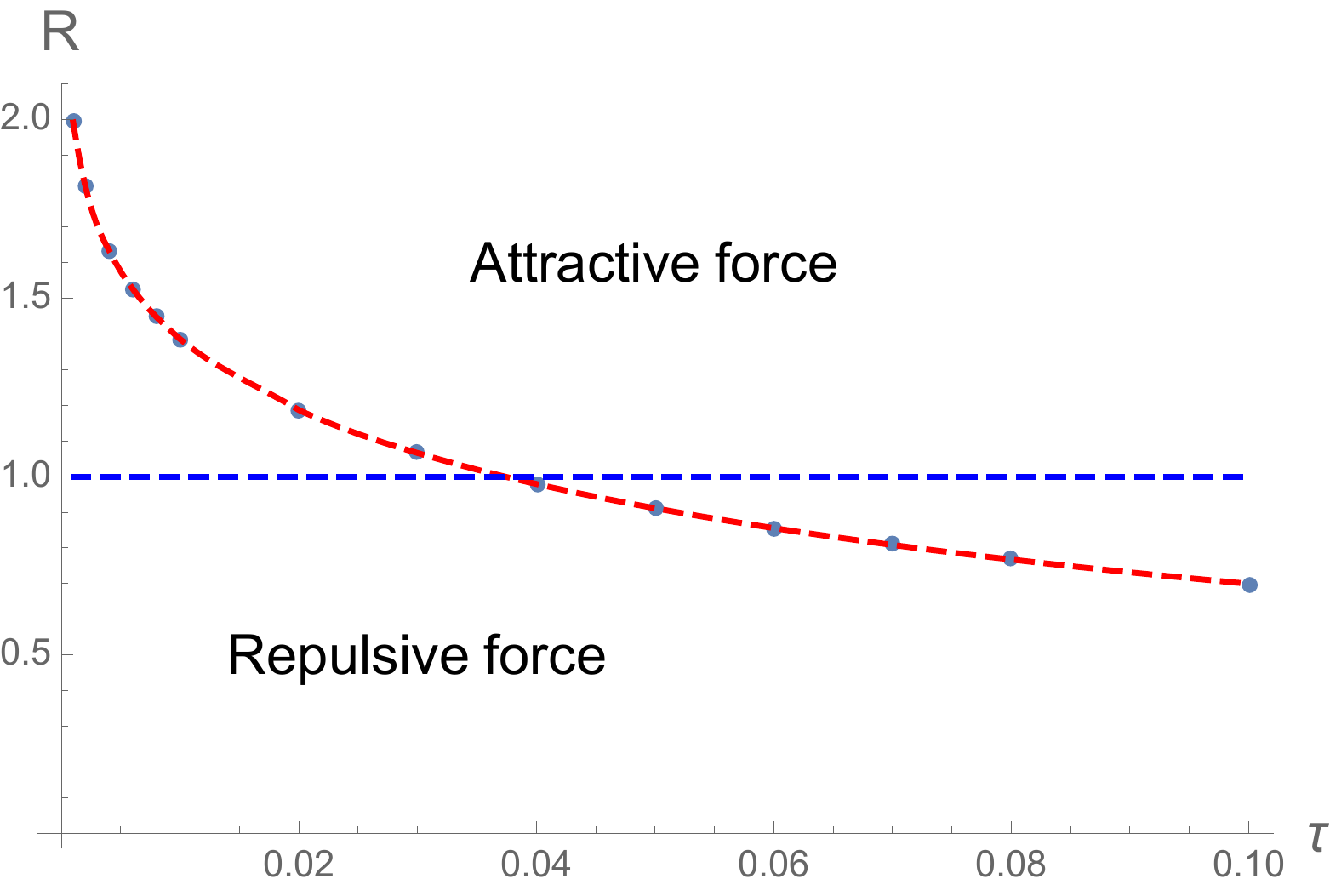}
\caption{Numerical computation of the dividing line between conductors attracting and repelling at infinitesimal distance.
The dashed line is just an interpolation, points on this line satisfy equation \eqref{limiteforze2}.
The computed points lie in the region $0.001 \leq \tau \leq 0.1$. The horizontal line is an help for the reader to distinguish
the case of odd charges, $R>1$ from the case of even charges, $R<1$. The dashed curve is given in \eqref{r0aspettato2}.
\label{p2diagFase}}
\end{center}
\end{figure}

The question 1) has in principle a simple answer. Using the known limiting behaviors \eqref{cg1suk.1} and \eqref{sez1.Kirch1}
one find
\be f_1 \mathop{\sim}_{\kappa\to 0} \frac{a^2}{C_1(2\tau)^2} B(\tau)\,;\qquad
f_2\mathop{\sim}_{\kappa\to 0} -\frac12 \label{limiteforze}\ee
then the limiting behavior of the force is fixed by the relation
\be R_0(\tau) = 2 \frac{a^2}{C_1(2\tau)^2} B(\tau) \label{limiteforze2}\ee
which determine the ratio $R$ for which $F$ change sign.
For a given $\tau $, for $R> R_0$ the force is attractive while for $R< R_0$ the force is repulsive. The ``phase diagram''
relative to this description is given in figure~\ref{p2diagFase}. The horizontal line divide the region $R<1$ (like charges) from
the region $R>1$ (odd charges).

Using the result \eqref{andamentoB1} of the previous section and the fact that for small $\tau$, $C_1(2 \tau) \to 2 a /\pi + {\cal O}(\tau)$
we can predict that in this regime
\be R_0(\tau) \mathop{\simeq}_{\tau\to 0} 2\frac{\pi^2}{4} \frac1{2\pi^2}(\log\frac1\tau + 1.127) = \frac14(\log\frac1\tau + 1.127)
\label{r0aspettato} \ee
In section \ref{sezsingolocond} it has been shown that for small $\tau$ (see equation \eqref{Csmalldist}):
\be C_1(\tau) \simeq \frac2\pi + \frac1{\pi^2} \tau\left(\log\frac1\tau + 2.852\right) + {\cal O}(\tau^2\log^2\tau)\,
\label{C1smalltau}\ee
then, using  \eqref{andamentoB1}, the relation \eqref{limiteforze2} can be approximated by
\be R_0(\tau) \simeq \frac14\dfrac{\left(\log \frac{1}{\tau}+ 1.127\right) + \tau \left(\log \frac{1}{\tau}+0.0957\right)}{
   \left(1+\frac1\pi \tau \left(\log \left(\frac{1}{2 \tau}\right)+2.852\right)\right)^2}\,.
   \label{r0aspettato2} \ee
   This value for $R_0(\tau)$ is reported as a dashed curve in figure \ref{p2diagFase}.
Formulas \eqref{r0aspettato}, \eqref{r0aspettato2} give a complete answer to the question 1) stated above in the most interesting case of small $\tau$, because the behavior for $\tau\to 0$ has been justified analytically in the discussion on $B(\tau)$ in the previous section. In particular, as was physically expected from the result for two flat discs, given an arbitrary ratio $R$, defined in
\eqref{forzaexp2}, it always exists a small enough $\tau$ such that the nearly-contact force is repulsive, for equal electrodes.

An interesting special case must be noticed. If a charged disc approaches an equal uncharged disc, i.e. $R=1$, from
\eqref{r0aspettato2} follows that the two discs repel if $\tau<0.037$ and attract if $\tau>0.037$. The fact that the dimension
of the bodies can affect the sign of the force is, in our opinion, a quite interesting effect.

Let us now discuss the second question raised above. To fix the ideas we consider a given system, i.e. with $\tau$ fixed.
The case $Q_1 = Q_2$ is a special case, as the leading order of the attractive part of the force vanishes and, as
next to leading orders vanish at least like $\kappa\log\kappa$, the force is repulsive at short distances. This is a particular case of a general feature: when the conductors acquire 
the charge that they would have in the configuration in which they touch, the force is repulsive at short distances, see for example
\cite{paf2}. We exclude this particular case, corresponding to $R = 0$, from the following discussion.

To have an idea of the possible behavior of forces it is convenient to distinguish the case of odd charges, i.e. $R>1$ and even charges, i.e. $R<1$.

For odd charges, $R>1$ the force at large distance is attractive, then if the term $f_2$ prevails the force is attractive also at small distances and nothing peculiar happens.

This is the standard situation expected in similar cases, and it is what happens, for example, for spheres: the attractive polarization forces dominate at short distances and give an attraction in this region. If, instead, $f_1$ overcomes $R f_2$, and this can happens only if $R$ is not too large, we have a repulsive force at small distances and a stable equilibrium point where $f$
vanishes.

For even charges, i.e. $R<1$, the force is repulsive at large distances, then if $f_1$, repulsive, prevails  a couple of zeros for $f$
can appears, in such a case the stationary point nearest to $\kappa = 0$ is stable, the second one is unstable.

The various possibilities can be detected by computing the values
\be R_{max}= \text{Max}\left(- \frac{f_1}{f_2}\right) \,;\qquad R_{min}= \text{Min}\left(- \frac{f_1}{f_2}\right)\,.
\label{maxmin}\ee
We are concentrated only in the small distance behavior, then we look at the possible values of $\kappa$ realizing 
\eqref{maxmin} in the range $0\leq \kappa\leq 1$. 

If $R_{max} >1$ and $R_{min}<1$ both possibilities sketched above are possible. If $R_{max} < 1$ only the second scenario is possible.
A sufficient condition for the existence of a repulsive force at short distance is evidently $-f_2(0)/f_1(0) > 0$ and this value coincides with $R_{max}$ or $R_{min}$ depending on $\tau$.

\begin{figure}[!ht]
\begin{center}
\includegraphics[width=0.49\textwidth]{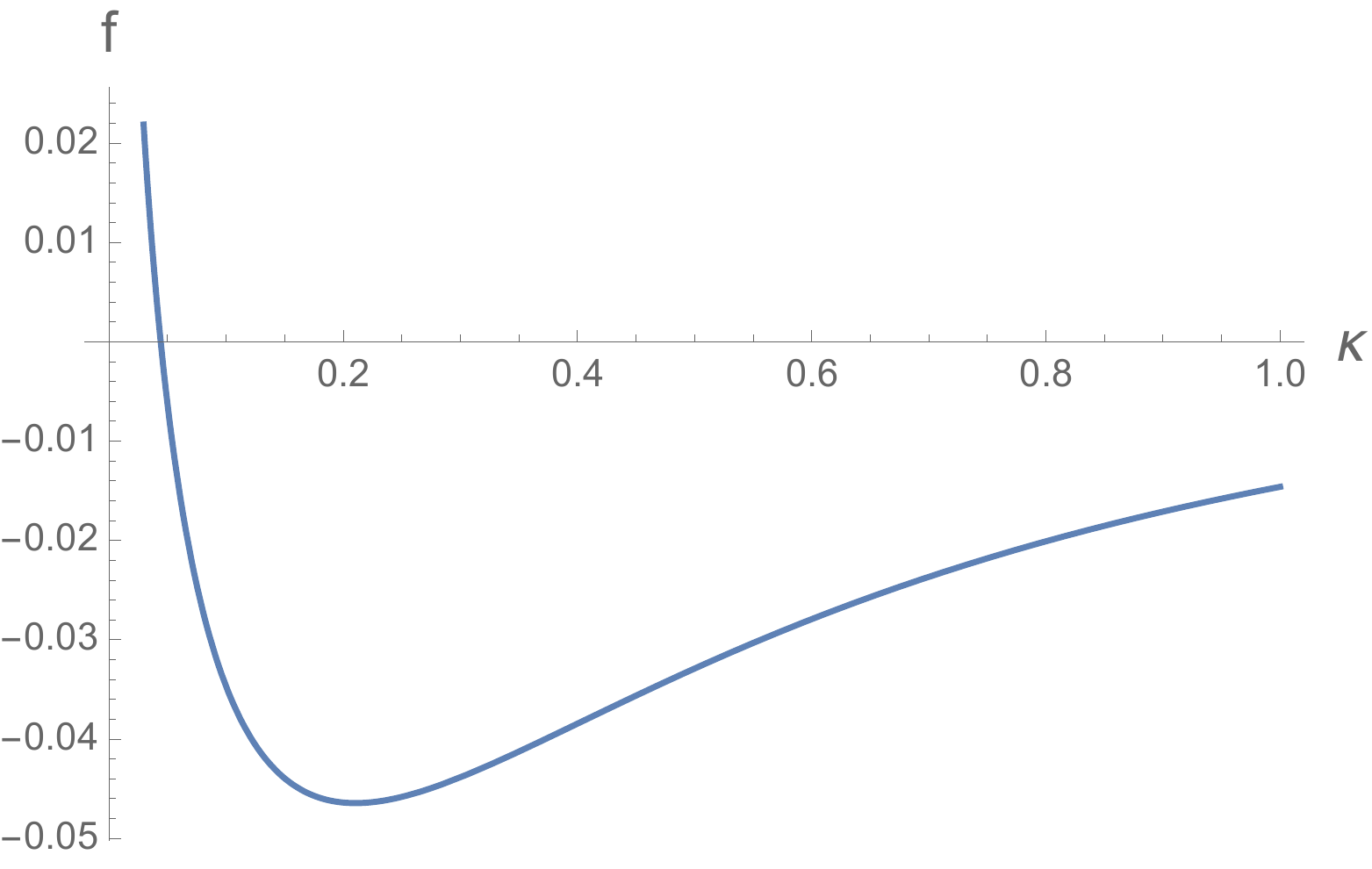}
\includegraphics[width=0.49\textwidth]{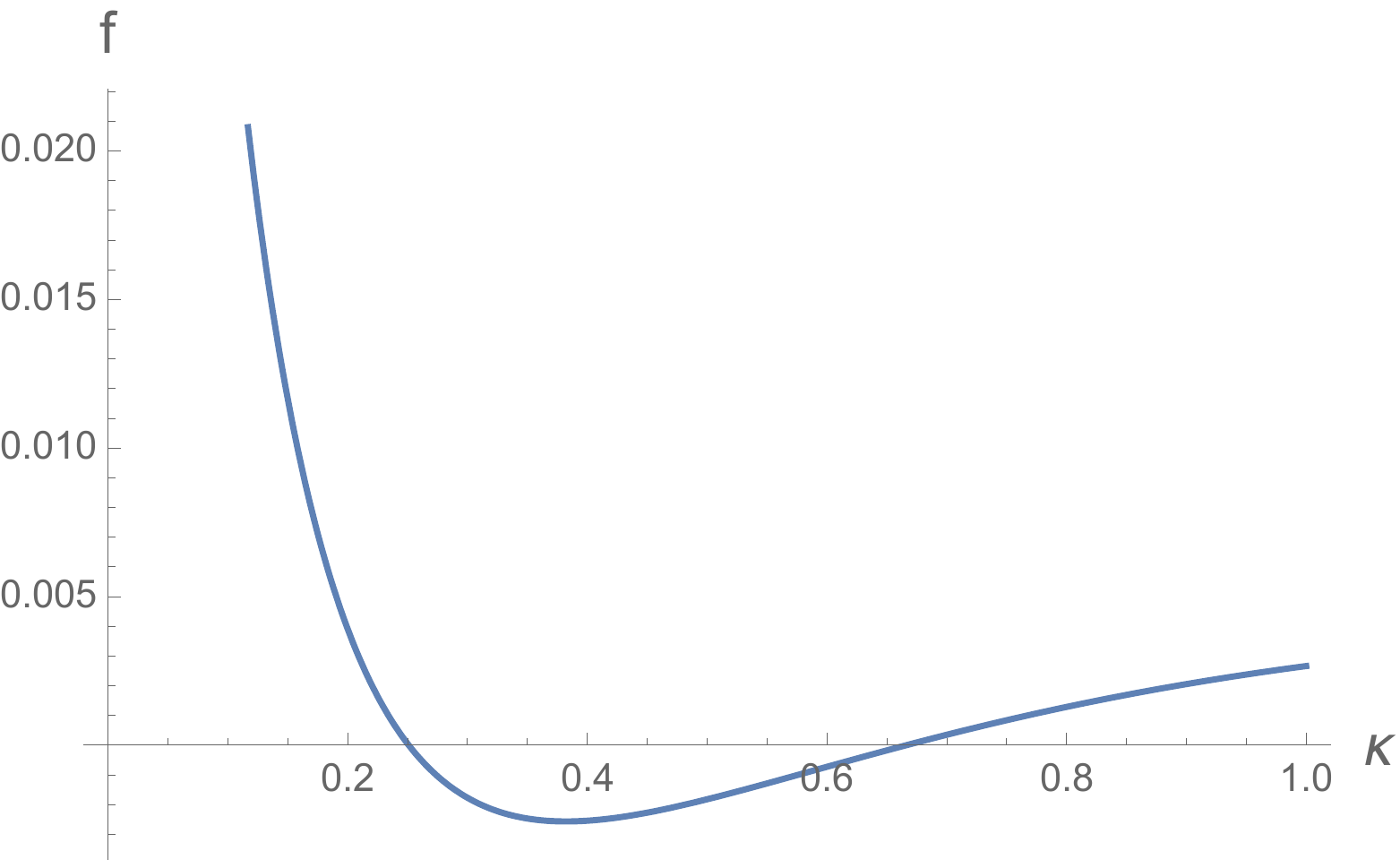}
\caption{
Graphs for the dimensionless force $f$ for $\tau=0.01$. 
First panel: $R=1.1$. Second panel: $R=0.93$.
\label{figuraF1}}
\end{center}
\end{figure}

Let us give a couple of examples. For $\tau = 0.01$ one find, from \eqref{maxmin}
\be R_{max} \simeq 1.38\,;\quad\text{for}\quad \kappa=0;\qquad R_{min} \simeq 0.918\,;\quad\text{for}\quad \kappa = 0.409\,.\label{valoriR}\ee
These values have been obtained interpolating the values in tables \ref{tabtot1}-\ref{tabtot5} and solving \eqref{maxmin}. As announced the value for $R_{max}$ is attained  for $\kappa = 0$. 
For $R< R_{min}$ the force is always repulsive, while for $R> R_{max}$ is always attractive.
Two typical graphs of $f$ for $R$ in the intermediate region are reported in figure~\ref{figuraF1}.

In the first case $R=1.1>1$, the charges are of opposite sign, the force is attractive at large distances and repulsive at short distances, the stable equilibrium point shown in the figure (where $f=0$) is at $\kappa\simeq 0.044$. In the second panel it is shown the force for $R=0.93$, In this case we have two equilibrium points, the first, stable, at $\kappa\simeq 0.251$ the second, unstable, at $\kappa\simeq 0.667$. These features can be verified also by plotting the potential energy as a function of $\kappa$.

\begin{figure}[!ht]
\begin{center}
\includegraphics[width=0.49\textwidth]{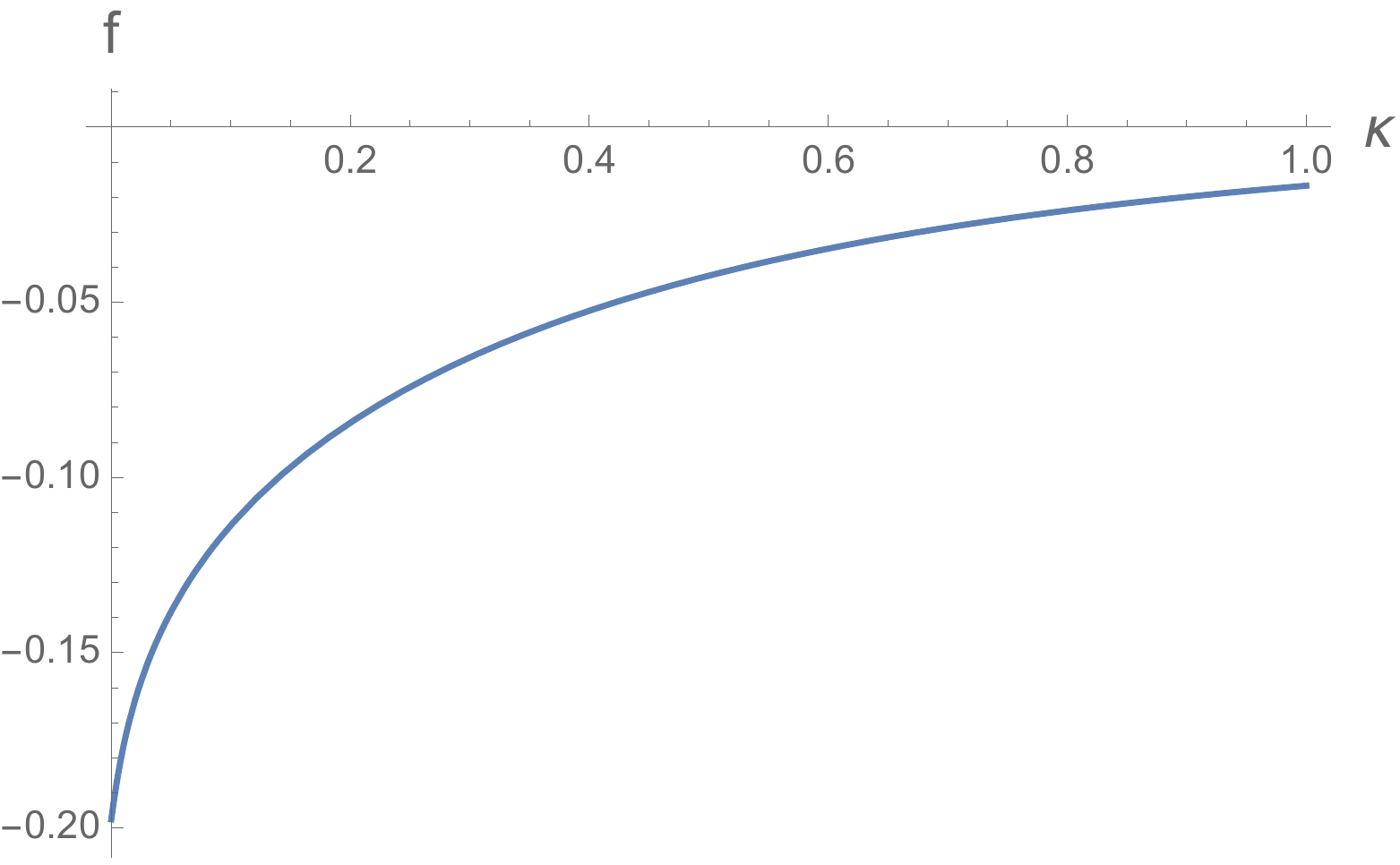}
\includegraphics[width=0.49\textwidth]{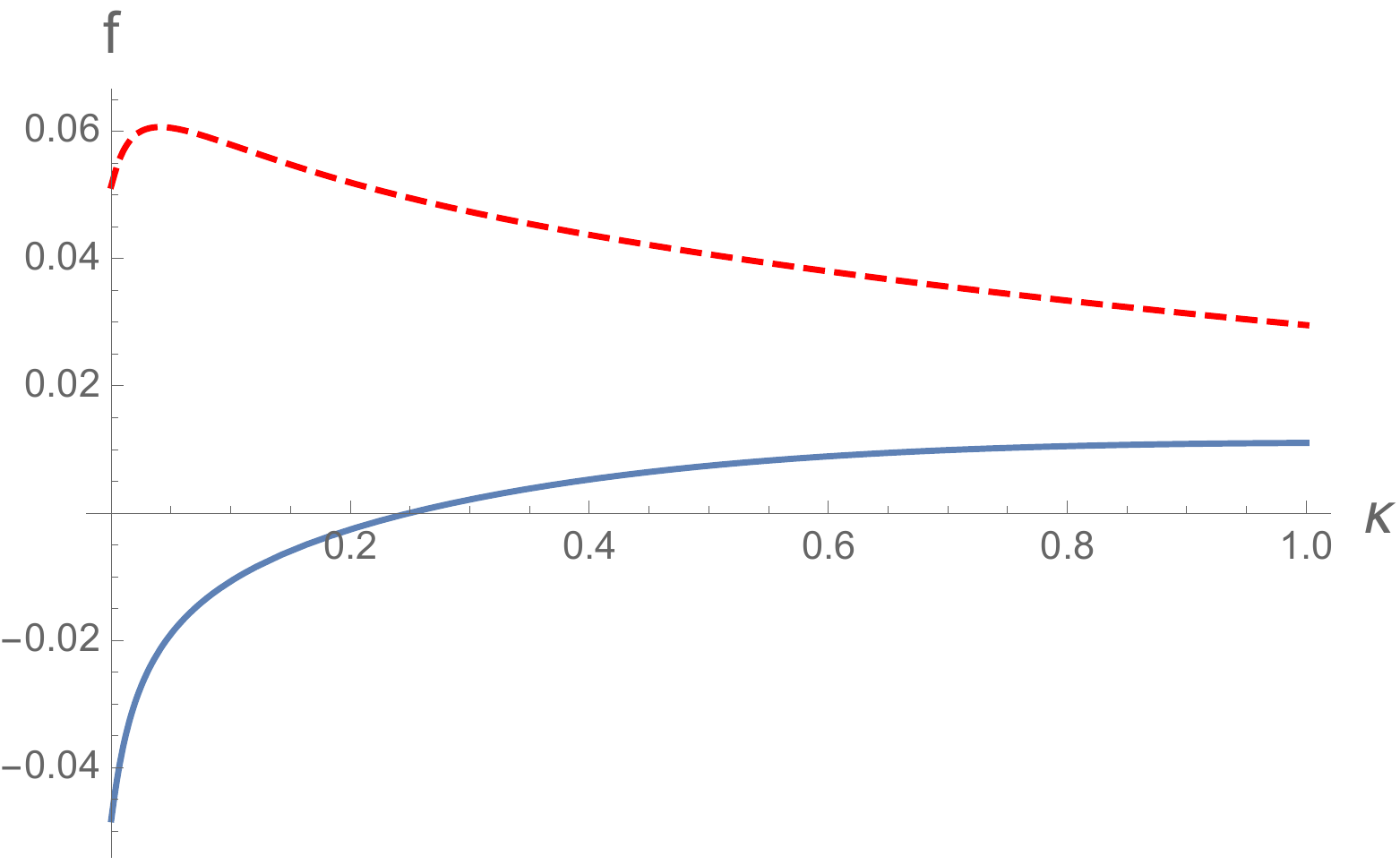}
\caption{
Graphs for the dimensionless force $f$ for $\tau=0.1$. 
First panel: $R=1.1$. Second panel: $R=0.8$ (lower continuous curve) and $R=0.6$ (upper dashed curve).\label{figuraF2}}
\end{center}
\end{figure}

As a second example we take $\tau = 0.1$, we expect in this case stronger attractive polarization forces. In effect
the function $-f_2(\kappa)/f_1(\kappa)$ is a monotonic increasing function, with a minimum at the boundary with value $R_{min} = 0.7$,
then for all values $R> 0.7$ the force is attractive at small distances. This implies that for $R>1$ (unlike charges) the force is always attractive, while for $0.7 < R < 1$ (like charges) an unstable equilibrium point develops. For $R< 0.7$ the force is everywhere repulsive. The different cases are shown in figure~\ref{figuraF2}.

From these examples it is clear that the more interesting cases are obtained for small $\tau$, when the electrodes are more similar to two flat discs. 
\section{Long distance behavior\label{sezlongdist}}
In this section we show how from data at long distances one can extract the intrinsic parameters $C_1, \alpha, D$ of the single cylinder.

\begin{figure}[!ht]
\begin{center}
\includegraphics[width=0.49\textwidth]{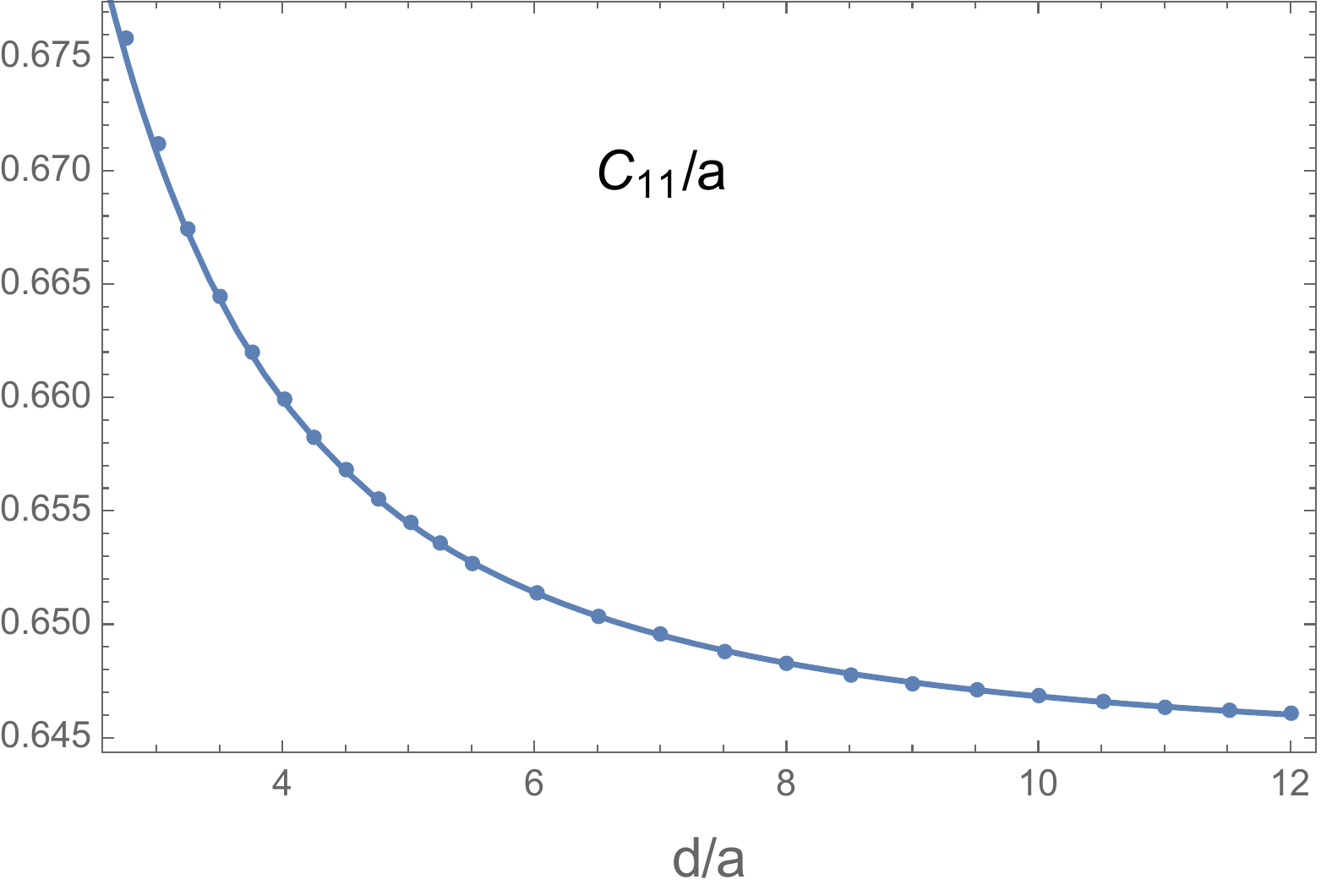}
\includegraphics[width=0.49\textwidth]{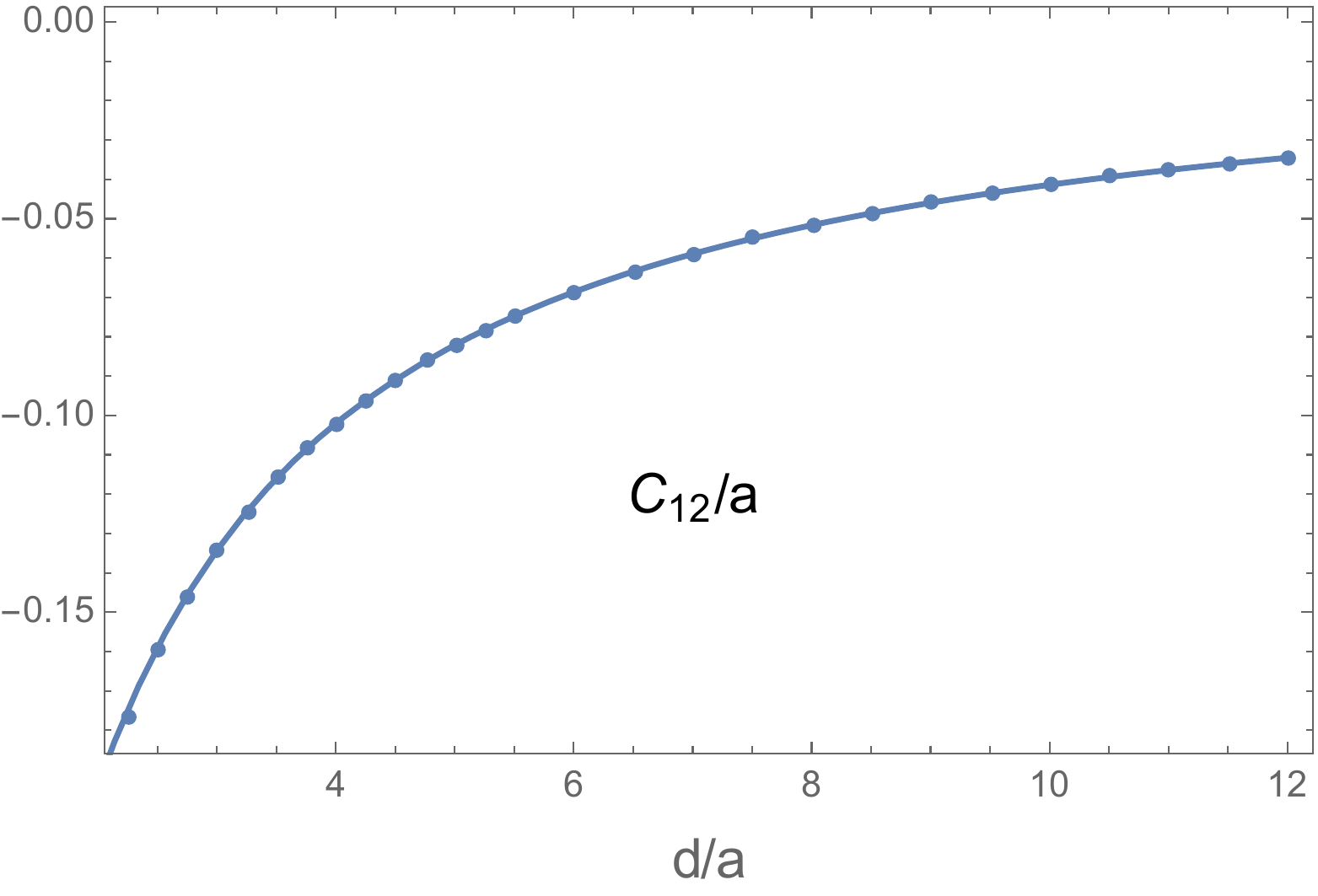}
\caption{
\label{c11c12as}}
\end{center}
\end{figure}

In the asymptotic region, $d\gg a$, the capacitance coefficients can be expanded in powers of $1/d$ (see \cite{LL}
for the leading terms and
 and \cite{mac} for the rest):
\begin{subequations}\label{sviluppoas.1}
\begin{align}
C_{11} &\simeq C_1\left(1 + \frac{C_1^2}{d^2} + \frac1{d^4}\Bigl( C_1^4 + 2 C1^2 D + C_1 \alpha\Bigr)\right)\\
C_{12} &\simeq - \frac{C_1^2}{d}\left(1 + \frac{D}{d^2} + \frac{C_1^2}{d^2}\right)
\end{align}
\end{subequations}
$d$ is the distance between the centers of the two cylinders in our case, $d = \ell+b$.
$D$ and $\alpha$ are the $zz$ components of quadrupole per unit charge and polarizability respectively.
The coefficients $C_{11}, C_{12}$ are given in terms of the computed $C_{g_1}, C$ by
\[ C_{11} = \frac12 C_{g_1} + C\,;\qquad C_{12} = \frac12 C_{g_1} - C \,.\]
To display one possible procedure for the extraction the parameters we consider two discs with thickness $\tau = b/a = 0.01$.
The intrinsic parameters computed directly with the methods of section \ref{sezsingolocond} are
\be C_1/a = 0.6441727\,;\quad \alpha/a^3 = 0.00256\,;\qquad D/a^2 = -0.676131 \label{valoriparam}\ee
The computed data are given in table~\ref{tabas}
\begin{table}[!ht]
\[
\begin{array}{lcc|lcc}
\kappa=\ell/a & C/a & C_{g_1}/a & \kappa=\ell/a & C/a & C_{g_1}/a \\[3pt]
 1.5 & 0.49467 & 0.47806 & 5.25 & 0.365857 & 0.575343 \\
 1.75 & 0.468046 & 0.491348 & 5.5 & 0.363709 & 0.578027 \\
 2. & 0.448208 & 0.502905 & 6. & 0.359979 & 0.582824 \\
 2.25 & 0.432903 & 0.513006 & 6.5 & 0.356855 & 0.586983 \\
 2.5 & 0.420768 & 0.521879 & 7. & 0.354202 & 0.590622 \\
 2.75 & 0.410931 & 0.529714 & 7.5 & 0.351921 & 0.593831 \\
 3. & 0.402811 & 0.536669 & 8. & 0.34994 & 0.59668 \\
 3.25 & 0.396004 & 0.542874 & 8.5 & 0.348203 & 0.599228 \\
 3.5 & 0.390221 & 0.548436 & 9. & 0.346669 & 0.601518 \\
 3.75 & 0.385253 & 0.553446 & 9.5 & 0.345304 & 0.603588 \\
 4. & 0.380942 & 0.557978 & 10. & 0.344082 & 0.605468 \\
 4.25 & 0.377168 & 0.562094 & 10.5 & 0.342981 & 0.607182 \\
 4.5 & 0.373838 & 0.565848 & 11. & 0.341985 & 0.608752 \\
 4.75 & 0.37088 & 0.569283 & 11.5 & 0.341079 & 0.610194 \\
 5. & 0.368235 & 0.572437 & 12. & 0.340251 & 0.611525 \\
\end{array}
\]
\caption{Values of $C, C_{g_1}$ for a system of two parallel discs with thickness $b/a = \tau = 0.01$. $\kappa = \ell/a$ is the distance between the nearest faces.
\label{tabas}}
\end{table}
\begin{table}[!ht]
\begin{center}
\begin{tabular}{l|ccc}
 & $C_1/a$ & $ D/a^2$ & $\alpha/a^3$ \\ \hline\\
 from: $M_{11}$ & 0.644172(1) & & 0.0026(1)\\
 from: $M_{12}$ &      & -0.6758(2) & \\ \hline \\
 from: $C_{11}$ & 0.644172(1) & & 0.0030(3) \\
 from: $C_{12}$ & 0.644172(1) & -0.6758(2) & \\ \hline \\
 direct calc. & 0.6441727 & -0.676131 & 0.0025619
\end{tabular}
\caption{Results for the intrinsic parameters $C_1, D, \alpha$ for a couple of discs with thickness $\tau = b/a = 0.01$.
\label{tabris1}}
\end{center}
\end{table}
and the qualitative agreement with \eqref{sviluppoas.1} is shown in figure~\ref{c11c12as}.
Table~\ref{tabas} was computed using $N=10$ polynomials in each variable ($r$ and $z$), i.e. each submatrix in \eqref{sistfin0} was $10\times10$. 

To extract the coefficients from the data one can directly perform a fit with polynomials in $1/d$ of the form \eqref{sviluppoas.1},
but a more efficient way is to consider the potential coefficients defined in \eqref{sez1.1}, which in our case are
\begin{subequations}\label{sviluppoas.2}
\begin{align}
M_{11} &= \frac{C_{11}}{C_{11}^2 - C_{12}^2} \simeq \frac{1}{C_1} - \frac{\alpha}{d^4}\\
M_{12} &= - \frac{C_{12}}{C_{11}^2 - C_{12}^2}\simeq \frac1d + \frac{D}{d^3}
\end{align}
\end{subequations}
Using the potential coefficients we can consider the contributions of $\alpha$ and $D$ separately.
We used a fit given by \eqref{sviluppoas.2} and add a further power in $1/d$. The fit was performed on points with $\kappa>7$.
The results are summarized in table~\ref{tabris1}. The estimated errors in the fit are shown in parenthesis.
The agreement with \eqref{valoriparam} is excellent, in our opinion.
In practical cases the measures of $C_{11}$ and $C_{12}$ are obtained with different instruments in different experimental conditions, see for example \cite{sensori}, then it can be problematic to combine experimental data to obtain $M_{11}$ and $M_{12}$. In these cases
one can perform first a fit for $C_{12}$, extracting $C_1$ and $D$ then insert this value of $D$ to extract $C_1$ and $\alpha$ from $C_{11}$.
The procedure can be iterated to obtain the maximum consistency of the data but we limit here to show the results for the simple 
scheme described above, see second part of table~\ref{tabris1}.
The source of the bigger error on $\alpha$ is clearly the difficulty in disentangling the $\alpha$ and $D$ parameters in $C_{11}$ when $\alpha$ is small, as in the case here considered.

The analysis can be repeated step by step for other thickness and in general it is easier for larger thickness. As an example for two discs with thickness $\tau = 0.2$ we have computed the values of $C, C_{g_1}$ given in table~\ref{tabas2}.
The result of the analysis is given in table~\ref{tabris2}.
\begin{table}[!ht]
\[
\begin{array}{lcc|lcc}
\kappa=\ell/a & C/a & C_{g_1}/a & \kappa=\ell/a & C/a & C_{g_1}/a \\[3pt]
 1.5 & 0.570937 & 0.538149 & 5.25 & 0.419552 & 0.645068 \\
 1.75 & 0.540442 & 0.552011 & 5.5 & 0.416914 & 0.648212 \\
 2. & 0.517514 & 0.564245 & 6. & 0.412323 & 0.653859 \\
 2.25 & 0.499681 & 0.575077 & 6.5 & 0.408464 & 0.658783 \\
 2.5 & 0.485441 & 0.584703 & 7. & 0.405177 & 0.663114 \\
 2.75 & 0.473825 & 0.593293 & 7.5 & 0.402345 & 0.666951 \\
 3. & 0.464181 & 0.600988 & 8. & 0.39988 & 0.670373 \\
 3.25 & 0.456055 & 0.607912 & 8.5 & 0.397715 & 0.673443 \\
 3.5 & 0.449121 & 0.614166 & 9. & 0.3958 & 0.676212 \\
 3.75 & 0.44314 & 0.619837 & 9.5 & 0.394093 & 0.678721 \\
 4. & 0.43793 & 0.624999 & 10. & 0.392562 & 0.681007 \\
 4.25 & 0.433355 & 0.629715 & 10.5 & 0.391182 & 0.683096 \\
 4.5 & 0.429305 & 0.634038 & 11. & 0.389932 & 0.685013 \\
 4.75 & 0.425698 & 0.638012 & 11.5 & 0.388794 & 0.686779 \\
 5. & 0.422465 & 0.641678 & 12. & 0.387753 & 0.68841 \\
\end{array}\]
\caption{Values of $C, C_{g_1}$ for a system of two parallel discs with thickness $b/a = \tau = 0.2$. $\kappa = \ell/a$ is the distance between the nearest faces.
\label{tabas2}}
\end{table}
\begin{table}[!h]
\begin{center}
\begin{tabular}{l|ccc}
 & $C_1/a$ & $ D/a^2$ & $\alpha/a^3$ \\ \hline\\
 from: $M_{11}$ & 0.729365(1) & & 0.06664(1)\\
 from: $M_{12}$ &      & -0.7262(2) & \\ \hline \\
 from: $C_{11}$ & 0.729365(1) & & 0.067(1) \\
 from: $C_{12}$ & 0.729364(1) & -0.7258(2) & \\ \hline \\
 direct calc. & 0.7293653 & -0.7265308 & 0.06664155
\end{tabular}
\caption{Results for the intrinsic parameters $C_1, D, \alpha$ for a couple of discs with thickness $\tau = b/a = 0.2$.
\label{tabris2}}
\end{center}
\end{table}

We give this relatively long discussion on long distance results for several reasons:
\begin{itemize}
\item[a)] The check of the asymptotic expansion \eqref{sviluppoas.1} is a quite severe test on any numerical computation of capacitance coefficients. In the cases showed above, for example, the extraction of the small parameter $\alpha$ was impossible with low precision data.
\item[b)] The expansion is very sensitive to physical and geometrical parameters. For example the absence of odd (even) powers in $1/d$ for $C_{11}$ ($C_{12}$) is due to the fact that $d$ is the distance between the \textit{centers} of the electrodes, in the notation used in this paper $d = a(\kappa + \tau)$.
\item[c)] Practical measures of the capacities are inevitably affected by a variety of disturbances, in particular offsets. The property $C_{12}\to 0$ at large distances clearly helps to eliminate the offsets in measures of this quantity. For $C_{11}$ one can use
the fact that the asymptotic value ($C_1$) must be the same value which gives the leading term in the dependence on $d$, $C_1^3/d^3$.
\end{itemize}
All these  observations do not depend on the particular form of the electrodes, for example in \cite{sensori} they have been used
for square-plates.

\section{Series of capacitors\label{sezseries}}
This last section is devoted to a classical problem, that often arises teaching electrostatics in undergraduate courses,
at least in the mind of the teacher, see also\cite{serie2,smytheLib}. In elementary courses it is always assumed that a series of two identical capacitors in series gives rise to a total mutual capacitance $C_T = C/2$ where $C$ is the mutual capacitance of a single capacitor. Clearly this formula neglects edge effects and can be interesting  to apply the machinery developed in this paper to construct an approximate solution for the problem and test it.

\begin{figure}[!ht]
\begin{center}
\includegraphics[width=0.5\textwidth]{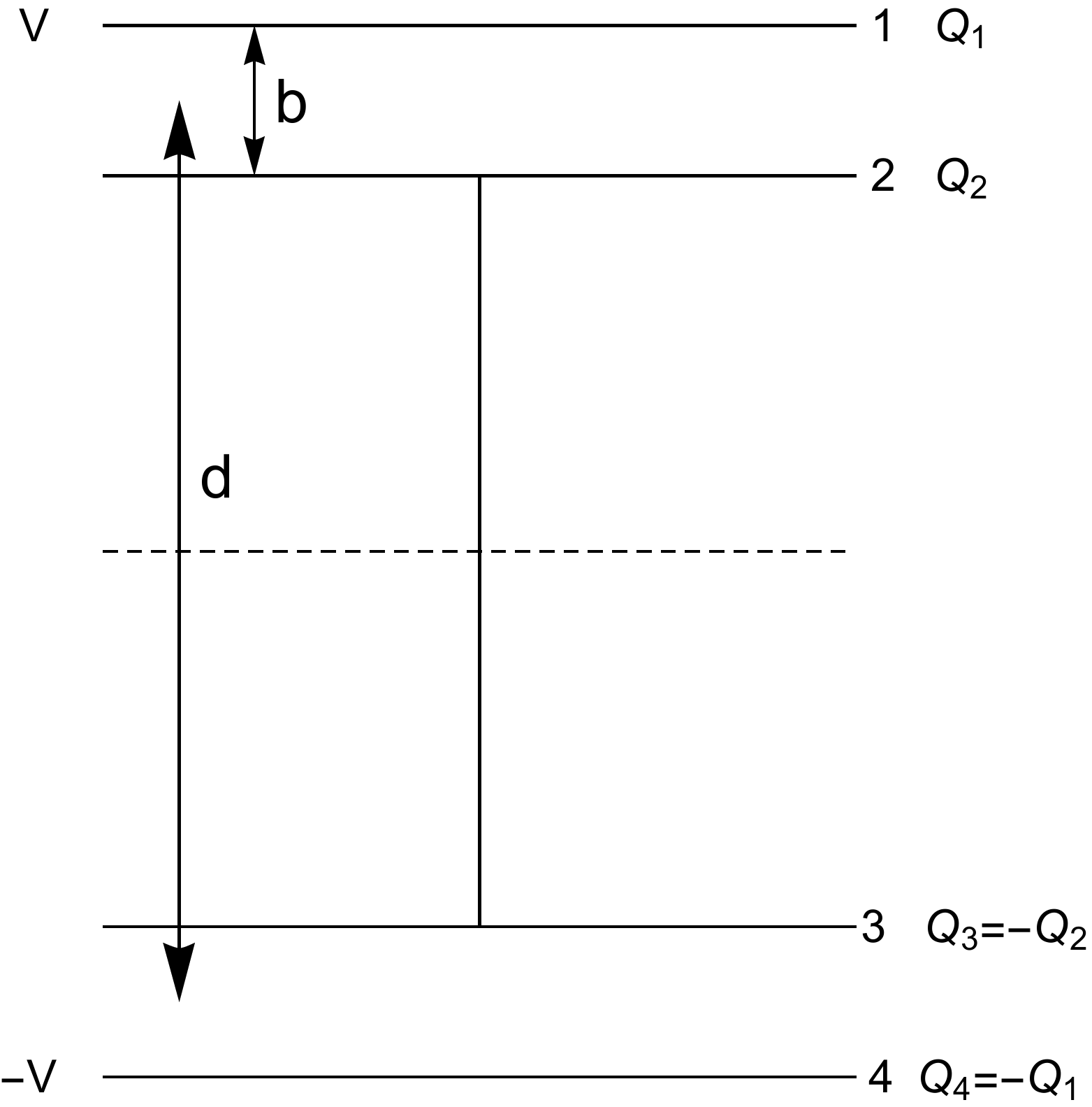}
\caption{
The geometrical variables for a system with four parallel flat discs and a connection wire, i.e. two capacitor in series.\label{schema4D}}
\end{center}
\end{figure}

A discussion similar to the one presented below is
presented in \cite{serie1}, following a different scheme, here we consider literally the textbook's scheme: two identical capacitors, each composed by two flat discs of radius $a$ connected by a tiny wire. The distance between the center of the capacitors is $d$. The scheme is shown in figure~\ref{schema4D}.

The tiny wire plays an essential role and we define the system ideal if we can neglect its contribution to the capacity. By symmetry the charges are related as shown in figure~\ref{schema4D}. What we call ``operationally'' total capacity of the system is the ratio
\be C_T = \frac{Q_1}{\Delta V} \equiv \frac{Q_1}{2 V} \label{defCT}\ee
i.e. the same definition adopted if the constituent parts of the system are closed in a black box. Our problem is to express $C_T$
in terms of the capacitance coefficients of the separated capacitors, assuming that $d\gg b,\; d\gg a$.
The key observation is that by symmetry the median plane in figure \eqref{schema4D} has potential zero, 
as the tiny conducting  wire cross this plane his potential is also zero and by consequence the potential on the discs 2 and 3 in figure is null. Now we can forget, or just cut, the wire and write down the connection between charges and potentials as depicted in the figure.
As the only potentials different from 0 are on the discs 1 and 4:
\be
Q_1 = (C_{11} - C_{14}) V \,;\qquad \Rightarrow \qquad C_T = \frac12\left( C_{11} - C_{14}\right)\label{eq4D.1}\ee
It is known that the mutual induction coefficients like $C_{14}$ go to zero as $d\to \infty$, see for example \cite{mac2}, then in a crude approximation
\be C_T \simeq \frac12 C_{11}\label{eq4D.2}\ee
Here we make another reasonable assumption, i.e. that for large $d$, the coefficient $C_{11}$ does not depend in the first approximation on the existence of the other capacitor, then is the same as in the case of a simple two discs capacitor.
The formula \eqref{eq4D.1} is exact, equation \eqref{eq4D.2} is our proposed approximation. Clearly the derivation is valid for a couple of identical capacitors of any form, the circular nature of the constituents has played no role up to now.

For $b/a=\tau\to 0$, which is a quite good approximation for commercial capacitors, we recover the naive formula. In effect
\be C_{11} = \frac 12 {C_{g_1}} + C \simeq \frac14 C_1 + C\label{eq4D.2bb}\ee
and for $\tau\to 0$, $C_{g_1} \to C_1/2$ while $C$ diverges, then $C_T\to C/2$, as expected. The problem is if
\eqref{eq4D.2} is a sensible improvement with respect the naive formula. In any case we note that
\eqref{eq4D.2bb} gives a systematic deviation from the naive formula, the deviation does not vanish in the limit $d\to\infty$ and in particular for a disc
\be C_T \to \frac12 C + \frac18 \frac{2 a}{\pi} = \frac12 C + \frac14 \frac{a}{\pi}\label{eq4D.2bb1}\ee

With our machinery is quite easy to compute $Q_1$ for the configuration shown in the figure. A short manipulation of the formulas gives the following set of linear equations
\be \sum_j {\cal M}_{ij} b_j = \frac1{2^p \Gamma(p+1)} \delta_{i,1} \,;\qquad {\cal M} = \begin{pmatrix} M^{(1,1)} & M^{(1,2)} \\ M^{(2,1)} & M^{(2,2)}\end{pmatrix}\label{matrice4D}\ee
where $p=1/2$.

Putting
\be A_{mn} = \frac{2\pi}{4n-3}\,\delta_{mn}\,;\quad
B_{mn}(x) = 4 a\,\int_0^\infty d\omega j_{2m-2}(\omega a) j_{2n-2}(\omega a) e^{-\omega x}
\ee
where $j_n(x)$ are spherical Bessel functions, we have, in the notations of figure~\ref{schema4D} for the parameters:
\begin{align}
M^{(1,1)}_{mn} &= A_{mn} - B_{mn}(d+b)\,;\qquad M^{(2,2)} = A_{mn} - B_{mn}(d - b)\\
M^{(1,2)}_{mn} &= B_{mn}(\tau) - B_{mn}(d)\,;\qquad M^{(2,1)}_{mn} = M^{(1,2)}_{nm}
\end{align}
The solution of the system determines the capacity as
\be C_T = \frac12 2\pi \frac1{2^p \Gamma(p+1)}\, b_1 \ee

\begin{table}[!ht]
\[
\begin{array}{l||ccccc||c|c}
\tau \Bigl\backslash \ell/a  & 5 & 10 & 15 & 25 & 100 & C/2a & C_{11}/2a \\ \hline 
0.1 & 1.56405 & 1.55996 & 1.55837 & 1.557 & 1.55535 & 1.46949 & 1.55478 \\
 0.01 & 12.8905 & 12.8867 & 12.8852 & 12.884 & 12.8825 & 12.8016 & 12.882 \\
 0.001 & 125.479 & 125.475 & 125.474 & 125.473 & 125.471 & 125.391 & 125.471 
 \end{array}
\]
\caption{The computed capacitance, defined as the charge on the exterior plate divided by the potential difference, for a system 
composed by two capacitors in series. The ``thickness'' (distance between near discs) is $b = \tau a$, the length of the connecting wire
is $\ell$, then $d = \ell+b$.
\label{tab4D}}
\end{table}

In table~\ref{tab4D} we give the numerical results, the naive approximation and the estimation \eqref{eq4D.2} for several values of
$\tau$ and $\ell = d-\tau$ (the distance between nearest surfaces in the different copacitors). 
We remember that for circular discs we have an analytical expression for $C_{11}$ at small $\tau$, in the general case it must 
be computed or measured.

The numerical values have been obtained by using a numerical integration routine for the matrix elements. The values reported in the last row have an estimated possible error of 2 on the last digit, while the digits of the first two rows appear to be exact.

It appears that the substitution
$C\to C_{11}$ in the formula for capacitance in series is a quite good approximation for the edge effects. 

In view of the fact that the approximation appears to be slightly lower than the computed values 
it is tempting to consider the further correction $-C_{14}$ in \eqref{eq4D.1}, computed perturbatively as $C_1^2/d$ for large $d$. But this would be incorrect as, we repeat, the approximation consists in identifying the coefficient $C_{11}$ in the case of 4 discs with the same coefficient in the case of 2 disc, the value actually reported in the table. The corrections to this identification are expected to be of order $1/d$, then the inclusion of an approximate form for $C_{14}$ appears inappropriate. In effect from the numerical table the agreement of the last column with the computed values of $C_T$ is so good that we suspect a partial cancellation 
of the corrections of order $1/d$.

At the risk of being pedantic we emphasize that these conclusions are theoretical, a direct comparison with real measurements
would be  appropriate. The main problem is the wire: the reader has surely noticed that the final equation
\eqref{eq4D.1} does not depend in any way from the exact type of connection between the two capacitors. In effect we can 
incorporate the discs 2, and 3 and the wire in a unique conductor (as they in effect are). The conductor is symmetric
by reflection with respect the intermediate symmetry plane, then is surely at zero potential for the same reasons explained above.
Equation \eqref{eq4D.1} then immediately follows from the general relations \eqref{sez1.1}. The ``ideality'' of a tiny wire
consists in the fact that we can eliminate the wire without affecting appreciably the equilibrium configuration, and in particular
that the coefficients $C_{11}$ etc. are the coefficients relative to the problem of 4 discs, without the wire. This is surely false, for example, if the intermediate part is deformed in a thick cylinder of radius $a$ and length $\ell$. 
 We think that the methods used in this
paper can be extended to deal with the more realistic case of a real wire and it would be interesting to have an idea of its
effects on $C_T$.

The general formula for two different capacitors connected by an ``ideal wire'' can be deduced from \eqref{sez1.1} imposing that
the inner part of the system has total charge null and, by definition of mutual capacitance, $Q_1 = Q = - Q_4$.
Here we limit to quote the result in the case of two different capacitors, $A$ and $B$, composed with identical components, like two discs, two squares etc:
\be C_T = \dfrac{C_A C_B}{C_A + C_B} + \frac12 \dfrac{C_{g_1}^{(A)} C_{g_1}^{(B)}}{C_{g_1}^{(A)}+ C_{g_1}^{(B)}}\,.
\label{cseriex}\ee
Here $C_A, C_B$ are the usual mutual capacitances. Equation \eqref{cseriex} reduces to \eqref{eq4D.2}, \eqref{eq4D.2bb}
for equal capacitors. We note that the usual ``inverse sum rule'' underestimates the capacity for a quantity
non vanishing as the distance goes to infinity, this is true also in the general case of arbitrary capacitors.
Let us note, for completeness, that if the same capacitors are connected in parallel the capacitance is, in the same hypothesis of ideal wires and distant capacitors, $C_{\text{parall}} = C_A + C_B$, without corrections.

% **
% **
% **
% **
% **
% **
\section{Details on the matrix elements\label{sezmatrici}}
\subsection{Two conductors}
The cylinders have radius $a$ and length $2h$. $d$ is the distance between the centers, then the distance between the nearest surfaces is $d - 2 h = \dist \equiv a \,\kappa$. In actual computations clearly one can put $a=1$. The signs $\pm$ apply to the computation of $C_g$ or $C$ respectively, as explained in previous sections.

\begin{subequations}\label{matriciX}
\begin{align}
&A^{(0)}_{mn} = a\int_0^\infty \!\!d\omega\,4 I_0(\omega)K_0(\omega)
\dfrac{J_{m-1+s}(\omega h)}{(\omega h)^s}
\dfrac{J_{n-1+s}(\omega h)}{(\omega h)^s} i^{m+n-2}\cdot X_{mn}\label{matriceA0cyl}\\
&\text{with}\qquad X_{mn}= \left\{
\begin{array}{ll}
\pm  i \sin(\omega d) & m+n\;\text{odd}\\[3pt]
(-1)^{m-1} \pm   \cos(\omega d) & m+n\;\text{even}
\end{array}
\right.\nonumber
\end{align}
\begin{align}
& A^{(1)}_{mn}  = 2\pi  (-1)^{m-1} a\int_0^\infty \!\!d\omega\, J_0(\omega a) \frac{J_{2n-2 +p}(\omega a)}{(\omega a)^p}
\left[ e^{-\omega h} \pm e^{- \omega(d-h)}\right] \dfrac{I_{m-1+s}(\omega h)}{(\omega h)^s}\\
& B^{(0)}_{mn} = A^{(1)}_{nm}\,;\qquad C^{(0)}_{mn} = A^{(2)}_{nm}\nonumber
\end{align}
\begin{align}
& A^{(2)}_{mn}  = 2\pi  a\int_0^\infty \!\!d\omega\, J_0(\omega a) \frac{J_{2n-2 +p}(\omega a)}{(\omega a)^p}
\left[  e^{-\omega h} \pm (-1)^{m-1} e^{- \omega(d+h)}\right] \dfrac{I_{m-1+s}(\omega h)}{(\omega h)^s}
\end{align}
\begin{align}
& B^{(1)}_{mn} = 
2\pi   a\int_0^\infty \!\!d\omega\, \frac{ J_{2m-2+p}(\omega a)}{(\omega a)^p} \frac{ J_{2n-2+p}(\omega a)}{(\omega a)^p}
\left[ 1 \pm e^{-\omega|d-2h|}\right]\\
& B^{(2)}_{mn} = 
2\pi   a\int_0^\infty \!\!d\omega\, \frac{ J_{2m-2+p}(\omega a)}{(\omega a)^p} \frac{ J_{2n-2+p}(\omega a)}{(\omega a)^p}
\left[ e^{-2 \omega h} \pm e^{-\omega|d|}\right]\,;\qquad
C^{(1)}_{mn} = B^{(2)}_{mn}
\\
& C^{(2)}_{mn} = 2\pi   a\int_0^\infty \!\!d\omega\,
 \frac{ J_{2m-2+p}(\omega a)}{(\omega a)^p} \frac{ J_{2n-2+p}(\omega a)}{(\omega a)^p}
\left[ 1 \pm e^{-\omega(d+2h)}\right]
\end{align}
\end{subequations}

We see that each matrix is composed by two parts, the second one depend on $d$ while the first one is the only which will be needed for the computation of the capacities of a single conductor (see below). Moreover in a computation performed on several distances this part of the matrix can be computed only once. Accordingly we will denote the separate parts of the first matrix with the notation $A^{(0,a)}, A^{(0,b)}$ and similarly for the other cases. \textit{All} the matrices elements of type $(a)$ can be computed analytically for generic values of the parameters $s, p$ of Gegenbauer end Jacobi polynomials.
Their expression is the following, we put $a=1$ to simplify the formulas:

\begin{align*}
A^{(0,a)}_{mn} &= \frac{1}{\pi h} \cdot\\
&\hskip-25pt G_{5,5}^{3,3}\left(\frac{1}{h^2}|
\begin{array}{c}
 \frac{1}{2},\frac{1}{2} (-m-n+3),-\frac{\left|
   m-n\right| }{2}+s+\frac{1}{2},\frac{1}{2} (m+n+4
   s-1),\frac{1}{2} (\left| m-n\right| +2 s+1) \\
 0,0,s,0,s+\frac{1}{2} \nonumber\\
\end{array}
\right)
\\
A^{(1,a)}_{mn} &= \frac{2\pi}{h^s}(-1)^{m-1}FJJIE(m-1+s,2n-2+p,p+s,h)
\,;\quad A^{(2,a)}_{mn} = (-1)^{m-1} A^{(1,a)}_{mn}\\
B^{(1,a)}_{mn} &= 2\pi FJJ(2 m - 2 + p, 2 n - 2 + p, 2 p)\,;
\quad
B^{(2,a)}_{mn} = 2\pi FJJE(2 m - 2 + p, 2 n - 2 + p, 2 p, 2 h)
\\
C^{(2,a)}_{mn} &= 2\pi FJJ(2 m - 2 + p, 2 n - 2 + p, 2 p)\\
B^{(1,b)}_{mn} &= 2\pi FJJE(2 m - 2 + p, 2 n - 2 + p, 2 p, d-2h)\\
B^{(2,b)}_{mn} &= 2\pi FJJE(2 m - 2 + p, 2 n - 2 + p, 2 p, d)\\
C^{(2,b)}_{mn} &= 2\pi FJJE(2 m - 2 + p, 2 n - 2 + p, 2 p, d+2h)
\end{align*}

Here $G$ is the Meijer G-function, the functions $FJJ, FJJE, FJJIE$ denote respectively integrals of two Bessel functions and a power,
two Bessel functions an exponential and a power, and finally two Bessel functions a Bessel $I$ function, an exponential and a power.
Their explicit general expression is given below, after the presentation of the matrices. We have used the notations of reference
\cite{GRR} and we used the \textit{Mathematica} software \cite{math} to perform some of the algebraic computations.
Previous matrices can be specialized to particular systems.\\
\textbf{Two hollow cylinders -}
In this case only the matrix $A^{(0)}_{mn}$ above, \eqref{matriceA0cyl}, is needed.
The natural parameter for Gegenbauer polynomials is $s=0$ in this case.\\ 
\textbf{Two flat discs -}
Only a combnation of matrices $B, C$ above appear. We give here the explicit numerical and analytical expression (with $a=1$ in the latter):
\begin{subequations}\label{matricedischi}
\begin{align}
&B_{mn} = 2\pi a\int_0^\infty d\omega \dfrac{J_{2m-2+p}(a\omega)}{(a\omega)^p}\dfrac{J_{2n-2+p}(a\omega)}{(a \omega)^p}
\left(1 \pm e^{-2 h\omega}\right) \equiv B^{(a)}_{mn} \pm B^{(b)}_{mn}
\\
& B_{mn} =
2\pi\Bigl\{FJJ(2m-2 +p,2n-2 +p,2p) \pm FJJE(2m-2 +p,2n-2 +p,2p, 2 h) \Bigr\}
\end{align}\end{subequations}
We remember that, in this case, with $a=1$, $d = 2 h$ is the distance between the discs.
The natural choice of the parameter $p$ is $p=1/2$. 
In this case the first part of the matrix, the one non depending on $h$, is diagonal and
$ B^{(a)}_{nn} = {2\pi}/{(4n-3)}$.
The system to be solved and the capacities are given by
\be \sum_j B_{ij} b_j = t_p \delta_{i1}\,;\qquad C_{g_1} = 2\pi t_p b_1\,;\qquad C = \frac12 2\pi t_p b_1 \label{capdischimat}\ee
depending on whether one chooses the plus or minus sign in \eqref{matricedischi}.

\subsection{A single conductor}
In the case of a single cylinder of radius $a$ and length $2 h$ it is convenient to write separately the contribution of even and odd Gegenbauer polyomials, in the first case we can compute the capacity and the quadrupole moment (for unit charge). In the second case we can compute the polarizability. According to our previous notations the length of the cylinder is $L = 2 h$.
In the following all matrix elements are given for cylinder of unit radius.

For the computation of $C$ and $D$ the system to be solved is given in \eqref{equazionepari}, where:
\begin{subequations}\label{uncilindro}
\begin{align}
& A^{(0)}_{mn} = a\int_0^\infty \!\!d\omega\, (-1)^{n+m}\left[ 4 I_0(\omega a) K_0(\omega a) 
\dfrac{J_{2m-2+s}(\omega h)}{(\omega h)^s}\dfrac{J_{2n-2+s}(\omega h)}{(\omega h)^s}\right]
\label{uncilindrovuoto}
\end{align}

\begin{align}
& A^{(1)}_{mn} = a\int_0^\infty  \!\!d\omega\, 4\pi e^{-\omega h} J_0(\omega a) \dfrac{I_{2m-2+s}(\omega h)}{(\omega h)^s}
\dfrac{J_{2n-2+p}(\omega h)}{(\omega h)^p} = 2 B^{(0)}_{nm}
\end{align}
\begin{align}
& B^{(1)}_{mn} = a\int_0^\infty \!\!d\omega\,2\pi (1+e^{-2 \omega h})\dfrac{J_{2m-2+p}(\omega a)}{(\omega a)^p}
\dfrac{J_{2n-2+p}(\omega a)}{(\omega a)^p}
\end{align}
\end{subequations}
The prefactor 2 in $B^{(0)}$ is due to our normalization to the charge of a single basis of the cylinder.

The analytical expression can obtained form matrix of type $(a)$ essentially by the replacement $2m\to 2m-1$ in the indices 
relative to Gegenbauer polynomials, in any case the explicit expression is
\begin{align*}
A^{(0)}_{m n} &= \frac1{\pi h}\cdot
G_{5,5}^{3,3}\left(\frac{1}{h^2}|
\begin{array}{c}
 \frac{1}{2},-m-n+\frac{5}{2},-\left| m-n\right|
   +s+\frac{1}{2},m+n+2 s-\frac{3}{2},\left| m-n\right|
   +s+\frac{1}{2} \\
 0,0,s,0,s+\frac{1}{2} \\
\end{array}
\right)\\
A^{(1)}_{m n} &= \frac{4\pi}{h^s} FJJIE(2m-2+s,2n-2+p,p+s,h)\\
B^{(1)}_{m n} &= 2\pi\Bigl\{ FJJ(2m-2+p,2n-2+p,2p) + FJJE(2m-2+p,2n-2+p,2p,2h)\Bigr\}
\end{align*}

For the computation of $\alpha$ the system to be solved is given in \eqref{equazionedispari}, where:

\begin{subequations}\label{equazionialpha1cyl}
\begin{align}
& A^{(0)}_{mn} = a\int_0^\infty \!\!d\omega\, (-1)^{n+m}\left[ 4 I_0(\omega a) K_0(\omega a) 
\dfrac{J_{2m-1+s}(\omega h)}{(\omega h)^s}\dfrac{J_{2n-1+s}(\omega h)}{(\omega h)^s}\right]
\label{equazionialpha1cylH}\\
& A^{(1)}_{mn} = a\int_0^\infty  \!\!d\omega\, 4\pi e^{-\omega h} J_0(\omega a) \dfrac{I_{2m-1+s}(\omega h)}{(\omega h)^s}
\dfrac{J_{2n-2+p}(\omega h)}{(\omega h)^p} = 2 B^{(0)}_{nm}\\
& B^{(1)}_{mn} = a\int_0^\infty \!\!d\omega\,2\pi (1-e^{-2 \omega h})\dfrac{J_{2m-2+p}(\omega a)}{(\omega a)^p}
\dfrac{J_{2n-2+p}(\omega a)}{(\omega a)^p}
\end{align}
\end{subequations}
The analytical results for the integrals are:
\[A^{(0)}_{mn} = \frac1{\pi h}
G_{5,5}^{3,3}\left(\frac{1}{h^2}|
\begin{array}{c}
 \frac{1}{2},-m-n+\frac{3}{2},-\left| m-n\right|
   +s+\frac{1}{2},m+n+2 s-\frac{1}{2},\left| m-n\right|
   +s+\frac{1}{2} \\
 0,0,s,0,s+\frac{1}{2} \\
\end{array}
\right)
\]
\[ A^{(1)} = \frac{4\pi}{h^s}\,FJJIE(2 m - 1 + s, 2 n - 2 + p, p + s, h) \]
\begin{align*}
& B^{(1)} = 2\pi\Bigl\{
FJJ(2 m - 2 + p, 2 n - 2 + p, 2 p) - 
 FJJE(2 m - 2 + p, 2 n - 2 + p, 2 p, 2 h)
\Bigr\}
\end{align*}
As in the previous section the matrices can be specialized to describe an hollow cylinder.
In both cases only the matrices $A^{(0)}_{mn}$ has to be used.
\subsection{Integrals\label{sezintegrali}}
\begin{align}
&FJJIE(\mu,\nu,\lambda,\alpha) = \int_0^\infty \!\! dt 
J_0(t) I_\mu(\alpha t) J_\nu(t) t^{-\lambda}\,e^{-\alpha t}\\
& = \frac{2^{-\lambda -5}}{\sqrt{\pi } \alpha ^{3/2}}\left\{
\frac{16 \alpha  \Gamma \left(\lambda +\frac{1}{2}\right) \Gamma
   \left(\frac{1}{4} (-2 \lambda +2 \nu +1)\right)}{\Gamma
   \left(\frac{1}{4} (2 \lambda -2 \nu +3)\right) \Gamma
   \left(\frac{1}{4} (2 \lambda +2 \nu +3)\right)^2}\cdot\right.\nonumber\\
   \,
  & \, _6F_5\left(\frac{\lambda }{2}+\frac{1}{4},\frac{\lambda
   }{2}+\frac{3}{4},\frac{1}{4}-\frac{\mu }{2},\frac{3}{4}-\frac{\mu
   }{2},\frac{\mu }{2}+\frac{1}{4},\frac{\mu
   }{2}+\frac{3}{4};\right.\nonumber\\
   &\hskip50pt\left.
   \frac{1}{2},\frac{\lambda }{2}-\frac{\nu
   }{2}+\frac{3}{4},\frac{\lambda }{2}-\frac{\nu
   }{2}+\frac{3}{4},\frac{\lambda }{2}+\frac{\nu
   }{2}+\frac{3}{4},\frac{\lambda }{2}+\frac{\nu
   }{2}+\frac{3}{4};-\frac{1}{\alpha ^2}\right)\nonumber\\
 & - \frac{(2 \mu -1) (2 \mu +1) \Gamma \left(\lambda +\frac{3}{2}\right)
   \Gamma \left(\frac{1}{4} (-2 \lambda +2 \nu -1)\right)}{\Gamma
   \left(\frac{1}{4} (2 \lambda -2 \nu +5)\right) \Gamma
   \left(\frac{1}{4} (2 \lambda +2 \nu +5)\right)^2}\cdot\nonumber\\
  & \, _6F_5\left(\frac{\lambda }{2}+\frac{3}{4},\frac{\lambda
   }{2}+\frac{5}{4},\frac{3}{4}-\frac{\mu }{2},\frac{5}{4}-\frac{\mu
   }{2},\frac{\mu }{2}+\frac{3}{4},\frac{\mu
   }{2}+\frac{5}{4}; \right.\nonumber\\
  & \hskip50pt\left.\frac{3}{2},\frac{\lambda }{2}-\frac{\nu
   }{2}+\frac{5}{4},\frac{\lambda }{2}-\frac{\nu
   }{2}+\frac{5}{4},\frac{\lambda }{2}+\frac{\nu
   }{2}+\frac{5}{4},\frac{\lambda }{2}+\frac{\nu
   }{2}+\frac{5}{4};-\frac{1}{\alpha ^2}\right)\nonumber\\
   & +  \frac{4^{\lambda -\nu +2} \alpha ^{\lambda -\nu +\frac{1}{2}} \Gamma
   \left(\lambda -\nu -\frac{1}{2}\right) \Gamma (-\lambda +\mu +\nu
   +1)}{\Gamma (\nu +1) \Gamma (\lambda +\mu -\nu )}\cdot\nonumber\\
   & \, _6F_5\left(\frac{\nu }{2}+\frac{1}{2},\frac{\nu }{2}+1,-\frac{\lambda
   }{2}-\frac{\mu }{2}+\frac{\nu }{2}+\frac{1}{2},-\frac{\lambda
   }{2}-\frac{\mu }{2}+\frac{\nu }{2}+1,-\frac{\lambda }{2}+\frac{\mu
   }{2}+\frac{\nu }{2}+\frac{1}{2},-\frac{\lambda }{2}+\frac{\mu
   }{2}+\frac{\nu }{2}+1;\right.\nonumber\\
   &\hskip50pt\left.\left. 1,-\frac{\lambda }{2}+\frac{\nu
   }{2}+\frac{3}{4},-\frac{\lambda }{2}+\frac{\nu }{2}+\frac{5}{4},\nu
   +1,\nu +1;-\frac{1}{\alpha ^2}\right)\right\}\nonumber
\end{align}

\be
\begin{split}
FJJ(\mu,\nu,\lambda) &= \int_0^\infty\!\! dt J_\mu(t) J_\nu(t) t^{-\lambda} \\
&= 
\frac{2^{-\lambda } \Gamma (\lambda ) \Gamma \left(\frac{1}{2} (-\lambda
   +\mu +\nu +1)\right)}{\Gamma \left(\frac{1}{2} (\lambda +\mu -\nu
   +1)\right) \Gamma \left(\frac{1}{2} (\lambda -\mu +\nu +1)\right)
   \Gamma \left(\frac{1}{2} (\lambda +\mu +\nu +1)\right)}
   \end{split}
   \ee
\begin{align}
&FJJE(\mu,\nu,\lambda,\alpha) = \int_0^\infty\!\! dt J_\mu(t) J_\nu(t) t^{-\lambda}\,e^{-\alpha t} 
%\\
 = 2^{-\mu -\nu } \Gamma (\mu +\nu +1) \alpha ^{\lambda -\mu -\nu -1}
   \Gamma (-\lambda +\mu +\nu +1) \nonumber\\
  &\, _4\tilde{F}_3\left(\frac{1}{2} (\mu
   +\nu +1),\frac{1}{2} (\mu +\nu +2),\frac{1}{2} (-\lambda +\mu +\nu
   +1),\frac{1}{2} (-\lambda +\mu +\nu +2);\right.\nonumber\\
  &\hskip20pt\left.\mu +1,\nu +1,\mu +\nu
   +1;-\frac{4}{\alpha ^2}\right)
\end{align}

Here $_pF_q$ is the generalized hypergeometric function, and $\tilde F$ its regularized form.

\section{Conclusion\label{sezconclusioni}}
In this work we have provided a semianalytical procedure for the computation of the capacity matrix in the case of two cylinders. The method is based on a Galerkin  expansion and in principle can be arbitrarily accurate. A parallel computation
using a Boundary Elements Method (BEM) has been used to confirm the procedure.

In the case of a single cylinder we give a quite complete numerical computation of the capacity and the basic physical parameters, quadrupole moments and polarizability. Several asymptotic expansions have been confirmed or improved.

The main result of this paper is the analysis of forces in the case of two electrodes. This study was possible due to the rather precise calculation of the different entries of the capacity matrix. A new and non-trivial scenario results from
this study. First of all the attractive or repulsive character of the forces at very short distance is shown to depends
on two parameters, the ratio of the two charges and the thickness: the two regimes are separated by a critical line.
Secondly as the thickness changes the dependence of the force on distance shows different qualitative behaviors, in general is not monotonic and one or more equilibrium positions can appear. We think that these conclusions can have some
importance in the physics of NEMS and in general in the control of the phenomena related to two close conductors. 

From the theoretical side probably the most relevant results are the necessity of a correction to the classical  Kirchhoff approximation
for circular capacitors and an explicit check of the relevance of quadrupole and polarizability in the long
distance behavior of the capacity coefficients.

The method can have different applications and, as an example, we treat the classical problem of the effective capacity of 
two capacitors in series.

There are several points in which this work can be improved. 

From the numerical point of view we think that better routines
for the computation of generalized hypergeometric series would help. Some integrals of product of Bessel functions have to be computed numerically,  a more systematic analysis of the validity of the Levin algorithm for this kind of computations
could improve the performances of the computation. 

On the theoretical side it would be useful to push the methods of asymptotic matching in order to compute analytically
the corrections to the Kirchhoff formula and extract a closed formula for the forces at short distances.

Finally an explicit experimental study of electrostatic forces at short distance could verify the rather complex scenario outlined in this paper and this would be relevant in experiments involving micro-conductors.

\newpage
% tabella 1

\begin{table}[!h]
\begin{tabular}{lccc|ccc}
  & \multicolumn{3}{c}{$C$} & \multicolumn{3}{c}{$C_{g_1}$}\\[3pt]
  \cline{2-7}\\[-5pt]
$\kappa$ & $\tau=0.001$  & $\tau=0.005$ & $\tau=0.01$ & $\tau=0.001$  & $\tau=0.005$ & $\tau=0.01$ \\[5pt]

0.0001 & 2501.223 & 2501.352 & 2501.411 & 0.3192679 & 0.3221191 &
   0.3251991 \\
 0.0003 & 834.3982 & 834.5216 & 834.5796 & 0.3193477 & 0.3221841 &
   0.3252575 \\
 0.0005 & 500.9901 & 501.1087 & 501.1659 & 0.3194259 & 0.3222489 &
   0.3253158 \\
 0.0007 & 358.0857 & 358.2000 & 358.2565 & 0.3195028 & 0.3223132 &
   0.3253738 \\
 0.0009 & 278.6861 & 278.7966 & 278.8524 & 0.3195785 & 0.3223772 &
   0.3254317 \\
 0.001 & 250.8942 & 251.0028 & 251.0583 & 0.3196160 & 0.3224091 & 0.3254606
   \\
 0.003 & 84.09039 & 84.17478 & 84.22449 & 0.3203235 & 0.3230306 & 0.3260289
   \\
 0.005 & 50.70057 & 50.77141 & 50.81677 & 0.3209783 & 0.3236269 & 0.3265818
   \\
 0.007 & 36.38006 & 36.44190 & 36.48382 & 0.3215990 & 0.3242030 & 0.3271215
   \\
 0.009 & 28.41873 & 28.47403 & 28.51315 & 0.3221944 & 0.3247624 & 0.3276494
   \\
 0.01 & 25.63082 & 25.68346 & 25.72136 & 0.3224843 & 0.3250366 & 0.3279094
   \\
 0.015 & 17.25995 & 17.30296 & 17.33600 & 0.3238727 & 0.3263620 & 0.3291745
   \\
 0.02 & 13.06796 & 13.10483 & 13.13437 & 0.3251823 & 0.3276249 & 0.3303900
   \\
 0.03 & 8.867306 & 8.896597 & 8.921384 & 0.3276353 & 0.3300106 & 0.3327039
   \\
 0.04 & 6.761012 & 6.785714 & 6.807349 & 0.3299276 & 0.3322550 & 0.3348949
   \\
 0.05 & 5.493850 & 5.515425 & 5.534785 & 0.3321003 & 0.3343908 & 0.3369886
   \\
 0.06 & 4.646916 & 4.666201 & 4.683826 & 0.3341780 & 0.3364387 & 0.3390020
   \\
 0.07 & 4.040478 & 4.058002 & 4.074253 & 0.3361768 & 0.3384128 & 0.3409470
   \\
 0.08 & 3.584575 & 3.600696 & 3.615825 & 0.3381084 & 0.3403234 & 0.3428326
   \\
 0.09 & 3.229173 & 3.244147 & 3.258339 & 0.3399814 & 0.3421782 & 0.3446656
   \\
 0.1 & 2.944223 & 2.958238 & 2.971635 & 0.3418026 & 0.3439835 & 0.3464516
   \\
 0.15 & 2.084626 & 2.095491 & 2.106210 & 0.3502895 & 0.3524127 & 0.3548093
   \\
 0.2 & 1.650705 & 1.659786 & 1.668890 & 0.3580009 & 0.3600877 & 0.3624390
   \\
 0.3 & 1.211879 & 1.218968 & 1.226225 & 0.3718390 & 0.3738829 & 0.3761789
   \\
 0.4 & 0.9895156 & 0.9954926 & 1.001691 & 0.3841757 & 0.3861973 & 0.3884624
   \\
 0.5 & 0.8546925 & 0.8599492 & 0.8654487 & 0.3954077 & 0.3974177 &
   0.3996653 \\
 0.6 & 0.7640662 & 0.7688134 & 0.7738120 & 0.4057606 & 0.4077658 &
   0.4100042 \\
 0.7 & 0.6989163 & 0.7032818 & 0.7079013 & 0.4153777 & 0.4173828 &
   0.4196177 \\
 0.8 & 0.6498148 & 0.6538824 & 0.6582036 & 0.4243559 & 0.4263645 &
   0.4286002 \\
 0.9 & 0.6114886 & 0.6153164 & 0.6193961 & 0.4327651 & 0.4347798 &
   0.4370196 \\
 1. & 0.5807536 & 0.5843839 & 0.5882633 & 0.4406580 & 0.4426811 & 0.4449277
   \\ \hline\hline
 \multicolumn{7}{c}{Extrapolated values}\\ \hline

0.0001 & 2501.235 & 2501.364 & 2501.423 & 0.3192680 & 0.3221191 &
   0.3251991 \\
 0.0003 & 834.4006 & 834.5242 & 834.5822 & 0.3193478 & 0.3221842 &
   0.3252575 \\
 0.0005 & 500.9907 & 501.1094 & 501.1666 & 0.3194260 & 0.3222489 &
   0.3253158 \\
 0.0007 & 358.0859 & 358.2003 & 358.2568 & 0.3195029 & 0.3223132 &
   0.3253738 \\
 0.0009 & 278.6862 & 278.7967 & 278.8525 & 0.3195786 & 0.3223772 &
   0.3254317 \\
 0.001 & 250.8942 & 251.0029 & 251.0584 & 0.3196166 & 0.3224093 & 0.3254607
   \\
 0.003 & 84.09040 & 84.17479 & 84.22449 & 0.3203235 & 0.3230307 & 0.3260289
   \\
 0.005 & 50.70057 & 50.77141 & 50.81677 & 0.3209783 & 0.3236269 & 0.3265819

\end{tabular}
\caption{ $C$ and $C_{g_1}$ for two cylinders at distances $d= \kappa a$ for three different
thickness $b = \tau a$. The lower part of the table contains the extrapolated values using the procedure
\eqref{proceduraN}.
\label{tabtot1}}
\end{table}

% tabella 2

\begin{table}[!h]
\begin{tabular}{lccc|ccc}
  & \multicolumn{3}{c}{$C$} & \multicolumn{3}{c}{$C_{g_1}$}\\[3pt]
  \cline{2-7}\\[-5pt]
$\kappa$ & $\tau=0.05$  & $\tau=0.075$ & $\tau=0.1$ & $\tau=0.05$  & $\tau=0.075$ & $\tau=0.1$ \\[5pt]

0.0001 & 2501.561 & 2501.604 & 2501.634 & 0.3447596 & 0.3551037 &
   0.3647013 \\
 0.0003 & 834.7292 & 834.7724 & 834.8047 & 0.3448029 & 0.3551434 &
   0.3647385 \\
 0.0005 & 501.3149 & 501.3580 & 501.3905 & 0.3448462 & 0.3551830 &
   0.3647757 \\
 0.0007 & 358.4049 & 358.4479 & 358.4805 & 0.3448894 & 0.3552226 &
   0.3648128 \\
 0.0009 & 279.0001 & 279.0431 & 279.0757 & 0.3449326 & 0.3552622 &
   0.3648500 \\
 0.001 & 251.2057 & 251.2487 & 251.2812 & 0.3449541 & 0.3552820 & 0.3648685
   \\
 0.003 & 84.36617 & 84.40859 & 84.44091 & 0.3453836 & 0.3556765 & 0.3652388
   \\
 0.005 & 50.95342 & 50.99530 & 51.02735 & 0.3458094 & 0.3560686 & 0.3656073
   \\
 0.007 & 36.61599 & 36.65736 & 36.68913 & 0.3462317 & 0.3564583 & 0.3659739
   \\
 0.009 & 28.64129 & 28.68216 & 28.71366 & 0.3466506 & 0.3568456 & 0.3663388
   \\
 0.01 & 25.84763 & 25.88825 & 25.91963 & 0.3468588 & 0.3570384 & 0.3665205
   \\
 0.015 & 17.45406 & 17.49354 & 17.52429 & 0.3478880 & 0.3579940 & 0.3674230
   \\
 0.02 & 13.24571 & 13.28414 & 13.31430 & 0.3488987 & 0.3589365 & 0.3683151
   \\
 0.03 & 9.022188 & 9.058769 & 9.087844 & 0.3508697 & 0.3607843 & 0.3700703
   \\
 0.04 & 6.900157 & 6.935140 & 6.963254 & 0.3527809 & 0.3625870 & 0.3717892
   \\
 0.05 & 5.621241 & 5.654827 & 5.682075 & 0.3546391 & 0.3643486 & 0.3734746
   \\
 0.06 & 4.765070 & 4.797422 & 4.823885 & 0.3564500 & 0.3660726 & 0.3751288
   \\
 0.07 & 4.151119 & 4.182368 & 4.208117 & 0.3582180 & 0.3677618 & 0.3767539
   \\
 0.08 & 3.688944 & 3.719202 & 3.744295 & 0.3599469 & 0.3694189 & 0.3783517
   \\
 0.09 & 3.328205 & 3.357565 & 3.382054 & 0.3616398 & 0.3710461 & 0.3799237
   \\
 0.1 & 3.038641 & 3.067182 & 3.091112 & 0.3632995 & 0.3726451 & 0.3814714
   \\
 0.15 & 2.162767 & 2.188078 & 2.209729 & 0.3711717 & 0.3802724 & 0.3888869
   \\
 0.2 & 1.718754 & 1.741771 & 1.761733 & 0.3784672 & 0.3873897 & 0.3958448
   \\
 0.3 & 1.267727 & 1.287639 & 1.305218 & 0.3917772 & 0.4004567 & 0.4086880
   \\
 0.4 & 1.038060 & 1.055924 & 1.071875 & 0.4038021 & 0.4123260 & 0.4204095
   \\
 0.5 & 0.8982849 & 0.9146746 & 0.9294246 & 0.4148419 & 0.4232614 &
   0.4312429 \\
 0.6 & 0.8040410 & 0.8193068 & 0.8331258 & 0.4250784 & 0.4334273 &
   0.4413376 \\
 0.7 & 0.7361124 & 0.7504873 & 0.7635589 & 0.4346320 & 0.4429340 &
   0.4507952 \\
 0.8 & 0.6847987 & 0.6984466 & 0.7109019 & 0.4435861 & 0.4518592 &
   0.4596880 \\
 0.9 & 0.6446630 & 0.6577041 & 0.6696406 & 0.4520019 & 0.4602602 &
   0.4680699 \\
 1. & 0.6124160 & 0.6249419 & 0.6364347 & 0.4599263 & 0.4681811 & 0.4759824
   \\ \hline\hline
 \multicolumn{7}{c}{Extrapolated values}\\ \hline

0.0001 & 2501.573 & 2501.616 & 2501.647 & 0.3447596 & 0.3551037 &
   0.3647013 \\
 0.0003 & 834.7319 & 834.7751 & 834.8073 & 0.3448029 & 0.3551434 &
   0.3647385 \\
 0.0005 & 501.3156 & 501.3587 & 501.3912 & 0.3448462 & 0.3551830 &
   0.3647757 \\
 0.0007 & 358.4051 & 358.4482 & 358.4807 & 0.3448894 & 0.3552226 &
   0.3648128 \\
 0.0009 & 279.0002 & 279.0432 & 279.0758 & 0.3449326 & 0.3552622 &
   0.3648500 \\
 0.001 & 251.2058 & 251.2488 & 251.2813 & 0.3449542 & 0.3552820 & 0.3648685
   \\
 0.003 & 84.36618 & 84.40860 & 84.44092 & 0.3453836 & 0.3556766 & 0.3652388
   \\
 0.005 & 50.95342 & 50.99530 & 51.02735 & 0.3458094 & 0.3560686 & 0.3656073

\end{tabular}
\caption{ $C$ and $C_{g_1}$ for two cylinders at distances $d= \kappa a$ for three different
thickness $b = \tau a$. The lower part of the table contains the extrapolated values using the procedure
\eqref{proceduraN}.
\label{tabtot2}}
\end{table}

% tabella 3

\begin{table}[!h]
\begin{tabular}{lccc|ccc}
  & \multicolumn{3}{c}{$C$} & \multicolumn{3}{c}{$C_{g_1}$}\\[3pt]
  \cline{2-7}\\[-5pt]
$\kappa$ & $\tau=0.125$  & $\tau=0.15$ & $\tau=0.175$ & $\tau=0.125$  & $\tau=0.15$ & $\tau=0.175$ \\[5pt]

0.0001 & 2501.663 & 2501.687 & 2501.705 & 0.3737611 & 0.3824042 &
   0.3907095 \\
 0.0003 & 834.8326 & 834.8557 & 834.8748 & 0.3737964 & 0.3824381 &
   0.3907421 \\
 0.0005 & 501.4176 & 501.4406 & 501.4608 & 0.3738318 & 0.3824719 &
   0.3907747 \\
 0.0007 & 358.5074 & 358.5304 & 358.5507 & 0.3738671 & 0.3825057 &
   0.3908073 \\
 0.0009 & 279.1026 & 279.1256 & 279.1459 & 0.3739023 & 0.3825395 &
   0.3908399 \\
 0.001 & 251.3082 & 251.3313 & 251.3514 & 0.3739200 & 0.3825564 & 0.3908562
   \\
 0.003 & 84.46754 & 84.49040 & 84.51058 & 0.3742719 & 0.3828937 & 0.3911814
   \\
 0.005 & 51.05378 & 51.07651 & 51.09662 & 0.3746224 & 0.3832298 & 0.3915055
   \\
 0.007 & 36.71539 & 36.73800 & 36.75802 & 0.3749715 & 0.3835647 & 0.3918286
   \\
 0.009 & 28.73975 & 28.76224 & 28.78218 & 0.3753190 & 0.3838984 & 0.3921507
   \\
 0.01 & 25.94563 & 25.96806 & 25.98796 & 0.3754923 & 0.3840648 & 0.3923114
   \\
 0.015 & 17.54988 & 17.57204 & 17.59172 & 0.3763534 & 0.3848924 & 0.3931109
   \\
 0.02 & 13.33951 & 13.36139 & 13.38087 & 0.3772062 & 0.3857129 & 0.3939043
   \\
 0.03 & 9.112340 & 9.133713 & 9.152811 & 0.3788876 & 0.3873334 & 0.3954733
   \\
 0.04 & 6.987099 & 7.008001 & 7.026741 & 0.3805387 & 0.3889281 & 0.3970196
   \\
 0.05 & 5.705325 & 5.725790 & 5.744194 & 0.3821616 & 0.3904982 & 0.3985444
   \\
 0.06 & 4.846587 & 4.866645 & 4.884733 & 0.3837579 & 0.3920452 & 0.4000486
   \\
 0.07 & 4.230311 & 4.249990 & 4.267779 & 0.3853291 & 0.3935701 & 0.4015332
   \\
 0.08 & 3.766019 & 3.785340 & 3.802849 & 0.3868765 & 0.3950741 & 0.4029990
   \\
 0.09 & 3.403339 & 3.422325 & 3.439568 & 0.3884015 & 0.3965580 & 0.4044467
   \\
 0.1 & 3.111987 & 3.130657 & 3.147647 & 0.3899051 & 0.3980229 & 0.4058773
   \\
 0.15 & 2.228888 & 2.246211 & 2.262111 & 0.3971343 & 0.4050866 & 0.4127922
   \\
 0.2 & 1.779578 & 1.795840 & 1.810860 & 0.4039488 & 0.4117709 & 0.4193578
   \\
 0.3 & 1.321150 & 1.335828 & 1.349509 & 0.4165852 & 0.4242158 & 0.4316244
   \\
 0.4 & 1.086460 & 1.099995 & 1.112688 & 0.4281666 & 0.4356644 & 0.4429471
   \\
 0.5 & 0.9429962 & 0.9556568 & 0.9675830 & 0.4389012 & 0.4463039 &
   0.4534951 \\
 0.6 & 0.8459005 & 0.8578656 & 0.8691745 & 0.4489255 & 0.4562589 &
   0.4633825 \\
 0.7 & 0.7756869 & 0.7870818 & 0.7978805 & 0.4583331 & 0.4656164 &
   0.4726902 \\
 0.8 & 0.7224922 & 0.7334088 & 0.7437773 & 0.4671917 & 0.4744397 &
   0.4814776 \\
 0.9 & 0.6807750 & 0.6912843 & 0.7012833 & 0.4755518 & 0.4827761 &
   0.4897895 \\
 1. & 0.6471769 & 0.6573333 & 0.6670116 & 0.4834526 & 0.4906632 & 0.4976610
   \\ \hline\hline
 \multicolumn{7}{c}{Extrapolated values}\\ \hline

0.0001 & 2501.676 & 2501.699 & 2501.717 & 0.3737611 & 0.3824042 &
   0.3907095 \\
 0.0003 & 834.8352 & 834.8583 & 834.8775 & 0.3737964 & 0.3824381 &
   0.3907421 \\
 0.0005 & 501.4183 & 501.4413 & 501.4615 & 0.3738318 & 0.3824719 &
   0.3907747 \\
 0.0007 & 358.5077 & 358.5307 & 358.5510 & 0.3738671 & 0.3825057 &
   0.3908073 \\
 0.0009 & 279.1027 & 279.1257 & 279.1460 & 0.3739023 & 0.3825395 &
   0.3908399 \\
 0.001 & 251.3083 & 251.3313 & 251.3514 & 0.3739200 & 0.3825564 & 0.3908562
   \\
 0.003 & 84.46755 & 84.49040 & 84.51058 & 0.3742719 & 0.3828937 & 0.3911814
   \\
 0.005 & 51.05378 & 51.07651 & 51.09662 & 0.3746224 & 0.3832298 & 0.3915055

\end{tabular}
\caption{ $C$ and $C_{g_1}$ for two cylinders at distances $d= \kappa a$ for three different
thickness $b = \tau a$. The lower part of the table contains the extrapolated values using the procedure
\eqref{proceduraN}.
\label{tabtot3}}
\end{table}

% tabella 4

\begin{table}[!h]
\begin{tabular}{lccc|ccc}
  & \multicolumn{3}{c}{$C$} & \multicolumn{3}{c}{$C_{g_1}$}\\[3pt]
  \cline{2-7}\\[-5pt]
$\kappa$ & $\tau=0.2$  & $\tau=0.225$ & $\tau=0.250$ & $\tau=0.2$  & $\tau=0.225$ & $\tau=0.250$ \\[5pt]

0.0001 & 2501.723 & 2501.741 & 2501.757 & 0.3987322 & 0.4065130 &
   0.4140833 \\
 0.0003 & 834.8925 & 834.9099 & 834.9258 & 0.3987637 & 0.4065437 &
   0.4141132 \\
 0.0005 & 501.4791 & 501.4959 & 501.5119 & 0.3987953 & 0.4065743 &
   0.4141431 \\
 0.0007 & 358.5690 & 358.5857 & 358.6013 & 0.3988269 & 0.4066050 &
   0.4141723 \\
 0.0009 & 279.1641 & 279.1808 & 279.1965 & 0.3988584 & 0.4066357 &
   0.4142021 \\
 0.001 & 251.3696 & 251.3864 & 251.4018 & 0.3988742 & 0.4066510 & 0.4142177
   \\
 0.003 & 84.52879 & 84.54546 & 84.56082 & 0.3991891 & 0.4069570 & 0.4145173
   \\
 0.005 & 51.11476 & 51.13137 & 51.14671 & 0.3995031 & 0.4072622 & 0.4148144
   \\
 0.007 & 36.77610 & 36.79265 & 36.80800 & 0.3998162 & 0.4075666 & 0.4151098
   \\
 0.009 & 28.80019 & 28.81669 & 28.83199 & 0.4001284 & 0.4078702 & 0.4154057
   \\
 0.01 & 26.00593 & 26.02241 & 26.03765 & 0.4002842 & 0.4080217 & 0.4155546
   \\
 0.015 & 17.60954 & 17.62588 & 17.64103 & 0.4010596 & 0.4087763 & 0.4162903
   \\
 0.02 & 13.39853 & 13.41475 & 13.42983 & 0.4018297 & 0.4095260 & 0.4170208
   \\
 0.03 & 9.170167 & 9.186155 & 9.201006 & 0.4033542 & 0.4110109 & 0.4184720
   \\
 0.04 & 7.043814 & 7.059566 & 7.074201 & 0.4048585 & 0.4124784 & 0.4199078
   \\
 0.05 & 5.761000 & 5.776548 & 5.791027 & 0.4063435 & 0.4139273 & 0.4213238
   \\
 0.06 & 4.901286 & 4.916607 & 4.930912 & 0.4078101 & 0.4153616 & 0.4227280
   \\
 0.07 & 4.284094 & 4.299217 & 4.313354 & 0.4092590 & 0.4167785 & 0.4241162
   \\
 0.08 & 3.818934 & 3.833870 & 3.847849 & 0.4106910 & 0.4181799 & 0.4254899
   \\
 0.09 & 3.455438 & 3.470189 & 3.484018 & 0.4121063 & 0.4195665 & 0.4268499
   \\
 0.1 & 3.163309 & 3.177890 & 3.191569 & 0.4135062 & 0.4209384 & 0.4281966
   \\
 0.15 & 2.276867 & 2.290681 & 2.303703 & 0.4202865 & 0.4275960 & 0.4347416
   \\
 0.2 & 1.824875 & 1.838052 & 1.850521 & 0.4267428 & 0.4339514 & 0.4410034
   \\
 0.3 & 1.362370 & 1.374542 & 1.386123 & 0.4388431 & 0.4458959 & 0.4528017
   \\
 0.4 & 1.124683 & 1.136087 & 1.146981 & 0.4500465 & 0.4569858 & 0.4637835
   \\
 0.5 & 0.9788971 & 0.9896904 & 1.000031 & 0.4605062 & 0.4673607 & 0.4740770
   \\
 0.6 & 0.8799353 & 0.8902274 & 0.9001128 & 0.4703277 & 0.4771181 &
   0.4837719 \\
 0.7 & 0.8081801 & 0.8180524 & 0.8275524 & 0.4795863 & 0.4863284 &
   0.4929347 \\
 0.8 & 0.7536856 & 0.7631990 & 0.7723684 & 0.4883376 & 0.4950438 &
   0.5016142 \\
 0.9 & 0.7108542 & 0.7200572 & 0.7289389 & 0.4966241 & 0.5033042 &
   0.5098484 \\
 1. & 0.6762877 & 0.6852185 & 0.6938469 & 0.5044790 & 0.5111415 & 0.5176676
   \\ \hline\hline
 \multicolumn{7}{c}{Extrapolated values}\\ \hline

0.0001 & 2501.736 & 2501.753 & 2501.769 & 0.3987322 & 0.4065130 &
   0.4140833 \\
 0.0003 & 834.8951 & 834.9126 & 834.9285 & 0.3987637 & 0.4065437 &
   0.4141132 \\
 0.0005 & 501.4798 & 501.4967 & 501.5126 & 0.3987953 & 0.4065743 &
   0.4141431 \\
 0.0007 & 358.5693 & 358.5860 & 358.6015 & 0.3988269 & 0.4066050 &
   0.4141723 \\
 0.0009 & 279.1642 & 279.1809 & 279.1966 & 0.3988584 & 0.4066357 &
   0.4142021 \\
 0.001 & 251.3697 & 251.3864 & 251.4019 & 0.3988742 & 0.4066510 & 0.4142177
   \\
 0.003 & 84.52880 & 84.54547 & 84.56083 & 0.3991891 & 0.4069570 & 0.4145173
   \\
 0.005 & 51.11477 & 51.13138 & 51.14671 & 0.3995031 & 0.4072622 & 0.4148144

\end{tabular}
\caption{ $C$ and $C_{g_1}$ for two cylinders at distances $d= \kappa a$ for three different
thickness $b = \tau a$. The lower part of the table contains the extrapolated values using the procedure
\eqref{proceduraN}.
\label{tabtot4}}
\end{table}

% tabella 5

\begin{table}[!h]
\begin{tabular}{lcc|cc}
  & \multicolumn{2}{c}{$C$} & \multicolumn{2}{c}{$C_{g_1}$}\\[3pt]
  \cline{2-5}\\[-5pt]
$\kappa$ & $\tau=0.275$  & $\tau=0.3$  & $\tau=0.275$  & $\tau=0.3$  \\[5pt]

0.0001 & 2501.771 & 2501.785 & 0.4214673 & 0.4286859 \\
 0.0003 & 834.9404 & 834.9540 & 0.4214956 & 0.4287136 \\
 0.0005 & 501.5250 & 501.5394 & 0.4215270 & 0.4287444 \\
 0.0007 & 358.6149 & 358.6292 & 0.4215554 & 0.4287722 \\
 0.0009 & 279.2108 & 279.2243 & 0.4215840 & 0.4288001 \\
 0.001 & 251.4163 & 251.4299 & 0.4215983 & 0.4288141 \\
 0.003 & 84.57527 & 84.58885 & 0.4218890 & 0.4290986 \\
 0.005 & 51.16103 & 51.17455 & 0.4221820 & 0.4293853 \\
 0.007 & 36.82230 & 36.83578 & 0.4224705 & 0.4296676 \\
 0.009 & 28.84627 & 28.85971 & 0.4227596 & 0.4299505 \\
 0.01 & 26.05191 & 26.06534 & 0.4229051 & 0.4300930 \\
 0.015 & 17.65521 & 17.66855 & 0.4236242 & 0.4307972 \\
 0.02 & 13.44391 & 13.45718 & 0.4243387 & 0.4314972 \\
 0.03 & 9.214923 & 9.228048 & 0.4257588 & 0.4328892 \\
 0.04 & 7.087959 & 7.100962 & 0.4271647 & 0.4342672 \\
 0.05 & 5.804628 & 5.817483 & 0.4285523 & 0.4356297 \\
 0.06 & 4.944366 & 4.957093 & 0.4299292 & 0.4369816 \\
 0.07 & 4.326666 & 4.339270 & 0.4312912 & 0.4383197 \\
 0.08 & 3.861024 & 3.873509 & 0.4326399 & 0.4396454 \\
 0.09 & 3.497062 & 3.509430 & 0.4339756 & 0.4409590 \\
 0.1 & 3.204486 & 3.216745 & 0.4352992 & 0.4422610 \\
 0.15 & 2.316048 & 2.327805 & 0.4417402 & 0.4486057 \\
 0.2 & 1.862378 & 1.873704 & 0.4479148 & 0.4546988 \\
 0.3 & 1.397190 & 1.407807 & 0.4595754 & 0.4662293 \\
 0.4 & 1.157429 & 1.167483 & 0.4704543 & 0.4770099 \\
 0.5 & 1.009977 & 1.019571 & 0.4806690 & 0.4871491 \\
 0.6 & 0.9096393 & 0.9188476 & 0.4903037 & 0.4967247 \\
 0.7 & 0.8367240 & 0.8456023 & 0.4994196 & 0.5057952 \\
 0.8 & 0.7812334 & 0.7898263 & 0.5080637 & 0.5144042 \\
 0.9 & 0.7375358 & 0.7458785 & 0.5162717 & 0.5225857 \\
 1. & 0.7022076 & 0.7103284 & 0.5240721 & 0.5303672 
   \\ \hline\hline
 \multicolumn{5}{c}{Extrapolated values}\\ \hline

0.0001 & 2501.783 & 2501.798 & 0.4214673 & 0.4286859 \\
 0.0003 & 834.9432 & 834.9568 & 0.4214956 & 0.4287136 \\
 0.0005 & 501.5257 & 501.5402 & 0.4215270 & 0.4287444 \\
 0.0007 & 358.6151 & 358.6294 & 0.4215554 & 0.4287722 \\
 0.0009 & 279.2109 & 279.2244 & 0.4215840 & 0.4288001 \\
 0.001 & 251.4164 & 251.4300 & 0.4215983 & 0.4288141 \\
 0.003 & 84.57528 & 84.58886 & 0.4218890 & 0.4290986 \\
 0.005 & 51.16103 & 51.17455 & 0.4221820 & 0.4293853

\end{tabular}
\caption{ $C$ and $C_{g_1}$ for two cylinders at distances $d= \kappa a$ for three different
thickness $b = \tau a$. The lower part of the table contains the extrapolated values using the procedure
\eqref{proceduraN}.
\label{tabtot5}}
\end{table}

\end{document}